\documentclass[a4paper,onecolumn,11pt,unpublished]{quantumarticle}
\pdfoutput=1
\usepackage[utf8]{inputenc}
\usepackage[english]{babel}
\usepackage[T1]{fontenc}
\usepackage{amsmath,amssymb,mathtools}
\usepackage{amsthm}
\usepackage{booktabs}
\usepackage{array}
\usepackage{tikz}
\usepackage{algorithm}
\usepackage{algpseudocode}
\usepackage[numbers,sort&compress]{natbib}
\usepackage{hyperref}
\hypersetup{hidelinks}%

\newtheorem{theorem}{Theorem}[section]
\newtheorem{lemma}[theorem]{Lemma}
\newtheorem{proposition}[theorem]{Proposition}
\newtheorem{corollary}[theorem]{Corollary}

\theoremstyle{definition}
\newtheorem{definition}[theorem]{Definition}

\theoremstyle{remark}
\newtheorem{remark}[theorem]{Remark}
\numberwithin{equation}{section}

\newcommand{\C}{\mathbb{C}}
\newcommand{\R}{\mathbb{R}}
\newcommand{\N}{\mathbb{N}}
\newcommand{\Z}{\mathbb{Z}}
\newcommand{\F}{\mathbb{F}}
\newcommand{\Hil}{\mathcal{H}}

\DeclareMathOperator{\rank}{rank}

\DeclareMathOperator{\Tr}{Tr}
\DeclareMathOperator{\Img}{Im}
\newcommand{\ket}[1]{\lvert #1\rangle}
\newcommand{\bra}[1]{\langle #1\rvert}
\newcommand{\braket}[2]{\langle #1\vert #2\rangle}
\newcommand{\GHZ}{\mathrm{GHZ}}
\newcommand{\TTN}{\mathrm{TTN}}
\newcommand{\TTNo}{\mathrm{TTN}^{\circ}}
\newcommand{\clus}{\mathcal{C}}
\newcommand{\cluso}{\mathcal{C}^{\circ}}
\newcommand{\Xn}{\mathcal{X}_n}

\newcommand{\Ufam}{\mathcal{U}}
\newcommand{\Nfam}{\mathcal{N}}
\newcommand{\GOSGS}{\textup{\textsc{gosgs}}}
\newcommand{\gaps}{\varepsilon^{(s)}_{*}}
\newcommand{\gaptot}{\varepsilon^{\mathrm{tot}}_{*}}
\newcommand{\gapsk}{\varepsilon^{(s,k)}_{*}}
\newcommand{\hull}[1]{{#1}^{\langle s\rangle}}
\newcommand{\dH}{d_{\mathrm{H}}}
\newcommand{\PH}{P_{\mathrm{H}}}
\newcommand{\acc}{\mathrm{acc}}
\newcommand{\rej}{\mathrm{rej}}
\newcommand{\RT}{\mathrm{RT}}
 
\begin{document}

\title{Learning the topology of tree tensor networks with optimal copy complexity}
\author{Ofek I. Cohen}
\affiliation{Tel Aviv University}
\maketitle

\begin{abstract}
How many copies of an unknown quantum state are needed to learn the tree on which it is a tree tensor network? We study exact pure states of $n$ qudits with Schmidt rank at most $\chi$ across every internal edge of an unknown tree of arbitrary degree, under a known gap $\varepsilon_0$ across every wrong merge of at most $s$ clusters. Such a gap determines the tree among the trees with at most one vertex of degree above $s+1$. On the resulting class, $\Theta(\log(n/\delta)/\varepsilon_0)$ copies are necessary and sufficient for success probability $1-\delta$, for fixed $d$, $\chi$ and $s$ with $2\le\chi\le d^2-1$, $\delta\le\frac13$ and $\varepsilon_0$ below a constant. For $d^2\le\chi\le d^3-1$ the learnable object is a coarsening of the tree, and the same rate holds for prime $d$, $s\ge4$ and $n\ge16$ divisible by four.
 \end{abstract}

\setcounter{tocdepth}{2}
\tableofcontents

\section{Introduction}\label{sec:intro}
A tree tensor network represents a pure state of $n$ sites by placing a tensor at each internal vertex of a tree whose leaves correspond to the sites~\cite{SDV06,HK09,Hack12}. On a given tree, the minimal bond dimension of an edge is the Schmidt rank of the state across the cut that the edge defines~\cite{HK09,Gra10,Hack12}, and the cost of storing and contracting the network grows with these ranks. The same state may require small bond dimensions on one tree topology and bond dimensions exponential in $n$ on another, so the topology matters. Existing algorithms that learn or test such states from copies take the tree as an input: tomography and agnostic learning on a known graph~\cite{CMS26}, proper agnostic learning on a known tree~\cite{CC26}, learners along a known chain~\cite{CPFS10,KMDS24,LCH25}, and the membership tester of Lovitz and Lowe~\cite{LL26}, who remark that little is known about learning more complicated tensor-network states. The tree is therefore the natural first thing to learn. The existing methods for choosing it work from a classical description of the state, by heuristics based on entanglement measures, mutual information or variational optimization~\cite{BG14,MVLN10,HUOH23,HOK25,WU24,SCLM26}, and they do not guarantee recovery. In experimental settings, moreover, a classical description may be unavailable, with access limited to measurements on prepared copies.

Caro, McHugh and Strelchuk state the missing step as a problem: recover, from copies of an unknown quantum state, a graph on which the state admits a representation with bounded bond dimension~\cite[\S1.4]{CMS26}. We address the case of trees: given copies of an unknown pure state that admits an exact representation on some tree with bond dimension at most $\chi$, we recover that tree, assuming a known positive lower bound on a gap, and we determine the number of copies that this needs, up to constant factors. The internal vertices of the tree may have any degree. This generality is needed to cover the models just cited. There every vertex of the tree holds a physical site, whereas here only the leaves do. To bring such a tree into our form, the site of each internal vertex is hung from it as a new leaf, which raises the degree of that vertex by one, while a leaf keeps its own site. A binary tree of those models therefore becomes a tree with vertices of degree four, and a generic state of those models is not a binary tree tensor network in our sense.

\subsection{The problem}\label{sec:s1-problem}
There are $n\ge4$ sites, each with the Hilbert space $\C^d$, $d\ge2$. An X-tree $T$ with leaf set $[n]:=\{1,\dots,n\}$ is a tree whose leaves are labeled by the sites and whose internal vertices have degree at least~$3$. Removing an edge partitions the leaves into two sets, the \emph{clusters} of $T$. For an integer $\chi\ge1$, $\TTNo(T,\chi)$ is the set of pure states whose Schmidt rank across every internal edge of $T$ is at most $\chi$ (Definition~\ref{def:ttn}), and we say that $T$ \emph{carries} each state of this set. Given copies of an unknown pure state $\psi$, the task is to output a tree $T$ that carries $\psi$. Contracting an internal edge of $T$ gives a \emph{coarsening} $T'$ of $T$ with $\TTNo(T,\chi)\subseteq\TTNo(T',\chi)$, since $T'$ imposes one rank constraint fewer. The object to learn is therefore the finest tree $T$ with $\psi\in\TTNo(T,\chi)$ within a class of candidate trees that is fixed below.

We learn the tree bottom-up. The learner keeps a partition of the sites into disjoint groups, called \emph{parts}, together with a subtree for each part whose leaves are the sites of that part. Initially, every site is a part on its own. At each step, the learner chooses several parts and applies a Schmidt-rank test to their union. If the test accepts, the learner replaces these parts by the single part formed by their union and joins their subtrees under a new vertex. Every accepted union thus becomes a cluster of the output tree, so the test should accept exactly the unions that are clusters of the true tree. When the test rejects every union that the learner examines, the learner stops and joins the subtrees of the remaining parts at one final vertex.

\paragraph{Quantitative identifiability assumption.}
Let $v$ be an internal vertex of the true tree, of degree $k$, and let $B_1,\dots,B_k$ be the sets of sites in the $k$ components that remain when $v$ is removed. A union of between two and $k-2$ of these sets is not a cluster of the true tree. Unless $v$ is the vertex formed last, $v$ must therefore be created by merging $k-1$ parts at once. We fix an integer $s\ge2$, the largest number of parts merged at once, so that the learner can create every vertex of degree up to $s+1$ and must leave a vertex of larger degree to the final step. We make a \emph{quantitative identifiability assumption} on the unions of at most $s$ clusters: $\psi$ has \emph{$s$-gap} at least $\varepsilon_0>0$, a number known to the learner. That is, there is a tree $T^*$ with $\psi\in\TTNo(T^*,\chi)$ such that, whenever the union $A$ of at most $s$ pairwise disjoint clusters of $T^*$ is neither a cluster nor $[n]$, $\psi$ is at squared distance at least $\varepsilon_0$ from every vector whose Schmidt rank across the cut between $A$ and its complement is at most $\chi$ (Definitions~\ref{def:gap} and~\ref{def:classes}).

Call an internal vertex of degree above $s+1$ a \emph{hub}. Among the trees with at most one hub, at most one satisfies the assumption, and it is the finest of these trees with $\psi\in\TTNo(T,\chi)$ (Theorem~\ref{thm:ident}). The candidate trees are therefore the trees with at most one hub. For binary trees and $s=2$ the assumption is a gap across every wrong merge of two clusters, and it is analogous to the promise in hidden-cut learning that each factor of an unknown product partition is far from any further product decomposition~\cite{BGW25,HEC26,LCEB26}, and to the lower bounds on edge strength in learning classical and quantum interaction graphs~\cite{SW12,FRS26}. Across a single cut, it is the promise of rank testing, rank at most $\chi$ or far from it~\cite{OW21,LL26}.
\subsection{Contributions}\label{sec:s1-contrib}
\paragraph{Identifiability.}
Let $\Xn(s)$ be the class of trees with at most one hub. A tree $T^*\in\Xn(s)$ that satisfies the assumption is the unique finest tree in $\Xn(s)$ that carries $\psi$, and no other tree of $\Xn(s)$ satisfies the assumption (Theorem~\ref{thm:ident}).

\paragraph{Optimal copy complexity.}
The learner $\GOSGS(s)$ of Algorithm~\ref{alg:gosgs} merges clusters from the bottom up whenever a one-sided rank test by weak Schur sampling accepts the union of at most $s$ current parts, examining small unions first, and stops when no union accepts. If every test answers correctly, it outputs $T^*$ with the smallest subtree containing all its hubs contracted to a vertex, which is $T^*$ itself exactly when $T^*\in\Xn(s)$ (Theorem~\ref{thm:core}). Run with bit-only Lüders tests on a single batch held in quantum memory, it needs $O((\chi^2+\chi\,(s\log(en/s)+\log(1/\delta)))/\varepsilon_0)$ copies in all (Theorem~\ref{thm:onebatch}), and it makes at most $N_T(s)=O(n^s)$ tests (Lemma~\ref{lem:counting}). We prove a matching information-theoretic lower bound for every procedure, even with adaptive collective measurements, on the class of gapped states of trees in $\Xn(s)$, when $2\le\chi\le d^2-1$, $\delta\le\frac13$ and $\varepsilon_0$ is below an explicit constant (Theorem~\ref{thm:lower}). Together, the least number $N_{\min}$ of copies of a procedure that knows $\varepsilon_0$ and $\delta$ and succeeds with probability $1-\delta$ on the class satisfies
\begin{equation*}
N_{\min}=\Theta\Bigl(\frac{\log(n/\delta)}{\varepsilon_0}\Bigr)
\end{equation*}
for fixed $d$, $\chi$ and $s$ (Corollary~\ref{cor:rate}). For $\chi\ge d^2$ every pair of sites has rank at most $\chi$, and the learnable object is the coarsening $T^{(k)}$ of the tree obtained by contracting every internal edge with a side of at most $k:=\lfloor\log_d\chi\rfloor$ sites. The learner, restricted to unions in a window of sizes, learns $T^{(k)}$ from one batch at the same rate when $T^{(k)}$ has at most one hub (Theorem~\ref{thm:largebond}), and for $d^2\le\chi\le d^3-1$, prime $d$, $s\ge4$ and $n\ge16$ divisible by four, the rate is again optimal (Theorem~\ref{thm:lower-k}).

\subsection{Related work}\label{sec:s1-related}
Lovitz and Lowe test membership in the class of a given tree with commuting rank tests on one batch~\cite{LL26}, and the quantum OR of Harrow, Lin and Montanaro shares one batch among one-sided tests of all cuts~\cite[Theorem~7]{HLM17}. Our tests are chosen adaptively and need not commute, and our lower bounds concern learning which tree carries the state, not testing a given tree. Learners on a known tree have costs that grow quickly with its maximum degree $\Delta$, the learner of Cedillo Vayson de Pradenne and Cotler~\cite{CC26} through an embedding into matrix product states whose bond dimension grows exponentially with $\Delta$, and the classical sketching learner of Tang et al.\ through a number of sketch functions growing as $n^{\Delta-1}$~\cite[Corollary~31]{THKY23}. In our learner the second largest degree $\Delta_2$ enters only the number of tests, $O(n^s)$ with $s\ge\Delta_2-1$, and the number of copies depends only logarithmically on that number.

Classically, Ballani and Grasedyck build the binary trees of the hierarchical tensor format bottom-up, merging a bounded number of parts at a time according to the ranks of a given tensor across cuts, without a recovery guarantee~\cite{BG14}. For bond dimension one, hidden-cut learning locates a product cut~\cite{BGW25}. To our knowledge, no learner was known that provably finds the tree of a gapped tree tensor network from a number of copies polynomial in $n$, and no lower bound on the number of copies that this task needs.
\section{Setting and notation}\label{sec:setting}
\subsection{States, cuts and tails}\label{sec:s2-states}
We consider $n\ge4$ \emph{sites} $[n]:=\{1,\dots,n\}$. For $n\le3$ there is only one tree of the kind defined below with leaf set $[n]$, the star, and nothing is to be learned. The Hilbert space of site $i$ is $\Hil_i:=\C^d$, $d\ge2$, with the standard basis $\ket0,\dots,\ket{d-1}$. For $A\subseteq[n]$, $\Hil_A:=\bigotimes_{i\in A}\Hil_i$, with $\Hil_\emptyset:=\C$, $\Hil:=\Hil_{[n]}$ and $A^c:=[n]\setminus A$. The \emph{product basis} of $\Hil_A$ consists of the vectors $\ket x:=\bigotimes_{i\in A}\ket{x_i}$ with $x\in\{0,\dots,d-1\}^A$. For $x\in\{0,\dots,d-1\}^A$ and $y\in\{0,\dots,d-1\}^{A^c}$ let $\ket{x,y}:=\bigotimes_{i=1}^n\ket{z_i}$, where $z_i:=x_i$ for $i\in A$ and $z_i:=y_i$ for $i\in A^c$. For $\varphi\in\Hil$, not necessarily normalized, the \emph{flattening} of $\varphi$ across the cut $A|A^c$ and its rank are
\begin{equation*}
\varphi^{(A)}:=\bigl(\braket{x,y}{\varphi}\bigr)_{x,y},\qquad r_\varphi(A):=\rank\varphi^{(A)}=r_\varphi(A^c),
\end{equation*}
where $x$ and $y$ run over $\{0,\dots,d-1\}^A$ and $\{0,\dots,d-1\}^{A^c}$. The flattening is the matricization of~\cite[Section~3]{Gra10}. A product vector $\varphi_1\otimes\dots\otimes\varphi_n$ has $r_\varphi(A)\le1$ for every $A$.

The unknown state is a unit vector $\psi\in\Hil$, and $r(A):=r_\psi(A)$ is its \emph{Schmidt rank} across the cut $A|A^c$. The reduced state $\rho_A:=\Tr_{A^c}\ket\psi\bra\psi$ has the eigenvalues $\alpha^A_1\ge\dots\ge\alpha^A_{d^{|A|}}$, and we put $\alpha^A_j:=0$ for $j>d^{|A|}$. By the Schmidt decomposition, $\rho_A$ and $\rho_{A^c}$ have the same nonzero eigenvalues, with the same multiplicities, and differ only in how many eigenvalues are zero. Moreover, $r(A)=\rank\rho_A$. For a density matrix $\rho$ with eigenvalues $\alpha_1\ge\alpha_2\ge\cdots$ and an integer $\chi\ge1$, the \emph{$\chi$-tail} is $\varepsilon(\rho;\chi):=\sum_{i>\chi}\alpha_i$, and $\varepsilon_A(\chi):=\varepsilon(\rho_A;\chi)$, written $\varepsilon_A(\psi;\chi)$ when the state must be named. Thus $\varepsilon_A(\chi)=\varepsilon_{A^c}(\chi)\le1-\alpha^A_1<1$, and $\varepsilon_A(\chi)=0$ if and only if $r(A)\le\chi$.

\begin{lemma}[Eckart--Young \cite{EY36}]\label{lem:kyfan}
Let $\psi\in\Hil$ be a unit vector, $A\subseteq[n]$, $\chi\ge1$, and let $\varphi\in\Hil$ satisfy $r_\varphi(A)\le\chi$. Then
\begin{equation*}
\lVert\psi-\varphi\rVert^2\ge\varepsilon_A(\chi),
\end{equation*}
with equality if $\varphi$ is the sum of the $\chi$ largest terms of a Schmidt decomposition of $\psi$ across $A|A^c$. Hence $\varepsilon_A(\chi)=\min\{\lVert\psi-\varphi\rVert^2:\ \varphi\in\Hil,\ r_\varphi(A)\le\chi\}$, and $|\varepsilon_A(\psi;\chi)^{1/2}-\varepsilon_A(\psi';\chi)^{1/2}|\le\lVert\psi-\psi'\rVert$ for all unit vectors $\psi,\psi'\in\Hil$.
\end{lemma}

\begin{proof}
The vector norm is the Frobenius norm of the flattening, so the inequality and its equality case are the Eckart--Young theorem~\cite{EY36}, which holds over $\C$ with the same proof. The last claim holds because the distance to a set is $1$-Lipschitz.
\end{proof}

The proof uses only the flattening, so the lemma holds for unit vectors in any finite tensor product of finite-dimensional spaces.

\subsection{Trees, clusters and hubs}\label{sec:s2-trees}
Following phylogenetics, where $X$ denotes a finite set of labels, an \emph{X-tree} with leaf set $[n]$ is a finite tree whose vertices of degree~$1$, the \emph{leaves}, are labeled bijectively by $[n]$, and whose other vertices, the \emph{internal} vertices, have degree at least~$3$. Semple and Steel~\cite[Definition~2.1.2]{SS03} call these trees phylogenetic $X$-trees, with $X=[n]$. Two X-trees are \emph{equal} if a graph isomorphism maps each leaf to the leaf with the same label. $\Xn$ is the set of X-trees with leaf set $[n]$, and an X-tree is \emph{binary} if all its internal vertices have degree~$3$. The \emph{star} has a single internal vertex. An edge is \emph{internal} if both of its ends are internal, and a \emph{leaf edge} if one of its ends is a leaf. For an edge $uv$ of $T$, $D(u\to v)$ denotes the set of leaves in the component of $T-uv$ that contains $v$. The sets $D(u\to v)$ and $D(v\to u)$ are the \emph{sides} of $uv$, and the unordered pair $\{D(u\to v),D(v\to u)\}$, a partition of $[n]$ into two parts, is its \emph{split}~\cite[Definition~3.1.1]{SS03}. A \emph{cluster} of $T$ is a side of an edge of $T$, and $\clus(T)$ is the set of clusters. For the hierarchical tensor format, Grasedyck calls the two sides of an edge a mode cluster and its complementary cluster~\cite[Section~3]{Gra10}. An \emph{internal cluster} is a side of an internal edge, and $\cluso(T)$ is the set of internal clusters. A \emph{branch} of an internal vertex $v$ is a set $D(v\to x)$ with $x$ a neighbor of $v$. The branches of $v$ partition $[n]$. By Lemma~\ref{lem:A-trees}, the sides of distinct edges are distinct, the singletons and their complements are clusters of every X-tree, $\cluso(T)=\{A\in\clus(T):2\le|A|\le n-2\}$ is closed under complements, any two clusters $Y,Z$ are \emph{compatible}, that is, one of $Y\cap Z$, $Y\setminus Z$, $Z\setminus Y$ and $[n]\setminus(Y\cup Z)$ is empty~\cite[Definition~3.1.3]{SS03}, and $T$ is determined by $\cluso(T)$. An X-tree has at most $n-2$ internal vertices and at most $2n-3$ edges, with equality exactly for binary trees.

\paragraph{Coarsening.}
Contracting an internal edge $uv$, that is, identifying $u$ with $v$ and removing the edge, gives an X-tree whose clusters are those of $T$ except the two sides of $uv$. We write $T'\preceq T$, and call $T'$ a \emph{coarsening} of $T$, if $T'$ arises from $T$ by contracting a set of internal edges. Equivalently, $\clus(T')\subseteq\clus(T)$ (Lemma~\ref{lem:A-trees}(e)). The relation $\preceq$ is a partial order on $\Xn$, the star is a coarsening of every tree, and the binary trees are the maximal elements. This is the partial order of Semple and Steel~\cite[Section~3.2]{SS03}, who say that $T$ refines $T'$.

\paragraph{Degrees and hubs.}
$\Delta_2(T)$ is the second largest degree of an internal vertex of $T$, that is, the largest degree among the internal vertices other than one vertex of maximum degree, with $\Delta_2(T):=0$ for the star. Throughout, $s\ge2$ is a fixed integer, the largest number of parts that the learner merges at once. A \emph{hub} of $T$ is an internal vertex of degree at least $s+2$, and
\begin{equation*}
\Xn(s):=\{T\in\Xn:\ T\text{ has at most one hub}\}=\{T\in\Xn:\ \Delta_2(T)\le s+1\}.
\end{equation*}
Thus every binary tree lies in $\Xn(2)$, every tree of maximum degree at most $s+1$ lies in $\Xn(s)$, and so does every tree with a single vertex of large degree, such as the star. The \emph{hub hull} of $T$ is the minimal subtree of $T$ that contains all hubs, which has no edges if $T$ has at most one hub, and the \emph{hull contraction} $\hull{T}$ is $T$ with all edges of its hub hull contracted. Thus $\hull{T}\preceq T$, with $\hull{T}=T$ if and only if $T\in\Xn(s)$.
\subsection{Tree tensor networks, union families and gaps}\label{sec:s2-ttn}
\begin{definition}[Tree tensor networks]\label{def:ttn}
For $T\in\Xn$ and an integer $\chi\ge1$,
\begin{gather*}
\TTN(T,\chi):=\{\varphi\in\Hil:\ r_\varphi(A)\le\chi\ \text{for all}\ A\in\clus(T)\},\\
\TTNo(T,\chi):=\{\varphi\in\Hil:\ r_\varphi(A)\le\chi\ \text{for all}\ A\in\cluso(T)\}.
\end{gather*}
\end{definition}

The set $\TTNo(T,\chi)$ leaves the ranks $r_\varphi(\{i\})\le d$ of the leaf edges unconstrained, so $\TTN(T,\chi)\subseteq\TTNo(T,\chi)$, with equality if $\chi\ge d$. Although Definition~\ref{def:ttn} uses only ranks, the name \emph{tree tensor network} is justified: by the theory of hierarchical tensor formats~\cite{HK09,Gra10,Hack12}, which allows dimension trees of any degree~\cite[Section~3]{Gra10}, $\TTN(T,\chi)$ is exactly the set of vectors that tree tensor networks on $T$ with all bond dimensions at most $\chi$ can represent, with a tensor of order $k$ at each internal vertex of degree $k$. We say that $T$ \emph{carries} $\varphi$ if $\varphi\in\TTNo(T,\chi)$, and that a cut $A|A^c$ is \emph{cheap} for $\varphi$ if $r_\varphi(A)\le\chi$. If $T'\preceq T$, then $\TTN(T,\chi)\subseteq\TTN(T',\chi)$ and $\TTNo(T,\chi)\subseteq\TTNo(T',\chi)$, since $T'$ imposes fewer constraints. In particular the star carries every vector, and the set of trees that carry $\varphi$ is closed under coarsening. The object to learn is therefore a \emph{finest} tree that carries $\psi$ within a class of trees, that is, a tree of the class that carries $\psi$ and is a coarsening of no other such tree. Such a tree need not be unique, as a state can have two incompatible cheap cuts, and Section~\ref{sec:ident} shows that it is unique within $\Xn(s)$ under the $s$-gap.

\paragraph{Union families.}
A bottom-up learner keeps a partition of $[n]$, initially into singletons, and replaces several of its parts by their union. To decide which parts to merge, it examines unions of at most $s$ parts. For $T\in\Xn$, the \emph{$s$-union family} $\Ufam_s(T)$ consists of the sets $A\ne[n]$ that are the union of $k$ pairwise disjoint clusters of $T$ for some $2\le k\le s$. The whole set $[n]$ is excluded because it is not a cut: it is, for example, the union of a cluster and its complement, and the learner never examines it, since at least two parts remain after every merge. The \emph{$s$-wrong-merge family} consists of the sets of $\Ufam_s(T)$ that are not clusters:
\begin{equation*}
\Nfam_s(T):=\Ufam_s(T)\setminus\clus(T).
\end{equation*}
As long as all parts of the learner's partition are clusters of $T$, each union it examines lies in $\Ufam_s(T)$, and a union in $\Nfam_s(T)$ is a \emph{wrong merge}: after it, the learner can no longer output $T$. The families grow with $s$. Their union $\Nfam_\infty(T):=\bigcup_{s\ge2}\Nfam_s(T)$ is the set of all nonempty proper subsets of $[n]$ that are not clusters: such a set has at least two elements, as singletons are clusters, so it is the union of its at least two singletons. Every $A\in\Nfam_s(T)$ has $2\le|A|\le n-2$, and $\Nfam_2(T)$ contains a pair of leaves (Lemma~\ref{lem:A-trees}(f)). For $s\ge3$ the decomposition of a member of $\Ufam_s(T)$ into disjoint clusters need not be unique, which is why the counting of Section~\ref{sec:learner} counts tests by the tested set, not by its decomposition.

\begin{definition}[Gaps]\label{def:gap}
Let $T\in\Xn$, $\chi\ge1$, and let $\psi\in\Hil$ be a unit vector. The \emph{$s$-gap} and the \emph{total gap} of $\psi$ with respect to $T$ are
\begin{equation*}
\gaps(\psi,T):=\gaps(\psi,T;\chi):=\min_{A\in\Nfam_s(T)}\varepsilon_A(\chi),\qquad\gaptot(\psi,T):=\min_{A\in\Nfam_\infty(T)}\varepsilon_A(\chi).
\end{equation*}
\end{definition}

By Lemma~\ref{lem:kyfan}, $\varepsilon_A(\chi)$ is the least squared distance from $\psi$ to a vector of Schmidt rank at most $\chi$ across $A|A^c$. So, for a number $\varepsilon_0>0$, the inequality $\gaps(\psi,T)\ge\varepsilon_0$ says exactly that $\psi$ is at squared distance at least $\varepsilon_0$ from every vector of rank at most $\chi$ across some set of $\Nfam_s(T)$. Section~\ref{sec:s1-problem} states the quantitative identifiability assumption in this form. The $s$-gap is nonincreasing in $s$, $\gaptot\le\gaps\le\varepsilon^{(2)}_*$, and $\gaps(\psi,T)<1$. The minimum runs over $\Nfam_s(T)$ and not over all cuts because, while its parts are clusters of $T$, the learner of Section~\ref{sec:learner} tests only clusters of $T$ and sets of $\Nfam_s(T)$ (Theorem~\ref{thm:core}). A gap on these sets therefore suffices for it. The total gap is the promise of identifiability across all cuts. If $T$ carries $\psi$ and the total gap of $\psi$ with respect to $T$ is positive, the cuts that are cheap for $\psi$ and have at least two sites on each side are exactly those of the internal clusters of $T$. So $\psi$ determines $T$, whatever its degrees. The $s$-gap is a weaker promise, and Theorem~\ref{thm:ident} shows that it determines the tree within $\Xn(s)$, the trees with at most one hub. Section~\ref{sec:ident} then defines the gapped classes.
\subsection{Procedures}\label{sec:s2-procedures}
The cost of a procedure in copies is the number of copies of $\psi$ that it consumes. A measurement is \emph{collective} if it acts jointly on several copies. A \emph{register} $\mathsf X$ is a finite-dimensional quantum system with Hilbert space $\Hil_{\mathsf X}$ and states
\begin{equation*}
\rho_{\mathsf X}\in\mathcal D(\Hil_{\mathsf X}):=\{\rho\in\mathcal L(\Hil_{\mathsf X}):\rho\ge0,\ \Tr\rho=1\},
\end{equation*}
where $\mathcal L(\Hil_{\mathsf X})$ is the set of linear operators on $\Hil_{\mathsf X}$. Two registers $\mathsf X$ and $\mathsf Y$ form the register $\mathsf{XY}$ with $\Hil_{\mathsf{XY}}:=\Hil_{\mathsf X}\otimes\Hil_{\mathsf Y}$~\cite[Section~2.1]{Wat18}. For example, the sites of $A$ in one copy of $\psi$ form its \emph{$A$-register}, and the $A$-registers of $m$ copies form a register $\mathsf X$ with $\Hil_{\mathsf X}=\Hil_A^{\otimes m}$ and $\rho_{\mathsf X}=\rho_A^{\otimes m}$. A \emph{quantum instrument}~\cite[Section~2.3.2]{Wat18} from a register $\mathsf X$ to a register $\mathsf X'$ with a finite outcome set $O$ is a family $\{\Phi_o\}_{o\in O}$ of completely positive maps $\Phi_o:\mathcal L(\Hil_{\mathsf X})\to\mathcal L(\Hil_{\mathsf X'})$ whose sum is trace preserving. Applied to $\omega\in\mathcal D(\Hil_{\mathsf X})$, it has outcome $o$ with probability $\Tr\Phi_o(\omega)$ and then leaves the state $\Phi_o(\omega)/\Tr\Phi_o(\omega)$. For an orthogonal projector $\Pi$ on $\Hil_{\mathsf X}$, the \emph{Lüders instrument}~\cite{Lue51} $\{\Pi,1-\Pi\}$ from $\mathsf X$ to $\mathsf X$ has the outcomes $1$ and $0$, with $\Phi_1(\omega)=\Pi\omega\Pi$ and $\Phi_0(\omega)=(1-\Pi)\omega(1-\Pi)$.

\begin{definition}[Procedures]\label{def:procedures}
Let $\mathcal O\supseteq\Xn$ be a finite set of outputs.
An \emph{$N$-copy procedure}, $N\ge0$, has rounds $j=1,\dots,R$ with finite outcome sets $O_j$. Its register $\mathsf X$ initially holds $N$ copies of $\psi$, with $\Hil_{\mathsf X}=\Hil^{\otimes N}$ and $\rho_{\mathsf X}=\ket\psi\bra\psi^{\otimes N}$, and it may be enlarged by ancillas and reduced by partial traces. In round $j$ the procedure applies a quantum instrument $\{\Phi^{(j)}_{o_j|o_{<j}}\}_{o_j\in O_j}$, chosen as a function of the earlier outcomes $o_{<j}:=(o_1,\dots,o_{j-1})$. At the end it outputs $\mathrm{out}(o_1,\dots,o_R)\in\mathcal O$. The record $o$ has probability $\Tr\Phi_o(\ket\psi\bra\psi^{\otimes N})$, where $\Phi_o:=\Phi^{(R)}_{o_R|o_{<R}}\circ\dots\circ\Phi^{(1)}_{o_1}$.
\end{definition}

The model allows collective measurements on all copies, ancillas, quantum memory, adaptive intermediate measurements, classical randomness and classical post-processing.

\subsection{The one-sided rank test}\label{sec:s2-ranktest}
Let $D,m\ge1$ and let $\rho$ be a density matrix on $\C^D$ with eigenvalues $\alpha_1\ge\dots\ge\alpha_D\ge0$. The symmetric group $\mathfrak S_m$ is the group of bijections $\pi:[m]\to[m]$. For $\pi\in\mathfrak S_m$, $R_\pi$ is the linear map on $(\C^D)^{\otimes m}$ with
\begin{equation*}
R_\pi(u_1\otimes\dots\otimes u_m):=u_{\pi^{-1}(1)}\otimes\dots\otimes u_{\pi^{-1}(m)}\qquad\text{for all }u_1,\dots,u_m\in\C^D,
\end{equation*}
which moves the factor in position $j$ to position $\pi(j)$. So $R_{\pi\sigma}=R_\pi R_\sigma$ for all $\pi,\sigma\in\mathfrak S_m$, and $\pi\mapsto R_\pi$ is a representation of the group $\mathfrak S_m$ on $(\C^D)^{\otimes m}$ by unitary operators. For a partition $\lambda\vdash m$, a Young diagram with $m$ boxes and $\ell(\lambda)$ rows, let $\mathrm{ch}^\lambda$ be the irreducible character of $\mathfrak S_m$ labeled by $\lambda$, $\mathrm{ch}^\lambda:\mathfrak S_m\to\C$, $\mathrm{ch}^\lambda(\pi):=\Tr U_\lambda(\pi)$ for every $\pi\in\mathfrak S_m$, where $U_\lambda$ is the corresponding irreducible representation. Let $f^\lambda:=\mathrm{ch}^\lambda(\mathrm{id})$ and $\Pi_\lambda:=\frac{f^\lambda}{m!}\sum_{\pi\in\mathfrak S_m}\mathrm{ch}^\lambda(\pi)R_\pi$, the orthogonal projector onto the $\lambda$-isotypic component of $\pi\mapsto R_\pi$. \emph{Weak Schur sampling} $\mathrm{WSS}_m$~\cite{CHW07,OW21} is the projective measurement $\{\Pi_\lambda\}_{\lambda\vdash m}$, a collective measurement of the $m$ copies. For an integer $\chi\ge1$, put $\Pi_{\mathrm{acc}}:=\sum_{\lambda\vdash m,\ \ell(\lambda)\le\chi}\Pi_\lambda$ and $\Pi_{\mathrm{rej}}:=1-\Pi_{\mathrm{acc}}$. The \emph{rank test} $\mathrm{RT}(\chi,m)$ is the Lüders instrument $\{\Pi_{\mathrm{acc}},\Pi_{\mathrm{rej}}\}$. It is \emph{bit-only}: it reveals only whether it accepts. Running $\mathrm{WSS}_m$ and accepting if and only if $\ell(\lambda)\le\chi$, the rank tester of O'Donnell and Wright \cite[Section~6.1]{OW21}, accepts with the same probability, but in general leaves a different post-measurement state. The \emph{acceptance probability} is $a_m(\rho):=\Tr(\Pi_{\mathrm{acc}}\rho^{\otimes m})$. For $A\subseteq[n]$, $\mathrm{RT}(\chi,m)$ \emph{on $A$} acts on the $A$-registers of $m$ copies, with $D=d^{|A|}$, and as the identity on all other registers. On $m$ fresh copies of $\psi$ it accepts with probability $a_m(A):=a_m(\rho_A)$.

For $\varepsilon_0\in(0,1]$ and $\delta'\in(0,1)$, the \emph{thresholds} of the rank test are
\begin{align*}
m_R(\chi,\varepsilon_0,\delta')&:=(\chi+1)\Bigl\lceil\frac{2(\chi+1)}{\varepsilon_0}\ln\frac{\chi+1}{\delta'}\Bigr\rceil,\\
m_R^+(\chi,\varepsilon_0,\delta')&:=\Bigl\lceil\frac{2(\chi+1)}{\varepsilon_0}\max\Bigl\{2\chi,\,8\ln\frac1{\delta'}\Bigr\}\Bigr\rceil\ \le\ \Bigl\lceil\frac{4\chi(\chi+1)+16(\chi+1)\ln\frac1{\delta'}}{\varepsilon_0}\Bigr\rceil.
\end{align*}

\begin{lemma}[One-sided rank test]\label{lem:ranktest}
Let $D,m,\chi\ge1$, let $\rho$ be a density matrix on $\C^D$, and $\varepsilon:=\varepsilon(\rho;\chi)$.
\begin{enumerate}
\item[(a)] \emph{Completeness.} If $\rank\rho\le\chi$, then $a_m(\rho)=1$.
\item[(b)] \emph{Spectrum.} $a_m(\rho)$ depends only on the nonzero eigenvalues of $\rho$.
\item[(c)] \emph{Soundness.} If $\varepsilon>0$ and $0<\mu\le\varepsilon/(2\chi+1)$, then
\begin{equation*}
a_m(\rho)\le\Pr[\mathrm{Bin}(m,\mu)\le\chi]\le(\chi+1)(1-\mu)^{\lfloor m/(\chi+1)\rfloor}\le(\chi+1)e^{-\mu\lfloor m/(\chi+1)\rfloor}.
\end{equation*}
If $0<\varepsilon_0\le\varepsilon$ and $\delta'\in(0,1)$, then $a_m(\rho)\le\delta'$ for every $m\ge m_R(\chi,\varepsilon_0,\delta')$ and for every $m\ge m_R^+(\chi,\varepsilon_0,\delta')$.
\item[(d)] \emph{Implementation.} With the rotations of the quantum Fourier transform over $\mathfrak S_m$ counted as exact gates, $\mathrm{WSS}_m$ is implemented by a circuit of $\mathrm{poly}(m)\cdot O(\log D)$ gates, and so is the Lüders instrument $\mathrm{RT}(\chi,m)$, by a circuit that computes the bit $[\ell(\lambda)\le\chi]$ into an ancilla, measures it and uncomputes.
\end{enumerate}
\end{lemma}

Appendix~\ref{app:ranktest} proves the lemma and gives the exact sources. Items (a) and (c) are due to, or refine, O'Donnell and Wright~\cite{OW21}, and (d) uses Harrow's construction~\cite{Har05}. The threshold $m_R^+$ is $O((\chi^2+\chi\log\frac1{\delta'})/\varepsilon_0)$, and for $\chi\ge2$ and constant $\delta'\le\frac13$ it is optimal up to a constant factor among one-sided tests that work in every dimension, by the lower bound of O'Donnell and Wright~\cite[Lemma~6.2(ii) and Theorem~1.11]{OW21}.

\paragraph{The rank test on a cut.}
Let $A\subseteq[n]$, and apply the lemma with $\rho=\rho_A$ and $D=d^{|A|}$. By (a), if $r(A)\le\chi$, then $a_m(A)=1$, so $\mathrm{RT}(\chi,m)$ on $A$ accepts $\psi^{\otimes m}$ with certainty. It then leaves $\psi^{\otimes m}$ unchanged, because an orthogonal projector $\Pi$ and a unit vector $\varphi$ with $\lVert\Pi\varphi\rVert=1$ satisfy $\Pi\varphi=\varphi$. By (b), $a_m(A)=a_m(A^c)$, as $\rho_A$ and $\rho_{A^c}$ have the same nonzero eigenvalues. The test on $A$ may therefore be applied to the smaller of $A$ and $A^c$. Then $D\le d^{\lfloor n/2\rfloor}$, and by (d) a test costs $\mathrm{poly}(m)\cdot O(n\log d)$ gates.

\section{Short unions: identifiability and the graded scan}\label{sec:ident}
For binary trees, identifiability rests on one fact: every binary tree $T'\ne T$ has a cluster that $T$ lacks, and a minimal such cluster is the union of two clusters common to both trees. In a general tree the vertex below such a cluster may have many children, so the union has many parts, and the hub condition is what keeps their number small.

\subsection{Short unions and the identifiability theorem}\label{sec:s3-ident}
Root an X-tree $T'$ at an internal vertex $\rho$. For each vertex $u\ne\rho$ let $A_u$ be the set of leaves of the subtree below $u$, that is, $A_u=D(p\to u)$ where $p$ is the parent of $u$. Every cluster of $T'$ is some $A_u$ or its complement.

\begin{lemma}[Short unions]\label{lem:shortunion}
Let $T\in\Xn$, and let $T'\in\Xn$ be rooted at an internal vertex $\rho$. Let $u\ne\rho$ be an internal vertex of $T'$ such that $A_u$ is not a cluster of $T$ but $A_x$ is a cluster of $T$ for every child $x$ of $u$. Then $A_u$ is the disjoint union of the $\deg(u)-1$ sets $A_x$, which are clusters of both $T$ and $T'$, so $A_u\in\Nfam_{\deg(u)-1}(T)\cap\cluso(T')$. Such a vertex $u$ exists whenever some $A_v$ with $v$ internal is not a cluster of $T$: take such a $v$ farthest from $\rho$.
\end{lemma}

\begin{lemma}[Rooting at the hub]\label{lem:hubroot}
Let $T\in\Xn$ and $T'\in\Xn(s)$ with $\clus(T')\not\subseteq\clus(T)$. Then $\Nfam_s(T)\cap\cluso(T')\ne\emptyset$.
\end{lemma}

The proofs are in Appendix~\ref{app:trees}. For the second lemma, $T'$ is rooted at its hub if it has one, so that every other internal vertex has degree at most $s+1$, and the union of Lemma~\ref{lem:shortunion} has at most $s$ parts.

\begin{theorem}[Identifiability on $\Xn(s)$]\label{thm:ident}
Let $n\ge4$, $s\ge2$, $\chi\ge1$ and $T\in\Xn(s)$.
\begin{enumerate}
\item[(a)] If $\varphi\in\Hil$ satisfies $r_\varphi(A)>\chi$ for every $A\in\Nfam_s(T)$, then $\varphi\notin\TTNo(T',\chi)$, a fortiori $\varphi\notin\TTN(T',\chi)$, for every $T'\in\Xn(s)$ that is not a coarsening of $T$.
\item[(b)] Let $\psi\in\TTNo(T,\chi)$ be a unit vector with $\gaps(\psi,T)>0$. Then every $T'\in\Xn(s)$ that carries $\psi$ is a coarsening of $T$, so $T$ is the unique finest tree in $\Xn(s)$ that carries $\psi$. If moreover $\psi\in\TTN(T,\chi)$, then $T$ is also the unique finest $T'\in\Xn(s)$ with $\psi\in\TTN(T',\chi)$.
\item[(c)] If $T_1,T_2\in\Xn(s)$ both satisfy the hypotheses of (b) for the same $\psi$, then $T_1=T_2$.
\end{enumerate}
\end{theorem}

\begin{proof}
(a) If $T'\in\Xn(s)$ is not a coarsening of $T$, then $\clus(T')\not\subseteq\clus(T)$, and Lemma~\ref{lem:hubroot} gives $A\in\Nfam_s(T)\cap\cluso(T')$. As $r_\varphi(A)>\chi$, $\varphi\notin\TTNo(T',\chi)\supseteq\TTN(T',\chi)$. (b) For $A\in\Nfam_s(T)$, $\varepsilon_A(\chi)\ge\gaps(\psi,T)>0$, so $r(A)>\chi$, and (a) applies, while $T$ carries $\psi$ by hypothesis. (c) By (b) for $T_1$, $T_2\preceq T_1$, and by symmetry $T_1\preceq T_2$. Two trees that are coarsenings of each other have the same clusters, so they are equal by Lemma~\ref{lem:A-trees}(e).
\end{proof}

For binary trees and $s=2$, the binary trees are the maximal elements of $\preceq$, so (a) excludes every binary $T'\ne T$. The hypothesis $\gaps>0$ in (b) cannot be dropped, since a product vector is carried by every tree.

\begin{definition}[Gapped classes]\label{def:classes}
For $d\ge2$, $\chi\ge1$, $s\ge2$, $n\ge4$ and $\varepsilon_0>0$, let
\begin{equation*}
\mathcal S^{d,\chi}_{n,s}(\varepsilon_0):=\bigl\{\psi\in(\C^d)^{\otimes n}:\ \lVert\psi\rVert=1,\ \exists\,T\in\Xn(s):\ \psi\in\TTN(T,\chi)\ \text{and}\ \gaps(\psi,T;\chi)\ge\varepsilon_0\bigr\},
\end{equation*}
and let $\mathcal S^{d,\chi,\circ}_{n,s}(\varepsilon_0)$ be the same set with $\TTNo(T,\chi)$ in place of $\TTN(T,\chi)$. For $\psi$ in either class the witness $T$ is unique, by Theorem~\ref{thm:ident}(c), and the same for both classes. We denote it by $T(\psi)$. A procedure \emph{succeeds} on $\psi$ if it outputs $T(\psi)$, and every other output is a failure.
\end{definition}

Thus $\mathcal S^{d,\chi}_{n,s}(\varepsilon_0)\subseteq\mathcal S^{d,\chi,\circ}_{n,s}(\varepsilon_0)$, with equality if $\chi\ge d$. Membership in these classes is the quantitative identifiability assumption of Section~\ref{sec:s1-problem}. In the statements below the unknown tree is written $T^*$: ``$\psi\in\TTNo(T^*,\chi)$ with $\gaps(\psi,T^*)\ge\varepsilon_0$'' means that $T^*$ is a witness of the promise, and when $T^*\in\Xn(s)$ it is $T(\psi)$. If $\chi\ge d^2$, or if $\chi=1$ and $\psi\in\TTN(T,1)$, then $\gaps(\psi,T)=0$ for every unit vector $\psi$ (Remark~\ref{rem:range}). Section~\ref{sec:s6-coarse} treats $\chi\ge d^2$.

\subsection{The graded scan}\label{sec:learner}
Bottom-up learners keep a partition $\mathcal S$ of $[n]$ and merge several of its parts at a time. Suppose every part is a cluster of $T^*$, and let $A$ be the union of $k$ parts with $2\le k\le|\mathcal S|-2$, so that at least two parts stay outside $A$. Then $A$ is either an internal cluster of $T^*$, of Schmidt rank at most $\chi$ if $\psi\in\TTNo(T^*,\chi)$, or a member of $\Nfam_k(T^*)$, with tail at least $\gaps(\psi,T^*)$ if $k\le s$ (Definitions~\ref{def:ttn} and~\ref{def:gap}). The rank test of Lemma~\ref{lem:ranktest} separates the two cases with one-sided error: it accepts every set of rank at most $\chi$ with certainty (part~(a)), and it accepts a set whose tail is at least $\varepsilon_0$ with probability at most $\delta'$ once $m\ge m_R(\chi,\varepsilon_0,\delta')$ (part~(c)). The graded one-sided greedy sibling learner, $\GOSGS(s)$, uses this test as a membership oracle in a bottom-up agglomeration. It examines all unions of $k$ parts before any union of $k+1$ parts, merges at the first accepted union, and stops when no union accepts. If every test answers correctly, the output is the hull contraction of $T^*$, whatever the order among unions of the same number of parts. Examining unions of fewer parts first ensures that each merge creates an internal vertex of $T^*$ with the same degree as in $T^*$.

Recall that $\RT(\chi,m)$ on $A\subseteq[n]$ accepts $m$ fresh copies of $\psi$ with probability $a_m(A)=a_m(\rho_A)$, and that $a_m(A)=a_m(A^c)$ (Lemma~\ref{lem:ranktest}(b)). The algorithm has a further parameter $k_0\ge1$, the \emph{window}: it tests or merges a union $W$ only if $k_0+1\le|W|\le n-k_0-1$, and skips the other unions. In this section $k_0=1$, which excludes nothing, since a union of at least two parts with at least two parts remaining has $2\le|W|\le n-2$. The window is used in Section~\ref{sec:s6-learner}.

$\GOSGS(s)$ (Algorithm~\ref{alg:gosgs}) keeps a partition $\mathcal S$ of $[n]$, initially into singletons, and for each part $P$ a rooted tree $\Gamma_P$ with leaf set $P$ that records how $P$ was formed. A \emph{round} with $|\mathcal S|$ parts examines the unions of $k$ distinct parts for $k=2,\dots,\min\{s,|\mathcal S|-2\}$, all unions of $k$ parts before any union of $k+1$ parts. The number $k$ of parts in a union is its \emph{arity}. Within one arity, the order is chosen by a \emph{rule}, a deterministic function of $\mathcal S$ and of all verdicts so far. Examining a union $W$ in the window triggers a \emph{test} of $W$ unless $W$ already has a stored verdict. Verdicts are stored per set. A test is one \emph{run} of $\RT(\chi,m)$ on the registers of $W$ or of $W^c$, in the $m$ copies of a batch that Section~\ref{sec:s4-onebatch} specifies. Its verdict is $\acc$ if the run accepts and $\rej$ otherwise. The round ends at the first examined union with verdict $\acc$: its parts are replaced by their union $J$, and $\Gamma_J$ is a new root whose children are the roots of the merged parts. If the round examines all unions of all arities up to $\min\{s,|\mathcal S|-2\}$ without an acceptance, the learner stops and outputs the X-tree $O$ obtained by joining a new vertex to the roots of all $\Gamma_P$, $P\in\mathcal S$. As $k\le|\mathcal S|-2$, at least three parts remain after a merge, so the final vertex has degree at least $3$ and $O$ is an X-tree. With three parts left, no union is examined, and the final vertex has degree three.

\begin{algorithm}[ht]
\caption{$\GOSGS(s)$ with window $k_0$ and test $\RT(\chi,m)$, with a rule}\label{alg:gosgs}
\begin{algorithmic}[1]
\State $\mathcal S\gets\{\{1\},\dots,\{n\}\}$; let $\Gamma_{\{i\}}$ be the single vertex $i$; no verdict is stored
\While{$|\mathcal S|\ge4$}
\State $\mathrm{found}\gets\textup{false}$
\For{$k=2,\dots,\min\{s,|\mathcal S|-2\}$ \textbf{while not} found}
\For{each $k$-subset $\{P_1,\dots,P_k\}$ of $\mathcal S$, ordered by the rule, \textbf{while not} found}
\State $W\gets P_1\cup\dots\cup P_k$
\If{$k_0+1\le|W|\le n-k_0-1$}
\If{$W$ has no stored verdict}\Comment{a test of $W$}
\State run $\RT(\chi,m)$ on the registers of $W$ or of $W^c$\label{alg:gosgs-run}
\State store the verdict $\acc$ on $W$ if the run accepts, and $\rej$ otherwise
\EndIf
\If{$W$ has verdict $\acc$}
\State $\mathcal S\gets(\mathcal S\setminus\{P_1,\dots,P_k\})\cup\{W\}$;\quad $\Gamma_W\gets$ a new root with children the roots of $\Gamma_{P_1},\dots,\Gamma_{P_k}$
\State $\mathrm{found}\gets\textup{true}$
\EndIf
\EndIf
\EndFor
\EndFor
\If{not found}\ \textbf{break}
\EndIf
\EndWhile
\State \Return the X-tree $O$ obtained by joining a new vertex to the roots of $\Gamma_P$, $P\in\mathcal S$
\end{algorithmic}
\end{algorithm}

\paragraph{The deterministic core.}
Call a verdict on a set $W\in\cluso(T^*)\cup\Nfam_s(T^*)$ \emph{correct} if it is $\acc$ for $W\in\cluso(T^*)$ and $\rej$ for $W\in\Nfam_s(T^*)$.

\begin{theorem}[Deterministic core]\label{thm:core}
Let $T^*\in\Xn$, $s\ge2$ and $k_0=1$, and run Algorithm~\ref{alg:gosgs} with any rule and any source of verdicts such that every test of a set in $\cluso(T^*)\cup\Nfam_s(T^*)$ returns the correct verdict. Then every part of every round is a cluster of $T^*$, every tested set lies in $\cluso(T^*)\cup\Nfam_s(T^*)$, every merge creates an internal vertex of $T^*$ with the same degree as in $T^*$, no set is tested twice, and the output is
\begin{equation*}
O=\hull{T^*},
\end{equation*}
the hull contraction of $T^*$. In particular, $O=T^*$ if and only if $T^*\in\Xn(s)$, and $O\preceq T^*$ always.
\end{theorem}

The proof is in Appendix~\ref{app:learner}. It works in the contracted tree of a partition of $[n]$ into clusters of $T^*$ with at least three parts, in which each part is replaced by a leaf (Lemma~\ref{lem:A-trees}(g)). Call an internal vertex of this tree \emph{extremal} if exactly one of its neighbors is internal. Unless the contracted tree is a star, it has at least two extremal vertices. At most $s$ parts are adjacent to an extremal vertex that is not a hub, and their union is a cluster of $T^*$. When a round stops and the contracted tree is not a star, every extremal vertex is a hub, so the internal vertices that remain form the hub hull.

\begin{lemma}[Counting]\label{lem:counting}
In any run of Algorithm~\ref{alg:gosgs}, with any rule, any window and any verdicts, each set is tested at most once, and the number of tests is at most
\begin{equation*}
N_T(s):=\sum_{k=2}^{s}\Bigl[\binom nk+\binom{n-1}k\Bigr]\ \le\ 2\sum_{k=2}^s\binom nk\ \le\ 2\Bigl(\frac{en}{s}\Bigr)^s\qquad(2\le s\le n).
\end{equation*}
For $s=2$, $N_T(2)=(n-1)^2$.
\end{lemma}

For $s\ge n$ we have $N_T(s)=N_T(n)$, since the binomial coefficients with $k>n$ vanish, and the algorithm never examines an arity above $n-2$. The bounds below that contain $s\log(en/s)$ are therefore stated for $s\le n$, and for larger $s$ they hold with $n$ in place of $s$.

All witnesses of the promise have the same hull contraction (Corollary~\ref{cor:witness}).
\section{Copies}\label{sec:copies}
This section determines the number of copies needed to learn the tree. Section~\ref{sec:s4-onebatch} runs the learner of Section~\ref{sec:learner} on one batch held in quantum memory. For $2\le\chi\le d^2-1$, Section~\ref{sec:lower} proves a lower bound for every procedure that matches the one-batch rate. For $\chi\ge d^2$, Sections~\ref{sec:s6-coarse} and~\ref{sec:s6-learner} show that the same learner, restricted to unions in a window of sizes, learns a coarsening $T^{(k)}$ of the tree.
\subsection{Learning from one batch}\label{sec:s4-onebatch}
Running every test on its own $m$ fresh copies would cost up to $N_T(s)\,m$ copies. Instead, the learner holds one batch of $m$ copies in a quantum memory and performs all its tests on this batch, after acceptances and rejections alike. The analysis below requires the tests to be bit-only Lüders instruments. For $S\subseteq[n]$, $\Pi^S_{\acc}$ denotes the accept projector of $\RT(\chi,m)$ on the $S$-registers of the $m$ copies of the batch, extended by the identity on all other registers, $\Pi^S_{\rej}:=1-\Pi^S_{\acc}$, and $\Phi:=\psi^{\otimes m}$.

\begin{definition}[Hypothesis (H$_{\mathrm{SB}}$): bit-only Lüders tests on one batch]\label{hyp:sb}
A learner satisfies (H$_{\mathrm{SB}}$) if it runs Algorithm~\ref{alg:gosgs} and the following hold.
\begin{enumerate}
\item[(i)] It prepares one batch $\mathsf B$ of $m$ copies of $\psi$, in the state $\Phi$, holds it in a quantum memory until its last test, and uses no other copies.
\item[(ii)] Each run in line~\ref{alg:gosgs-run}, a test of a set $W$, applies the instrument $\{\Pi^S_{\acc},\Pi^S_{\rej}\}$ to the $S$-registers of all $m$ copies of $\mathsf B$, where the side $S\in\{W,W^c\}$ is a function of the earlier verdicts.
\item[(iii)] Only the verdict is revealed: after either outcome the batch is left in the state given by Lüders' rule, every ancilla is returned to its initial state, and $\mathsf B$ is kept.
\end{enumerate}
\end{definition}

Lemma~\ref{lem:ranktest}(d) realizes (ii) and (iii) exactly: compute the bit $[\ell(\lambda)\le\chi]$ coherently, measure only this bit, and run the circuit backwards (Lemma~\ref{lem:B-uncompute}). The accept projectors of different tests need not commute. After a rejection the batch is in general no longer $\Phi$, so Lemma~\ref{lem:ranktest}(a), which makes a cluster test accept $\Phi$ with certainty, no longer guarantees that a later test of a cluster accepts. The proof follows the run that receives the correct verdict at every test. Its probability is the squared norm of a product of projectors applied to $\Phi$, which the quantum union bound of Gao~\cite{Gao15}, in the form of O'Donnell and Venkateswaran~\cite{OV22}, bounds below by terms evaluated on $\Phi$ alone. Only the cluster terms, which are evaluated on $\Phi$, vanish. The others are acceptance probabilities of sets of $\Nfam_s(T^*)$ on fresh copies. The good-path form of the quantum union bound is~\cite[Lemma~2.7 and Corollary~2.8]{BO24}, Gao applies it to sequential decoding~\cite{Gao15}, and Harrow, Lin and Montanaro share one batch among one-sided tests of all cuts through a quantum OR procedure~\cite[Theorem~7 and Corollary~11]{HLM17}.

For $W\subseteq[n]$, the \emph{rank-correct verdict} is $c(W):=\acc$ if $r(W)\le\chi$, and $c(W):=\rej$ otherwise. The \emph{correct-verdict history} $\mathfrak h$ is the run of Algorithm~\ref{alg:gosgs}, with the same rule, in which every test of a set $W$ returns $c(W)$. Let $\Nfam_s(\mathfrak h)$ be the family of the sets of $\Nfam_s(T^*)$ that $\mathfrak h$ tests, and $L_{\mathfrak h}:=\sum_{A\in\Nfam_s(\mathfrak h)}a_m(A)$. For $\delta,\varepsilon_0\in(0,1]$ let
\begin{equation*}
m^{(s)}_\star:=m_R\Bigl(\chi,\varepsilon_0,\frac\delta{4N_T(s)}\Bigr)=(\chi+1)\Bigl\lceil\frac{2(\chi+1)}{\varepsilon_0}\ln\frac{4(\chi+1)N_T(s)}{\delta}\Bigr\rceil.
\end{equation*}

\begin{theorem}[One batch suffices]\label{thm:onebatch}
Let $n\ge4$, $d\ge2$, $\chi\ge1$ and $s\ge2$ be integers and $\delta,\varepsilon_0\in(0,1]$. The learner knows $n$, $d$, $\chi$, $s$, $\varepsilon_0$ and $\delta$, receives copies of a unit vector $\psi\in(\C^d)^{\otimes n}$, and runs Algorithm~\ref{alg:gosgs} with $k_0=1$ under hypothesis (H$_{\mathrm{SB}}$) of Definition~\ref{hyp:sb}, with any rule. Let $\psi\in\TTNo(T^*,\chi)$ for a tree $T^*\in\Xn$ that is unknown to the learner, with $\gaps(\psi,T^*)\ge\varepsilon_0$.
\begin{enumerate}
\item[(a)] \emph{The correct-verdict history.} For every $m\ge1$, $\mathfrak h$ tests pairwise distinct sets, at most $N_T(s)$ of them, all in $\cluso(T^*)\cup\Nfam_s(T^*)$, and outputs $\hull{T^*}$.
\item[(b)] \emph{Success probability.} For every $m\ge1$, the probability that all verdicts of the run are rank-correct, that is, that the run follows $\mathfrak h$, is at least $1-4L_{\mathfrak h}$. Hence
\begin{equation*}
\Pr\bigl[O=\hull{T^*}\bigr]\ \ge\ 1-4\min\Bigl\{\sum_{A\in\Nfam_s(T^*)}a_m(A),\ N_T(s)\max_{A\in\Nfam_s(T^*)}a_m(A)\Bigr\}.
\end{equation*}
\item[(c)] \emph{One batch suffices.} If $m\ge m^{(s)}_\star$ or $m\ge m_R^+(\chi,\varepsilon_0,\delta/(4N_T(s)))$, then $\Pr[O=\hull{T^*}]\ge1-\delta$, and the learner uses $m$ copies in total. Moreover, for $s\le n$,
\begin{equation*}
m^{(s)}_\star\ \le\ \frac{2(\chi+1)^2\bigl(\ln\frac{8(\chi+1)}{\delta}+s\ln\frac{en}{s}\bigr)}{\varepsilon_0}+\chi+1\ =\ O\Bigl(\frac{\chi^2\,(s\log(en/s)+\log(\chi/\delta))}{\varepsilon_0}\Bigr),
\end{equation*}
and
\begin{equation*}
m_R^+\Bigl(\chi,\varepsilon_0,\frac\delta{4N_T(s)}\Bigr)\ \le\ \frac{4(\chi+1)\bigl(\chi+4\ln\frac8\delta+4s\ln\frac{en}s\bigr)}{\varepsilon_0}+1,
\end{equation*}
which is $O\bigl((\chi^2+\chi\,(s\log(en/s)+\log(1/\delta)))/\varepsilon_0\bigr)$.
\item[(d)] \emph{Measurements.} Every run makes at most $N_T(s)$ tests, each a collective measurement of the $m$ copies of the batch, on at most $\lfloor n/2\rfloor$ sites of each copy if the smaller side is chosen.
\end{enumerate}
\end{theorem}

The proof is in Appendix~\ref{app:onebatch}. Part (a) is the correct-verdict history, part (b) is the quantum union bound applied along it, and parts (c) and (d) follow from the soundness of the rank test and the counting lemma. If the order of examination or the choice of sides also depends on classical randomness that is independent of the copies, the learner is a mixture of learners as in the theorem, so every bound of parts (b), (c) and (d) that does not depend on the rule holds for it as well. The same applies to the learner with a window in Theorem~\ref{thm:largebond} and to the learner with inexact tests in Lemma~\ref{lem:D-inexact}.

Theorem~\ref{thm:onebatch} does not cover inputs outside the promise, on which sets of rank larger than $\chi$ can be accepted with large probability and the run need not follow the correct-verdict history. It assumes exact tests, and Lemma~\ref{lem:D-inexact} bounds the effect of inexact ones. The constant $4$ of the quantum union bound cannot be lowered for general sequences of projectors~\cite[Section~3.6]{OV22}. For fixed $s$ the number of copies on one batch is $O((\chi^2+\chi\log(n/\delta))/\varepsilon_0)$ with the threshold $m_R^+$, which is the smaller one when the logarithm dominates, and $O(\chi^2\log(n\chi/\delta)/\varepsilon_0)$ with $m_R$. The degree enters the copies only through $\log N_T(s)\le\log2+s\log(en/s)$, and the number of measurements through $N_T(s)=O(n^s)$. If $s\ge\Delta_2(T^*)-1$, the output is $T^*$ itself. The largest degree therefore costs nothing: $s$, and with it the copies and the measurements, depends only on the second-largest degree.
\subsection{The lower bound for \texorpdfstring{$2\le\chi\le d^2-1$}{2 <= chi <= d2-1}}\label{sec:lower}
The learner of Theorem~\ref{thm:onebatch} learns the tree from $O((\chi^2+\chi\,(s\log(en/s)+\log(1/\delta)))/\varepsilon_0)$ copies held in one batch. We show that for every $(d,\chi)$ with $2\le\chi\le d^2-1$ and every $s\ge2$, every procedure needs $\Omega(\log(n/\delta)/\varepsilon_0)$ copies on the class $\mathcal S^{d,\chi}_{n,s}(\varepsilon_0)$ when $\delta\le\frac13$ and $\varepsilon_0$ is below a constant $\varepsilon_{d,\chi}$, so that the copy complexity is $\Theta(\log(n/\delta)/\varepsilon_0)$ up to a factor that depends only on $d$, $\chi$ and $s$ (Corollary~\ref{cor:rate}). The argument is a direct sum of two-point problems. Each block is a pair of states of four sites with different trees and overlap $1-\varepsilon$. Further sites in a graph state~\cite{HEB04} join $K=\lfloor n/4\rfloor$ blocks into one state on $n$ sites, keep the blocks independent, so that success probabilities multiply, and force a unique tree with gap exactly $\theta\varepsilon$, where the constant $\theta\in(0,1]$ is defined below. The trees are binary, so the family lies in every class, and its states have a gap across every cut that is not a cluster, so the $s$-gap and the total gap coincide. Each block is a two-state discrimination problem~\cite{Hel76}, and the argument over all $K$ blocks, indexed by the hypercube $\{1,2\}^K$, is a quantum analogue of Assouad's hypercube method~\cite{Yu97}. For $K=1$ the bound is the two-point Helstrom bound, whose trace-distance form Lovitz and Lowe call trivial~\cite[\S4.2, footnote~4]{LL26}.

Unless stated otherwise, $d\ge2$, $\chi$ is an integer with $2\le\chi\le d^2-1$, and $q:=\min\{d,\chi\}$, so that $\chi<q^2$. Appendix~\ref{app:lower} builds the family over an arbitrary finite abelian group of order $q$ with a symmetric nondegenerate bicharacter, as in the graph codes of Schlingemann and Werner~\cite{SW01}. Here the group is $\Z_q$, and in Section~\ref{sec:largebond} it is $\Z_d^2$.

\begin{definition}[The pair set and the block pair]\label{def:s5-hard}
\begin{enumerate}
\item[(a)] The \emph{pair set} is the set $P$ of all $(a,b)\in\Z_q^2$ with $a\ne b$, which has $M:=|P|=q(q-1)$ elements, so that $q+M=q^2$. Put
\begin{equation*}
\theta:=\frac{q+M-\chi}{M}=\frac{q^2-\chi}{q(q-1)}\in(0,1]\qquad\text{and}\qquad\bar\varepsilon:=\frac{1-\chi/q^2}{(\sqrt2+\sqrt\theta)^2},
\end{equation*}
and $\varepsilon_{d,\chi}:=\theta\bar\varepsilon$. Then $\bar\varepsilon<\frac12\le1-1/q=M/(M+q)$, as $(\sqrt2+\sqrt\theta)^2\ge2$ and $q\ge2$.
\item[(b)] For $0<\varepsilon<1$ the \emph{block pair} consists of the four-site states
\begin{equation*}
\psi_1:=\sqrt{1-\varepsilon}\,\GHZ_q+\sqrt{\varepsilon/M}\sum_{(a,b)\in P}\ket{aabb}\qquad\text{and}\qquad\psi_2:=W_{(23)}\psi_1,
\end{equation*}
where $\GHZ_q:=q^{-1/2}\sum_{a<q}\ket{aaaa}$ and $W_{(23)}$ exchanges the sites $2$ and $3$.
\end{enumerate}
\end{definition}

The two states of the block pair have the different quartet trees $12|34$ and $13|24$, the binary trees on four sites whose internal edges split the sites as $\{1,2\}|\{3,4\}$ and $\{1,3\}|\{2,4\}$. They have overlap $\braket{\psi_1}{\psi_2}=1-\varepsilon$, and for $\varepsilon\le M/(M+q)$ every set of the $2$-wrong-merge family of the tree of $\psi_j$ has $\chi$-tail exactly $\theta\varepsilon$ (Lemma~\ref{lem:E-block}). The \emph{hard states} $\Psi_x$, $x\in\{1,2\}^K$, on $n=4K+L$ sites, put the block pair $\psi_{x_b}$ on the $b$-th of $K$ disjoint triples of sites, the \emph{blocks}, and replace the fourth sites of all $K$ block pairs by an isometry into a $\Z_q$ graph state on the remaining $K+L$ sites, the \emph{spine}, which is attached to the blocks by controlled-$Z$ gates. The spine keeps the blocks independent, so that overlaps factorize, and it fixes the gap. The binary tree $T_x$ hangs the blocks and the spine sites from a path, and across the $4K$ \emph{$\varepsilon$-type sets} of $x$, which form the family $\mathcal E_x\subseteq\Nfam_2(T_x)$, the state has the spectra of the block pairs. Definition~\ref{def:E-hard} gives the construction in full.

For $\sigma\in[0,1]$ let $\PH(\sigma):=\frac12\bigl(1+\sqrt{1-\sigma^2}\bigr)$. Suppose one of two pure states $\varphi_1,\varphi_2$ with $\lvert\braket{\varphi_1}{\varphi_2}\rvert=\sigma$ is prepared, each with probability $\frac12$. By Helstrom's theorem~\cite{Hel76}, $\PH(\sigma)$ is the largest probability with which a measurement guesses correctly which one was prepared. For $\delta\in(0,1)$ and $K\ge1$ let $a_K:=2(1-\delta)^{1/K}-1$, let $G_K(\delta):=\ln\frac1{1-a_K^2}$ if $a_K>0$ and $G_K(\delta):=0$ otherwise, and let $N_K(\varepsilon,\delta):=G_K(\delta)/\bigl(2\ln\frac1{1-\varepsilon}\bigr)$. Thus $N_1(\varepsilon,\delta)=\ln\frac1{4\delta(1-\delta)}\big/\bigl(2\ln\frac1{1-\varepsilon}\bigr)$ for $\delta<\frac12$. Finally $c_0:=\ln(9/8)/\ln21$, and $\dH$ is the Hamming distance on $\{1,2\}^K$. In the theorem, procedures are those of Definition~\ref{def:procedures}, and they may know $K$, $L$, $\varepsilon$ and the promise on the input.

\begin{theorem}[Lower bound]\label{thm:lower}
Let $d\ge2$ and $2\le\chi\le d^2-1$, let $q:=\min\{d,\chi\}$, and let $M$, $\theta$ and $\bar\varepsilon$ be those of Definition~\ref{def:s5-hard}.
\begin{enumerate}
\item[(a)] \emph{The hard family.} Let $K\ge1$, $L\ge0$, $n=4K+L$ and $0<\varepsilon\le\bar\varepsilon$. For the hard states $\Psi_x$ and the trees $T_x$ of Definition~\ref{def:E-hard} and all $x,y\in\{1,2\}^K$, $\Psi_x$ is a unit vector in $\TTN(T_x,\chi)$, $\braket{\Psi_x}{\Psi_y}=(1-\varepsilon)^{\dH(x,y)}$, the trees $T_x$ are pairwise distinct binary trees, and $\varepsilon_A(\Psi_x;\chi)\ge\theta\varepsilon$ for every nonempty proper $A\subseteq[n]$ that is not a cluster of $T_x$, with equality on $\mathcal E_x$ and, if $\varepsilon<\bar\varepsilon$, only there. Hence $\gaptot(\Psi_x,T_x)=\gaps(\Psi_x,T_x)=\theta\varepsilon$, $\Psi_x\in\mathcal S^{d,\chi}_{n,s}(\theta\varepsilon)$ and $T(\Psi_x)=T_x$ for every $s\ge2$.
\item[(b)] \emph{On the hard family.} Let $K\ge1$, $L\ge0$, $n=4K+L$ and $\varepsilon,\delta\in(0,1)$. If an $N$-copy procedure outputs $T_x$ with probability at least $1-\delta$ on $\Psi_x^{\otimes N}$ for every $x$, then $1-\delta\le\PH\bigl((1-\varepsilon)^N\bigr)^K$, hence $N\ge N_K(\varepsilon,\delta)$. Some $\lceil N_K(\varepsilon,\delta)\rceil$-copy procedure does this, so $\lceil N_K(\varepsilon,\delta)\rceil$ is the exact optimum.
\item[(c)] \emph{On the class.} Let $n\ge4$, $s\ge2$, $K:=\lfloor n/4\rfloor$, $\delta\in(0,1)$ and $0<\varepsilon_0\le\varepsilon_{d,\chi}$. Every $N$-copy procedure that succeeds with probability at least $1-\delta$ on every $\psi\in\mathcal S^{d,\chi}_{n,s}(\varepsilon_0)$, in particular every one that does so on $\mathcal S^{d,\chi,\circ}_{n,s}(\varepsilon_0)$, satisfies $N\ge N_K(\varepsilon_0/\theta,\delta)$. If $\delta\le\frac13$, then
\begin{equation}\label{eq:s5-class}
N\ \ge\ \frac{c_0\ln(n/\delta)}{2\ln\frac{1}{1-\varepsilon_0/\theta}}\ \ge\ c\,\frac{\ln(n/\delta)}{\varepsilon_0},\qquad c:=\frac{c_0\,\theta\,(1-\bar\varepsilon)}{2}>0,
\end{equation}
where $c$ depends only on $d$ and $\chi$.
\end{enumerate}
\end{theorem}
\begin{proof}[Proof idea]
The full proof is in Appendix~\ref{app:lower}. (a)~$\Psi_x$ is the image of $\psi_{x_1}\otimes\dots\otimes\psi_{x_K}$, $K$ independent block pairs, under an isometry acting on their fourth sites (Lemma~\ref{lem:E-connector}), so overlaps factorize. The crossing $CZ$ edges of each cluster of $T_x$ share a vertex, so its rank is at most $q\le\chi$ (Lemma~\ref{lem:E-membership}). Across an $\varepsilon$-type set, $\Psi_x$ has the spectrum of the block pair across a wrong cut, $q$ eigenvalues $(1-\varepsilon)/q$ and $M$ eigenvalues $\varepsilon/M$, with $\chi$-tail $\theta\varepsilon$ (Lemma~\ref{lem:E-epstype}). Any other nonempty proper set that is not a cluster has two disjoint boundary edges with a unimodular adjacency block (Lemma~\ref{lem:E-startsets}). Measuring the sites away from them does not increase the tail (Lemma~\ref{lem:E-deletion}) and leaves at most two blocks, which at $\varepsilon=0$ form a graph state with flat reduced state of rank at least $q^2$ (Lemma~\ref{lem:E-graphcut}). A Lipschitz bound then gives the tail $(\sqrt{1-\chi/q^2}-\sqrt{2\varepsilon})^2\ge\theta\varepsilon$ (Lemmas~\ref{lem:E-perturb} and~\ref{lem:E-gap}). The argument never uses that the set is a union of two clusters, so the gap holds across every cut and the family lies in every class. (b)~The Gram matrix of the $\Psi_x^{\otimes N}$ equals that of the product states $\varphi_{x_1}\otimes\dots\otimes\varphi_{x_K}$, where $\varphi_1,\varphi_2$ are qubit states with $\braket{\varphi_1}{\varphi_2}=(1-\varepsilon)^N$. A product dual certificate and the product of the Helstrom measurements give the bound and its attainment (Proposition~\ref{prop:E-assouad}). (c)~With $\varepsilon=\varepsilon_0/\theta$ the family lies in the class, and $G_K(\delta)\ge c_0\ln(n/\delta)$ (Lemma~\ref{lem:E-constants}).
\end{proof}

In part~(c), $\varepsilon_{d,\chi}=(1-1/\chi)(3-2\sqrt2)$ for every $d\ge\chi$, in particular $\varepsilon_{d,2}=(3-2\sqrt2)/2$ for every $d\ge2$, and $\varepsilon_{2,3}=1/36$ for qubits with $\chi=3$. For $K=1$, part~(b) is the two-point Helstrom bound for the block pair. Known lower bounds concern other tasks: Lovitz and Lowe~\cite{LL26} bound the number of copies needed to test whether $\psi$ lies in the tree tensor network class of a given tree, and Ouyang and Tomamichel~\cite{OT22} the number needed to identify a graph state. Here the tree is unknown and must be output.

\begin{corollary}[Copy complexity]\label{cor:rate}
Let $d\ge2$, $2\le\chi\le d^2-1$, $s\ge2$, $n\ge4$, $0<\delta\le\frac13$ and $0<\varepsilon_0\le\varepsilon_{d,\chi}$, with $\theta$ and $\bar\varepsilon$ those of Definition~\ref{def:s5-hard}. Let $N_{\min}$ be the least $N$ for which some $N$-copy procedure (Definition~\ref{def:procedures}) that knows $\varepsilon_0$ and $\delta$ succeeds with probability at least $1-\delta$ on every state of $\mathcal S^{d,\chi}_{n,s}(\varepsilon_0)$. The bounds below hold also with $\mathcal S^{d,\chi,\circ}_{n,s}(\varepsilon_0)$ in place of $\mathcal S^{d,\chi}_{n,s}(\varepsilon_0)$. With $m^{(s)}_\star$ of Section~\ref{sec:s4-onebatch},
\begin{equation*}
\frac{c_0\theta(1-\bar\varepsilon)}{2}\cdot\frac{\ln(n/\delta)}{\varepsilon_0}\ \le\ N_{\min}\ \le\ m^{(s)}_\star\ \le\ C_{\chi,s}\,\frac{\ln(n/\delta)}{\varepsilon_0},
\end{equation*}
with $C_{\chi,s}:=2(\chi+1)^2\bigl(1+s+\ln(8(\chi+1))/\ln12\bigr)\bigl(1+1/\ln12\bigr)=O(\chi^2(s+\log\chi))$. Thus $N_{\min}=\Theta(\log(n/\delta)/\varepsilon_0)$ for every fixed $(d,\chi,s)$, up to a factor $O(\chi^2(s+\log\chi)/\theta)$.
\end{corollary}

\begin{proof}
The lower bound is Theorem~\ref{thm:lower}(c), which covers both classes. Let $\psi$ lie in either class. As $\TTN(T,\chi)\subseteq\TTNo(T,\chi)$, we have $\psi\in\TTNo(T(\psi),\chi)$ with $\gaps(\psi,T(\psi))\ge\varepsilon_0$ and $T(\psi)\in\Xn(s)$, which is the promise of Theorem~\ref{thm:onebatch} with $T^*=T(\psi)$ and $\hull{T^*}=T^*$. So its learner, run with $\varepsilon_0$ and $\delta$ on $m^{(s)}_\star$ copies, outputs $T(\psi)$ with probability at least $1-\delta$. It is an $m^{(s)}_\star$-copy procedure of Definition~\ref{def:procedures}. Indeed, a run makes at most $N_T(s)$ tests with binary verdicts. The tested set and its side are functions of the earlier verdicts, so each test is one round of the procedure, a L\"uders instrument chosen from the earlier outcomes. Identity instruments pad the run to $N_T(s)$ rounds, and the output is a function of the outcomes. Hence $N_{\min}\le m^{(s)}_\star$. The bound on $m^{(s)}_\star$ in terms of $C_{\chi,s}$ is Lemma~\ref{lem:E-corconst}.
\end{proof}

The upper bounds are achieved by the learner of Theorem~\ref{thm:onebatch}, which uses collective weak Schur sampling and holds one batch in quantum memory through bit-only L\"uders tests (Definition~\ref{hyp:sb}). The lower bound has no factor that grows with $\chi$ or with $s$, while the upper bound contains the factor $\chi^2(s+\log\chi)$. The factor $s$ comes from the union bound over $N_T(s)=O(n^s)$ tests.

The results so far need $\chi\le d^2-1$. For $\chi\ge d^2$, every pair of sites has rank at most $d^2\le\chi$, so the $s$-gap vanishes identically (Remark~\ref{rem:range}). The learnable object is then a coarsening of $T^*$.

\subsection{Small sets and the coarsening}\label{sec:s6-coarse}
Fix $d\ge2$ and $\chi\ge2$, and let $k:=\max\{1,\lfloor\log_d\chi\rfloor\}$, so that $d^k\le\chi<d^{k+1}$ when $\chi\ge d$. A set $A\subseteq[n]$ is \emph{small} if $\min\{|A|,n-|A|\}\le k$. When $\chi\ge d$, every small set has $r(A)\le d^{\min\{|A|,n-|A|\}}\le d^k\le\chi$ for every state. When $\chi<d$ we have $k=1$, and no internal edge has a small side. For $T\in\Xn$ the \emph{$k$-coarsening} $T^{(k)}$ is $T$ with every internal edge that has a small side contracted. It is an X-tree, $T^{(k)}\preceq T$, its internal clusters are exactly the internal clusters of $T$ that are not small, and $(T^{(k)})^{(k)}=T^{(k)}$. Two trees with the same $k$-coarsening cannot be told apart by any procedure under a promise that constrains only ranks: if $T_1^{(k)}=T_2^{(k)}$, then $\TTN(T^{(k)}_1,\chi)\subseteq\TTN(T_1,\chi)\cap\TTN(T_2,\chi)$. Indeed, every cluster of $T_i$ that is not a cluster of $T^{(k)}_i$ is a small internal cluster. Such clusters exist only for $k\ge2$, and then every state has rank at most $d^k\le\chi$ across them. So the learnable object is $T^{*(k)}$, and the gap can only be promised on sets that are not small:
\begin{equation*}
\Nfam^{(k)}_s(T):=\{A\in\Nfam_s(T):\ k+1\le|A|\le n-k-1\},\qquad\gapsk(\psi,T):=\min_{A\in\Nfam^{(k)}_s(T)}\varepsilon_A(\chi).
\end{equation*}
The family $\Nfam^{(k)}_s(T)$ can be empty, for example for the star when $s\le k$. The minimum over an empty family is $1$ by convention, so that the promise $\gapsk(\psi,T)\ge\varepsilon_0$ is then void, and sums and maxima of acceptance probabilities over an empty family are $0$.
For $\chi\le d^2-1$ we have $k=1$, small sets are singletons and their complements, $T^{(1)}=T$, $\Nfam^{(1)}_s=\Nfam_s$, and everything reduces to the previous sections. For $\chi\ge d$, and so for every $\chi\ge d^2$, each site has rank at most $d\le\chi$, so $\TTN(T,\chi)=\TTNo(T,\chi)$ and the classes $\mathcal S^{d,\chi}_{n,s}(\varepsilon_0)$ and $\mathcal S^{d,\chi,\circ}_{n,s}(\varepsilon_0)$ of Definition~\ref{def:classes} coincide.

\subsection{The learner and identifiability of the coarsening}\label{sec:s6-learner}
Algorithm~\ref{alg:gosgs} with window $k_0=k$ skips every union $W$ with $|W|\le k$ or $|W|\ge n-k$: it neither tests nor merges $W$, and stores no verdict for it. Appendix~\ref{app:learner} proves the deterministic core for every window: with verdicts that are correct on $\cluso(T^{*(k)})\cup\Nfam^{(k)}_s(T^{*(k)})$, every part is a cluster of $T^{*(k)}$ and the output is $(T^{*(k)})^{\langle s\rangle}$ (Theorem~\ref{thm:C-core}). The counting of Lemma~\ref{lem:counting} is unchanged, since skipped unions are not tested.

\begin{theorem}[Large bond dimension]\label{thm:largebond}
Let $d\ge2$, $\chi\ge2$, $k=\max\{1,\lfloor\log_d\chi\rfloor\}$, $s\ge2$, $\delta,\varepsilon_0\in(0,1]$, and let $\psi\in\TTNo(T^*,\chi)$ for a tree $T^*\in\Xn$ with $\gapsk(\psi,T^{*(k)})\ge\varepsilon_0$. Then Algorithm~\ref{alg:gosgs} with window $k$, run on one batch under (H$_{\mathrm{SB}}$) with $m\ge m^{(s)}_\star$ copies as in Theorem~\ref{thm:onebatch}, outputs $(T^{*(k)})^{\langle s\rangle}$ with probability at least $1-\delta$.
\end{theorem}

\begin{proof}
The proof of Theorem~\ref{thm:onebatch} in Appendix~\ref{app:onebatch} applies to every window, with $T^{*(k)}$ in place of $T^*$, $\Nfam^{(k)}_s(T^{*(k)})$ in place of $\Nfam_s(T^*)$ and Theorem~\ref{thm:C-core} in place of Theorem~\ref{thm:core}: tests of internal clusters of $T^{*(k)}$, which are internal clusters of $T^*$, accept surely, every tested set lies in $\cluso(T^{*(k)})\cup\Nfam^{(k)}_s(T^{*(k)})$ while all verdicts are correct, and the sets of $\Nfam^{(k)}_s(T^{*(k)})$ have tails at least $\varepsilon_0$.
\end{proof}

\begin{theorem}[Identifiability of the coarsening]\label{thm:ident-k}
Let $\chi\ge2$, $k=\max\{1,\lfloor\log_d\chi\rfloor\}$ and $s\ge2$, and let $\psi\in\TTNo(T,\chi)$ for some $T\in\Xn$ with $T^{(k)}\in\Xn(s)$ and $\gapsk(\psi,T^{(k)})>0$. Then $T^{(k)}$ is the unique finest tree among the trees of $\Xn(s)$ that carry $\psi$ and have no small internal cluster: every such tree is a coarsening of $T^{(k)}$. If $T_1,T_2\in\Xn$ both satisfy the hypotheses for the same $\psi$, then $T_1^{(k)}=T_2^{(k)}$.
\end{theorem}

\begin{proof}
Let $T'\in\Xn(s)$ carry $\psi$, have no small internal cluster, and satisfy $\clus(T')\not\subseteq\clus(T^{(k)})$. Lemma~\ref{lem:hubroot} gives $A\in\Nfam_s(T^{(k)})\cap\cluso(T')$. As $T'$ has no small internal cluster, $k+1\le|A|\le n-k-1$, so $A\in\Nfam^{(k)}_s(T^{(k)})$, and the gap gives $r(A)>\chi$, contradicting $\psi\in\TTNo(T',\chi)$. So every such $T'$ is a coarsening of $T^{(k)}$, and $T^{(k)}$ itself is such a tree, since $\psi\in\TTNo(T,\chi)\subseteq\TTNo(T^{(k)},\chi)$. Two trees that are coarsenings of each other are equal.
\end{proof}

For $k=\max\{1,\lfloor\log_d\chi\rfloor\}$, $s\ge2$ and $\varepsilon_0>0$ let $\mathcal S^{d,\chi}_{n,s,k}(\varepsilon_0)$ be the set of unit vectors $\psi$ for which some $T\in\Xn$ satisfies $\psi\in\TTNo(T,\chi)$, $T^{(k)}\in\Xn(s)$ and $\gapsk(\psi,T^{(k)})\ge\varepsilon_0$. By Theorem~\ref{thm:ident-k} the tree $T^{(k)}$ is determined by $\psi$. It is the \emph{target} of $\psi$, and a procedure \emph{succeeds} on $\psi$ if it outputs the target. For $k=1$ the class is $\mathcal S^{d,\chi,\circ}_{n,s}(\varepsilon_0)$ of Definition~\ref{def:classes}, and the target is $T(\psi)$. Theorem~\ref{thm:largebond} learns $\mathcal S^{d,\chi}_{n,s,k}(\varepsilon_0)$ from one batch, since $(T^{(k)})^{\langle s\rangle}=T^{(k)}$ when $T^{(k)}\in\Xn(s)$. For $s\le k$ the target is always the star, because every other tree without small internal clusters has two vertices of degree at least $k+2$, so only $s\ge k+1$ is of interest.
\section{Beyond \texorpdfstring{$d^2$}{d2}: a lower bound for \texorpdfstring{$d^2\le\chi\le d^3-1$}{d2 <= chi <= d3-1}}\label{sec:largebond}
The lower bound of Theorem~\ref{thm:lower} does not apply for $\chi\ge d^2$: the hard family needs rank at least $\chi+1$ across wrong cuts of four-site blocks, which is impossible when $\chi\ge d^2$. We build the family instead over the group $G:=\Z_d^2$, so that each site of the construction is a \emph{super-site} of dimension $q:=d^2\le\chi$, and we encode every super-site into four physical qudits by a code of distance two, so that no cut that splits a super-site and has at least three qudits on each side is cheap. Throughout this section $d$ is a prime, $d^2\le\chi\le d^3-1$, so that $k=2$, and $n=4n'$ for an integer $n'\ge4$.

The \emph{coarse family} is the hard family of Definition~\ref{def:E-hard} over $G=\Z_d^2$, with the bicharacter $\langle g,h\rangle:=\omega^{g\cdot h}$, $\omega:=e^{2\pi i/d}$. Its states $\Psi_x$, the \emph{coarse states}, live on $n'=4K+L$ super-sites and have the binary trees $T_x$. Its constants are $M=q(q-1)$, $\theta=(q^2-\chi)/(q(q-1))$ and $\bar\varepsilon=(1-\chi/q^2)/(\sqrt2+\sqrt\theta)^2$. Every statement of Appendix~\ref{app:lower} holds for this group, since the appendix is written for an arbitrary finite abelian group with a symmetric nondegenerate bicharacter, and the hypotheses $q\le\chi\le q^2-1$ hold because $d^2\le\chi\le d^3-1<d^4-1$. The \emph{code} is the isometry $E:\C^G\to(\C^d)^{\otimes4}$,
\begin{equation*}
E\ket a:=\ket{\varphi_a}:=d^{-1/2}\sum_{t\in\Z_d}\ket{t+a_1,\ t+a_2,\ t+a_1+a_2,\ t}\qquad(a=(a_1,a_2)\in G),
\end{equation*}
a $[[4,2,2]]_d$ code: its stabilizers are $X^{\otimes4}$ and $Z^{-1}\otimes Z^{-1}\otimes Z\otimes Z$, where $X\ket t=\ket{t+1}$ and $Z\ket t=\omega^t\ket t$ on a single qudit, its logical operators have weight at least two, and the Schmidt rank of each basis codeword $\ket{\varphi_a}$ across every nonempty proper subset of its four qudits is $d$ (Lemma~\ref{lem:F-code}). For $d=2$ it is the four-qubit code of~\cite{VGW96,GBP97}. The \emph{physical family} consists of the states $\tilde\Psi_x:=E^{\otimes n'}\Psi_x\in(\C^d)^{\otimes n}$, and the tree $\tilde T_x$ is $T_x$ with every leaf replaced by a vertex of degree $5$ whose four leaves are the physical qudits of the super-site. Thus $\tilde T_x^{(2)}=\tilde T_x$, and $\tilde T_x\in\Xn(s)$ for every $s\ge4$.

\begin{theorem}[Lower bound for $d^2\le\chi\le d^3-1$]\label{thm:lower-k}
Let $d$ be a prime, $d^2\le\chi\le d^3-1$, $k=2$, and let $q$, $\theta$ and $\bar\varepsilon$ be those of the coarse family. Put
\begin{equation*}
\bar\varepsilon':=\min\Bigl\{\bar\varepsilon,\ \frac{1-\chi/d^3}{(\sqrt3+\sqrt\theta)^2}\Bigr\}\qquad\text{and}\qquad\varepsilon'_{d,\chi}:=\theta\bar\varepsilon'.
\end{equation*}
\begin{enumerate}
\item[(a)] \emph{The family.} Let $n'\ge4$, $K:=\lfloor n'/4\rfloor$, $L:=n'-4K$, $n=4n'$ and $0<\varepsilon\le\bar\varepsilon'$. For all $x,y\in\{1,2\}^K$, $\tilde\Psi_x$ is a unit vector in $\TTN(\tilde T_x,\chi)$, the trees $\tilde T_x$ are pairwise distinct, $\braket{\tilde\Psi_x}{\tilde\Psi_y}=(1-\varepsilon)^{\dH(x,y)}$, and every $A\subseteq[n]$ with $3\le|A|\le n-3$ that is not a cluster of $\tilde T_x$ has $\varepsilon_A(\tilde\Psi_x;\chi)\ge\theta\varepsilon$, with equality on $\mathcal E_x$. In particular $\gapsk(\tilde\Psi_x,\tilde T_x)=\theta\varepsilon$ for every $s\ge2$, and $\tilde\Psi_x\in\mathcal S^{d,\chi}_{n,s,2}(\theta\varepsilon)$ with target $\tilde T_x$ for every $s\ge4$.
\item[(b)] \emph{Lower bound.} Let $s\ge4$, $n\in4\N$ with $n\ge16$, $\delta\in(0,1)$ and $0<\varepsilon_0\le\varepsilon'_{d,\chi}$. Every $N$-copy procedure that succeeds with probability at least $1-\delta$ on every $\psi\in\mathcal S^{d,\chi}_{n,s,2}(\varepsilon_0)$ satisfies $N\ge N_K(\varepsilon_0/\theta,\delta)$ with $K=\lfloor n/16\rfloor$, and for $\delta\le\frac13$
\begin{equation*}
N\ \ge\ \frac{c_0\,\theta\,(1-\bar\varepsilon')}{4}\cdot\frac{\ln(n/\delta)}{\varepsilon_0}.
\end{equation*}
\end{enumerate}
\end{theorem}

\begin{corollary}[Copy complexity for $d^2\le\chi\le d^3-1$]\label{cor:rate-k}
Let $d$ be prime, $d^2\le\chi\le d^3-1$, $n\in4\N$ with $n\ge16$, $s\ge4$, $0<\delta\le\frac13$ and $0<\varepsilon_0\le\varepsilon'_{d,\chi}$. The least number of copies with which some procedure succeeds with probability at least $1-\delta$ on every state of $\mathcal S^{d,\chi}_{n,s,2}(\varepsilon_0)$ is $\Theta(\log(n/\delta)/\varepsilon_0)$, with constants depending only on $d$, $\chi$ and $s$.
\end{corollary}

\begin{proof}
The lower bound is Theorem~\ref{thm:lower-k}(b). The upper bound is Theorem~\ref{thm:largebond} on one batch of $m^{(s)}_\star=O(\chi^2(s\log(en/s)+\log(\chi/\delta))/\varepsilon_0)$ copies, which is a procedure of Definition~\ref{def:procedures} by the argument of Corollary~\ref{cor:rate}.
\end{proof}

\begin{proof}[Proof idea of Theorem~\ref{thm:lower-k}]
The full proof is in Appendix~\ref{app:largebond}. The encoding is an isometry on each super-site, so inner products and the spectra of all \emph{coarse} cuts, those that are unions of super-sites, are those of the coarse family, and Appendix~\ref{app:lower} over $G$ gives everything about them: clusters are cheap, the $\varepsilon$-type sets have tail exactly $\theta\varepsilon$, and the other coarse non-clusters have tail at least $\theta\varepsilon$. The rest of the proof, which is new, concerns the \emph{fine} cuts, which split some super-site. At $\varepsilon=0$ the physical state is a stabilizer state whose group is generated by the encoded stabilizers of the coarse state and the code stabilizers. A counting identity (Lemma~\ref{lem:F-counting}) expresses the rank of a fine cut $A$ through the coarse stabilizer group and the profile of the code on the parts $A\cap s$, which is trivial on at most one qudit, two-dimensional on two and everything on three. From it, every fine set with $3\le|A|\le n-3$ has rank at least $d^3>\chi$ (Lemma~\ref{lem:F-fine0}), the case analysis using only that cheap coarse sets are \emph{star sets}, sets whose boundary edges share a vertex, in a connected graph without four-cycles (Lemma~\ref{lem:F-cheap}). For $\varepsilon>0$, a set that splits at most two blocks is compared with the $\varepsilon=0$ state after an isometric replacement of the other blocks, at distance $\sqrt{2\varepsilon}$, and a set that splits at least three blocks is reduced by deletion to three encoded block vectors, at distance $\sqrt{3\varepsilon}$ (Lemma~\ref{lem:F-fine}). The Lipschitz bound gives tails at least $(\sqrt{1-\chi/d^3}-\sqrt{3\varepsilon})^2\ge\theta\varepsilon$ for $\varepsilon\le\bar\varepsilon'$. The lower bound (b) is then Proposition~\ref{prop:E-assouad}, applied to the Gram matrix of the $\tilde\Psi_x^{\otimes N}$, with the constants of Lemma~\ref{lem:E-constants} at $n'=n/4$.
\end{proof}

\begin{remark}[Scope]\label{rem:scope-k}
The restrictions come from the code: $n\in4\N$ from its four physical qudits per super-site, $k=2$ from its two logical qudits, and prime $d$ from the field structure used in the counting identity. The constant $\theta$ equals $1$ at $\chi=d^2$ and decreases to $(d^4-d^3+1)/(d^4-d^2)$ at $\chi=d^3-1$, and the bounds of Corollary~\ref{cor:rate-k} differ by the factor $O(\chi^2(s+\log\chi)/\theta)$ of Corollary~\ref{cor:rate}. A three-qudit code with logical operators of weight one does not work, since a weight-three stabilizer then sits on a cluster plus one physical qudit and that set is cheap. Distance two removes exactly this obstacle.
\end{remark}
\section{Discussion}\label{sec:discussion}
\paragraph{Resources and limitations.}
The learners of Theorems~\ref{thm:onebatch} and~\ref{thm:largebond} assume exact pure states that are tree tensor networks on an unknown tree, and a known positive lower bound $\varepsilon_0$ on a gap. This gap is the $s$-gap in Theorem~\ref{thm:onebatch}, and in Theorem~\ref{thm:largebond} it is the gap $\gapsk$ of the coarsening. The measurements of these learners are experimentally demanding. Each rank test is built on weak Schur sampling of $m$ copies of up to $\lfloor n/2\rfloor$ sites, with $\mathrm{poly}(m)\cdot O(n\log d)$ gates (Lemma~\ref{lem:ranktest}(d)), and $m$ grows as $(\chi^2+\chi\,(s\log(en/s)+\log(1/\delta)))/\varepsilon_0$. The $mn$ qudits of the batch must be held coherently through up to $N_T(s)$ tests, each with its Schur-sampling ancilla uncomputed, and Theorem~\ref{thm:onebatch} covers tests that satisfy (H$_{\mathrm{SB}}$) exactly or up to a small error in diamond distance (Lemma~\ref{lem:D-inexact}). Whether a learner whose tests reveal the Young diagram can fail on one batch of $m\ge m^{(s)}_\star$ copies is open. The learner outputs $T^*$ only if $s\ge\Delta_2(T^*)-1$, and its number of tests is bounded by $N_T(s)=O(n^s)$, which is polynomial in $n$ only for bounded $s$. For states that are only close to tree tensor networks, a certified approximation, rather than exact recovery, is the meaningful goal. The lower bounds on copies need $\varepsilon_0$ below an explicit constant, and the lower bound for $\chi\ge d^2$ is proved for $d^2\le\chi\le d^3-1$, prime $d$, $s\ge4$ and $n\in4\N$ only.

\paragraph{Open problems.}
\emph{The dependence on $\chi$ and $s$.} The bounds of Corollary~\ref{cor:rate} differ by the factor $O(\chi^2(s+\log\chi)/\theta)$. Whether the number of copies must grow with $\chi$ or with $s$ is open.

\emph{Time for unbounded $\Delta_2$.} Whether some procedure learns every X-tree under the total gap in time polynomial in $n$ is open.

\emph{Large bond dimension.} For $\chi\ge d^3$, extending our construction would use a code with $k=\lfloor\log_d\chi\rfloor$ logical qudits, distance two and length $k+2$. Such a stabilizer code exists for every odd prime $d$, and for $d=2$ only when $k$ is even~\cite{Rai99}, and the trees of its family lie in $\Xn(s)$ only for $s\ge k+2$. Whether the rate $\Theta(\log(n/\delta)/\varepsilon_0)$ holds for every $\chi\ge d^2$ and $s\ge k+2$, for every $n\ge2k+2$ and for composite $d$, is open.

\emph{Noise tolerance.} Is there a learner that, with the number of copies of Theorem~\ref{thm:onebatch}, outputs $T^*$ with probability at least $1-\delta$ on copies of every unit vector $\psi$ with $\min_\vartheta\lVert\psi-e^{i\vartheta}\varphi\rVert^2\le c\,\varepsilon_0$ for some $\varphi$ that satisfies the promise, with a constant $c>0$ independent of $n$, $\delta$, $\chi$ and $s$?

\emph{Loopy graphs.} Learning the graph of a projected entangled-pair state, or of a tensor network on another loopy graph, from copies is the general case of the question of~\cite[\S1.4]{CMS26}, and it is open.
 
\appendix
\clearpage
\phantomsection\addcontentsline{toc}{section}{Appendices}
{\centering\sffamily\LARGE Appendices\par}
\bigskip
\noindent The appendices contain the proofs that are omitted from the main text. Appendix~\ref{app:trees} proves the facts on trees and the lemmas behind the identifiability theorem, Appendix~\ref{app:ranktest} the items of the rank-test lemma, Appendix~\ref{app:learner} the deterministic core for every window and the counting lemma, Appendix~\ref{app:onebatch} the one-batch theorem, Appendix~\ref{app:lower} the lower bound on copies, and Appendix~\ref{app:largebond} the encoded lower bound.

\section{Trees and identifiability}\label{app:trees}
This appendix has three subsections. The first proves Lemma~\ref{lem:A-trees}, the facts about X-trees used throughout. The second proves Lemmas~\ref{lem:shortunion} and~\ref{lem:hubroot}, which lie behind Theorem~\ref{thm:ident}. The third proves a rank inequality, Lemma~\ref{lem:A-rank}, and the range of the bond dimension, Remark~\ref{rem:range}, which uses it and Lemma~\ref{lem:A-trees}(f). By this remark, the $s$-gap vanishes identically for $\chi\ge d^2$. For $\chi=1$ it vanishes on $\TTN(T,1)$ but can be positive on $\TTNo(T,1)$.

\subsection{Tree facts}
We use standard facts about finite trees. A tree with $k$ vertices has $k-1$ edges, deleting an edge leaves exactly two components, each containing one of its ends, any two vertices are joined by a unique path, and a tree with at least two vertices has at least two vertices of degree~$1$.

X-trees whose leaves are labeled by another finite set, such as the set of parts of a partition, are defined as in Section~\ref{sec:s2-trees}. A \emph{split} of $[n]$ is a partition of $[n]$ into two nonempty sets, \emph{trivial} if one of them is a singleton. Two splits are \emph{compatible} if a set of one is disjoint from a set of the other. Equivalently, any set of one and any set of the other are compatible in the sense of Section~\ref{sec:s2-trees}. The \emph{split system} of an X-tree is the set of the splits of its edges.

A \emph{cherry} is a cluster with two elements. In phylogenetics, two leaves form a cherry if they are adjacent to a common vertex~\cite[p.~8]{SS03}. Our cherries are exactly those whose common vertex has degree~$3$. Two leaves adjacent to a vertex of degree~$3$ form a side of its third edge. Conversely, the elements of a cherry are adjacent to a common vertex of degree~$3$ by Lemma~\ref{lem:A-trees}(f). So the two notions agree on binary trees, whose internal vertices all have degree~$3$.

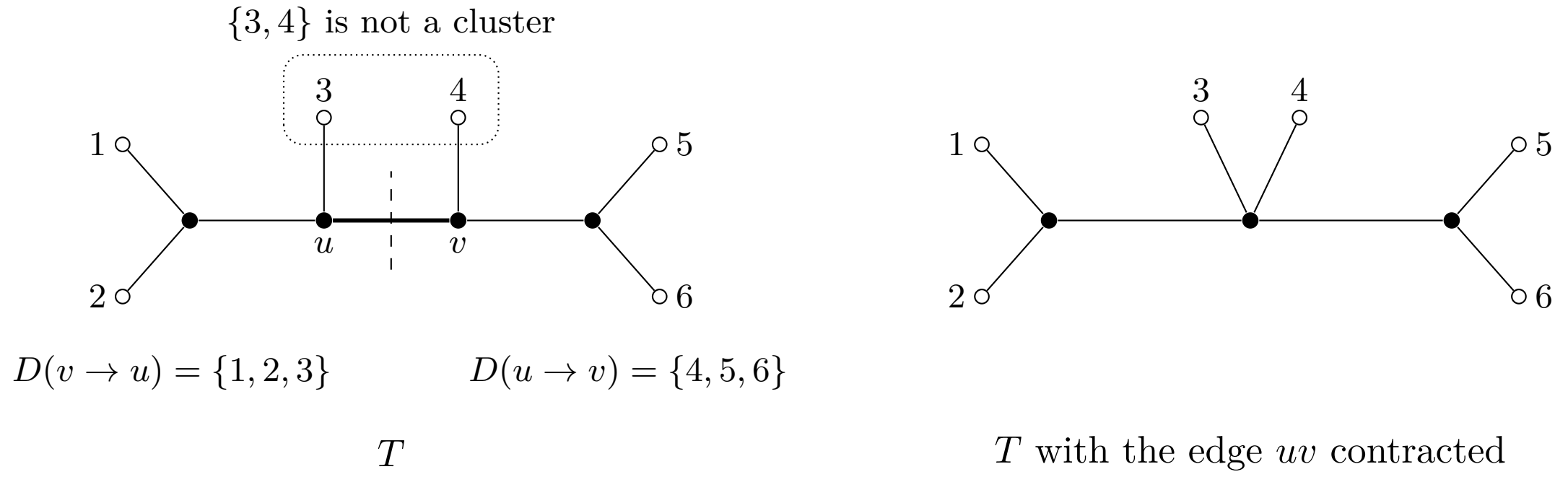
\begin{figure}[!hb]
\centering
\begin{tikzpicture}[x=0.82cm,y=0.82cm,
  iv/.style={circle,fill=black,inner sep=1.5pt},
  lf/.style={circle,draw=black,fill=white,inner sep=1.3pt},
  every label/.style={font=\footnotesize,inner sep=2pt}]
\begin{scope}
\node[iv] (a) at (0,0) {};
\node[iv,label=below:$u$] (u) at (1.5,0) {};
\node[iv,label=below:$v$] (v) at (3.0,0) {};
\node[iv] (b) at (4.5,0) {};
\node[lf,label=left:$1$] (l1) at (-0.75,0.85) {};
\node[lf,label=left:$2$] (l2) at (-0.75,-0.85) {};
\node[lf,label=above:$3$] (l3) at (1.5,1.15) {};
\node[lf,label=above:$4$] (l4) at (3.0,1.15) {};
\node[lf,label=right:$5$] (l5) at (5.25,0.85) {};
\node[lf,label=right:$6$] (l6) at (5.25,-0.85) {};
\draw (a)--(u) (v)--(b) (a)--(l1) (a)--(l2) (u)--(l3) (v)--(l4) (b)--(l5) (b)--(l6);
\draw[line width=1.1pt] (u)--(v);
\draw[dashed] (2.25,-0.55)--(2.25,0.55);
\draw[densely dotted,rounded corners=5pt] (1.05,0.85) rectangle (3.45,1.85);
\node[font=\footnotesize] at (2.25,2.2) {$\{3,4\}$ is not a cluster};
\node[font=\footnotesize] at (-0.2,-1.7) {$D(v\to u)=\{1,2,3\}$};
\node[font=\footnotesize] at (4.9,-1.7) {$D(u\to v)=\{4,5,6\}$};
\node[font=\small] at (2.25,-2.6) {$T$};
\end{scope}
\begin{scope}[shift={(9.6,0)}]
\node[iv] (a) at (0,0) {};
\node[iv] (x) at (2.25,0) {};
\node[iv] (b) at (4.5,0) {};
\node[lf,label=left:$1$] (l1) at (-0.75,0.85) {};
\node[lf,label=left:$2$] (l2) at (-0.75,-0.85) {};
\node[lf,label=above:$3$] (l3) at (1.7,1.15) {};
\node[lf,label=above:$4$] (l4) at (2.8,1.15) {};
\node[lf,label=right:$5$] (l5) at (5.25,0.85) {};
\node[lf,label=right:$6$] (l6) at (5.25,-0.85) {};
\draw (a)--(x)--(b) (a)--(l1) (a)--(l2) (x)--(l3) (x)--(l4) (b)--(l5) (b)--(l6);
\node[font=\small] at (2.25,-2.6) {$T$ with the edge $uv$ contracted};
\end{scope}
\end{tikzpicture}
\caption{Left: an X-tree $T$ with six leaves. The internal edge $uv$ has the sides $D(v\to u)=\{1,2,3\}$ and $D(u\to v)=\{4,5,6\}$, which are internal clusters, and the clusters $\{1,2\}$ and $\{5,6\}$ are cherries. The union $\{3,4\}$ of the clusters $\{3\}$ and $\{4\}$ is not a cluster, so it lies in $\Nfam_2(T)$. Right: contracting $uv$ gives a coarsening of $T$, whose clusters are those of $T$ other than $\{1,2,3\}$ and $\{4,5,6\}$.}
\label{fig:A-tree}
\end{figure}

Let $T\in\Xn$ with $n\ge3$. For a partition $\mathcal S$ of $[n]$ into clusters of $T$ with at least three parts, each part $P$ is the side $D(u_P\to v_P)$ of a unique edge $e_P=u_Pv_P$ by Lemma~\ref{lem:A-trees}(b). We write $C_P$ for the component of $T-e_P$ that contains $v_P$, whose leaves are the elements of $P$.

Lemma~\ref{lem:A-trees} has eight items. Items (a) and (b) are elementary facts on the vertices, the edges and the sides of the edges. Items (c) and (d) restate the splits-equivalence theorem~\cite[Theorem~3.1.4]{SS03}, and item (e) derives from it the description of coarsenings by clusters, which Section~\ref{sec:s2-trees}, the proof of Theorem~\ref{thm:ident} and Appendix~\ref{app:learner} use. Item (f) concerns cherries and the families $\Nfam_s(T)$. Item (g) describes the contracted tree, in which every part of a partition of $[n]$ into clusters of $T$ with at least three parts is replaced by a leaf. Appendix~\ref{app:learner} relies on this tree, and Figure~\ref{fig:C-contracted} there shows an example. Item (h) describes the X-tree that merges of parts build, as the learner of Section~\ref{sec:learner} does. Figure~\ref{fig:A-tree} illustrates sides, clusters, cherries and the contraction of an edge.

\begin{lemma}\label{lem:A-trees}
Let $T\in\Xn$ with $n\ge3$.
\begin{enumerate}
\item[(a)] \emph{Counting.} $T$ has at most $n-2$ internal vertices and at most $2n-3$ edges, with equality in both exactly when $T$ is binary. Each leaf lies on exactly one edge, and distinct leaves lie on distinct edges.
\item[(b)] \emph{Clusters.} The sides of the edges of $T$ are nonempty and pairwise distinct, so distinct edges have distinct splits. The clusters of the leaf edges are the singletons and their complements, $\cluso(T)=\{A\in\clus(T):2\le|A|\le n-2\}$, and $\cluso(T)$ is closed under complements. If $v$ is an internal vertex with neighbors $x_1,\dots,x_k$, then $D(x_1\to v)$ is the disjoint union of $D(v\to x_2),\dots,D(v\to x_k)$.
\item[(c)] \emph{Compatibility.} For $Y,Z\in\clus(T)$, at least one of $Y\cap Z$, $Y\setminus Z$, $Z\setminus Y$ and $[n]\setminus(Y\cup Z)$ is empty.
\item[(d)] \emph{Splits equivalence.} For $T'\in\Xn$, $T'=T$ if and only if $\cluso(T')=\cluso(T)$. More generally, a set of pairwise compatible splits of $[n]$ that contains all trivial splits $\{\{i\},[n]\setminus\{i\}\}$ is the split system of exactly one X-tree with leaf set $[n]$.
\item[(e)] \emph{Coarsening.} Contracting an internal edge $e$ of $T$ gives an X-tree whose clusters are those of $T$ other than the two sides of $e$. For $T'\in\Xn$, $T'$ arises from $T$ by contracting a set of internal edges if and only if $\clus(T')\subseteq\clus(T)$. Hence $\preceq$ is a partial order, two trees that are coarsenings of each other are equal, the star is a coarsening of every tree, and the binary trees are the maximal elements.
\item[(f)] \emph{Cherries and short unions.} If $\{i,j\}\in\clus(T)$ with $i\ne j$, then $i$ and $j$ are adjacent to a common vertex of degree~$3$. If $n\ge4$, distinct cherries are disjoint, and some pair of leaves lies in $\Nfam_2(T)$. Every $A\in\Nfam_s(T)$ satisfies $2\le|A|\le n-2$, and $\Nfam_s(T)\subseteq\Nfam_{s'}(T)$ for $s\le s'$.
\item[(g)] \emph{The contracted tree.} Let $\mathcal S$ be a partition of $[n]$ into clusters of $T$ with at least three parts. The subtrees $C_P$, $P\in\mathcal S$, are pairwise vertex-disjoint, and $u_P\notin C_Q$ for all $P,Q\in\mathcal S$. Deleting all $C_P$ and attaching a new leaf labeled $P$ to $u_P$ for every $P$ gives an X-tree $T|\mathcal S$ with leaf set $\mathcal S$, the \emph{contracted tree}. Its internal vertices are the internal vertices of $T$ that lie in no $C_P$, each with the same degree as in $T$. The clusters of $T|\mathcal S$ are sets of parts, and we read each of them as a subset of $[n]$, the union of its parts. Read in this way, the clusters of $T|\mathcal S$ are exactly the clusters of $T$ that are unions of parts. $T$ is recovered from $T|\mathcal S$ by replacing each leaf $P$ by $C_P$.
\item[(h)] \emph{Agglomeration.} Start from the partition of $[n]$ into singletons, where the rooted tree $\Gamma_{\{i\}}$ is the leaf $i$. Perform merges, each replacing $k\ge2$ parts $P_1,\dots,P_k$ of the current partition by $J:=P_1\cup\dots\cup P_k$, with $\Gamma_J$ a new root whose children are the roots of $\Gamma_{P_1},\dots,\Gamma_{P_k}$, such that at least three parts remain after each merge. Joining a new vertex to the roots of the final trees gives an X-tree $O$ with leaf set $[n]$, and $\cluso(O)$ consists of the merged sets and their complements, which are pairwise distinct.
\end{enumerate}
\end{lemma}

\begin{proof}
(a) Let $I$ be the number of internal vertices. A tree with $n+I$ vertices has $n+I-1$ edges, and the degree sum gives
\begin{equation*}
n+\sum_{v\text{ internal}}\deg v=2(n+I-1),
\end{equation*}
with $\sum\deg v\ge3I$. So $n+3I\le2n+2I-2$, that is $I\le n-2$, with equality exactly when all internal degrees are $3$, and the number of edges is $n+I-1\le2n-3$. Each leaf lies on exactly one edge, as it has degree~$1$. An edge joining two leaves would be a whole component, which forces $n=2$, so distinct leaves lie on distinct edges.

(b) \emph{Two edges.} Let $e$ be an edge and $f\ne e$ another one, which exists since the $n\ge3$ leaves lie on distinct edges by (a). Deleting $e$ and $f$ leaves three components, each containing a leaf, so the sides of $e$ are nonempty and no side of $e$ is a side of $f$. The edge $f$ lies in one component $X$ of $T-e$, and $X-f$ has two components. Let $K_0$ be the one containing the end of $e$ in $X$, let $K_f$ be the other one, and let $K_e$ be the component of $T-e$ other than $X$. Every edge other than $e$ and $f$ lies inside $K_e$, $K_0$ or $K_f$. So $e$ is the only edge leaving $K_e$, $f$ is the only edge leaving $K_f$, and $e$ and $f$ are the only edges leaving $K_0$.

If $b\in\{1,2\}$ edges leave a component $K$ of $T-e-f$, the degrees of its vertices sum to $2(|K|-1)+b<3|K|$, so some vertex of $K$ has degree at most~$2$ in $T$. As the internal vertices of $T$ have degree at least~$3$, this vertex is a leaf. Hence $L_e:=K_e\cap[n]$, $L_0:=K_0\cap[n]$ and $L_f:=K_f\cap[n]$ are nonempty, and they partition $[n]$. The sides of $e$ are $L_e$ and $L_0\cup L_f$, and those of $f$ are $L_f$ and $L_e\cup L_0$. As the three blocks are disjoint and nonempty, no side of $e$ is a side of $f$.

\emph{One edge.} The two sides of one edge are complementary and nonempty, hence distinct. The sides of the leaf edge at the leaf $i$ are $\{i\}$ and $[n]\setminus\{i\}$.

\emph{An internal vertex.} Removing an internal vertex $v$ with neighbors $x_1,\dots,x_k$ instead of an edge leaves the forest $T-v$, whose $k$ components contain $x_1,\dots,x_k$. Those containing $x_2,\dots,x_k$ are the components of $T-vx_i$ that avoid $v$, and together with $v$ they form the component of $T-x_1v$ that contains $v$. As $v$ is not a leaf,
\begin{equation*}
D(x_1\to v)=D(v\to x_2)\sqcup\dots\sqcup D(v\to x_k),
\end{equation*}
a disjoint union of $k-1\ge2$ nonempty sets. So the side of an edge at an internal end has at least two elements, while the side at a leaf end is a singleton. Hence an edge is internal if and only if both of its sides have at least two elements, which gives the description of $\cluso(T)$, and it is closed under complements.

(c) and (d) are the splits-equivalence theorem~\cite[Theorem~3.1.4]{SS03}: the splits of an X-tree are pairwise compatible, and a collection of pairwise compatible splits of $[n]$ that contains all trivial splits is the split system of exactly one X-tree with leaf set $[n]$, up to an isomorphism that preserves the leaf labels. The book states the theorem for trees in which any vertex may carry labels, and such a tree is an X-tree in our sense exactly when its split system contains all trivial splits~\cite[p.~46]{SS03}.

By (b), the split system of $T$ consists of the trivial splits and the splits $\{A,[n]\setminus A\}$ with $A\in\cluso(T)$, so $\cluso(T')=\cluso(T)$ implies $T'=T$, and a leaf-label-preserving isomorphism maps sides of edges to sides of edges, which gives the converse.

(e) \emph{One contraction.} Contracting an internal edge $uv$ identifies two vertices of degree at least $3$ into one of degree $\deg u+\deg v-2\ge4$ and leaves all other degrees unchanged, so the result is an X-tree. Every other edge keeps its two sides, and the split of $uv$ disappears, as by (b) no other edge has it. Hence contracting a set of internal edges removes exactly their splits, and $\clus(T')\subseteq\clus(T)$.

\emph{The converse.} Let $\clus(T')\subseteq\clus(T)$. The clusters of $T$ that are not clusters of $T'$ come in complementary pairs, by (b), each the split of an internal edge of $T$, since trivial splits belong to both trees. Contracting these edges gives an X-tree with the same clusters as $T'$, which equals $T'$ by (d).

\emph{Order and maximal elements.} Reflexivity and transitivity of $\preceq$ are clear, and antisymmetry follows: $\clus(T')=\clus(T)$ gives $T'=T$ by (d). The star has only trivial splits, which belong to every tree by (b), so it is a coarsening of every tree. A binary tree has $2n-3$ edges by (a), the largest possible number, and their splits are pairwise distinct by (b). So the split system of no X-tree properly contains that of a binary tree. A tree that is not binary is not maximal. If an internal vertex $v$ of $T$ has degree at least $4$, detach two neighbors of $v$ and join them to a new vertex adjacent to $v$. This construction gives an X-tree other than $T$ from which contracting the new edge at $v$ recovers $T$. For the partial order and its maximal elements see also~\cite[Proposition~3.2.1]{SS03}.

(f) Let $\{i,j\}=D(x\to v)$ for an edge $xv$. The component of $T-xv$ containing $v$ has the leaves $i,j$ only. As $D(x\to v)$ has two elements, $v$ is not a leaf, so it has degree at least $3$, and its neighbors other than $x$ lie in that component. By (b), the branches $D(v\to y)$ with $y\ne x$ are nonempty and partition $\{i,j\}$, and $y$ is a leaf when $D(v\to y)$ is a singleton. So $v$ has exactly the two further neighbors $i$ and $j$, and $\deg v=3$.

Let $n\ge4$. Two cherries $\{i,j\}$ and $\{i,l\}$ with $j\ne l$ would violate (c), since the four sets of (c) are then $\{i\}$, $\{j\}$, $\{l\}$ and $[n]\setminus\{i,j,l\}$, all nonempty. So the cherries are pairwise disjoint, at most $n/2$ of them, while there are $\binom n2>n/2$ pairs of leaves, each in $\Ufam_2(T)$, and a pair that is not a cherry lies in $\Nfam_2(T)$. For $n=3$ both claims fail: $T$ is the star (Section~\ref{sec:s2-states}), and its three pairs of leaves are clusters by (b), as complements of singletons.

A member of $\Nfam_s(T)$ is a union of at least two nonempty disjoint sets, so it has at least two elements. It is not $[n]$. And it is not a complement of a singleton, which is a cluster. The inclusion is clear from the definition.

(g) \emph{Disjoint subtrees.} Recall that $C_P$ is the component of $T-e_P$ containing $v_P$, where $e_P=u_Pv_P$ and $P=D(u_P\to v_P)$. We show that two subtrees $C_P$ and $C_Q$ with $P\ne Q$ are disjoint. The idea is that otherwise $C_Q$ would contain every vertex outside $C_P$, and with it the leaves of a third part. Let $P\ne Q$ be parts, and suppose that $C_P$ and $C_Q$ share a vertex. The edge $e_P$ is the only edge of $T$ joining $C_P$ to the other vertices. The subtree $C_Q$ is connected, meets $C_P$, and contains the leaves of $Q$, which are outside $C_P$ since $Q\cap P=\emptyset$. So $C_Q$ contains $e_P$ and $u_P$, and by symmetry $C_P$ contains $e_Q$. The vertices of $T$ outside $C_P$ form the other component of $T-e_P$, which is connected, contains $u_P$ and avoids $e_Q$, as $e_Q$ lies inside $C_P$. So these vertices lie in the component of $T-e_Q$ that contains $u_P$, which is $C_Q$. Their leaves are $[n]\setminus P\supsetneq Q$, as there are at least three parts. This contradicts the fact that the leaves of $C_Q$ are exactly $Q$. So the $C_P$ are pairwise disjoint.

\emph{The vertices $u_P$.} If $u_P\in C_Q$ for some $Q$, then $Q\ne P$, as $u_P\notin C_P$. Also $e_P\ne e_Q$, since otherwise $P$ and $Q$ would be the two sides of one edge, so $P\cup Q=[n]$ and no third part could exist. So $e_P$ is an edge of $T-e_Q$ at $u_P\in C_Q$, whence $v_P\in C_Q\cap C_P$, a contradiction.

\emph{The X-tree $T|\mathcal S$.} Deleting the pendant subtrees $C_P$, which are joined to the rest only by the edges $e_P$ at $u_P$, leaves a tree, and attaching the leaves $P$ at $u_P$ gives a tree whose leaves are these new vertices, since every leaf of $T$ lies in some $C_P$. A vertex $w$ of $T$ outside all $C_P$ has, for each neighbor $x\in C_P$, the edge $wx=e_P$ and $w=u_P$, so its neighbors in $C_P$ are replaced one for one by the leaves $P$, and its degree is unchanged. Hence $T|\mathcal S$ is an X-tree with leaf set $\mathcal S$ and the stated internal vertices.

\emph{Clusters of $T|\mathcal S$.} A cluster of $T$ is a side of an edge $e$. If $e$ lies inside some $C_P$, its sides are a subset of $P$ and the complement of this subset. The subset is nonempty and proper by (b), since $e\ne e_P$. So neither side is a union of parts. If $e=e_P$, its sides are $P$ and $[n]\setminus P$, the sides of the leaf edge at $P$ in $T|\mathcal S$. Otherwise $e$ is an edge of $T|\mathcal S$ with the same two sides, read as unions of parts, and these are unions of parts. Conversely every edge of $T|\mathcal S$ is of one of the last two kinds. Finally, replacing each leaf $P$ of $T|\mathcal S$ by $C_P$, joined to $u_P$ by the edge $e_P$, undoes the construction and gives $T$.

(h) By induction on the merges, starting from the trees $\Gamma_{\{i\}}$, every $\Gamma_J$ is a rooted tree with leaf set $J$ in which every vertex other than a leaf has at least two children. Joining a new vertex $c$ to the final roots gives a tree $O$ with leaf set $[n]$. In $O$, every leaf has one neighbor, and every merge root has its $k\ge2$ children and one further neighbor. The vertex $c$ has at least three neighbors, since $n\ge3$ and at least three parts remain after each merge. So $O$ is an X-tree. The edge from a merge root to its further neighbor joins two internal vertices, and its side at the root is the merged set $J$. Every internal edge of $O$ is of this kind, since the internal vertices of $O$ are $c$ and the merge roots. So $\cluso(O)$ consists of the merged sets and their complements, which are pairwise distinct by (b).
\end{proof}
\subsection{Identifiability}
The proofs of Lemmas~\ref{lem:shortunion} and~\ref{lem:hubroot} use the rooting of Section~\ref{sec:s3-ident}. There, $T'$ is rooted at an internal vertex $\rho$, and $A_u=D(p\to u)$ is the set of leaves below a vertex $u\ne\rho$ with parent $p$. The children of an internal vertex $u\ne\rho$ are its $\deg(u)-1$ neighbors other than $p$, where $\deg(u)$ is its degree in $T'$.

\begin{proof}[Proof of Lemma~\ref{lem:shortunion}]
The leaves below $u$ are partitioned by its children, so $A_u=\bigsqcup_xA_x$ over the $\deg(u)-1\ge2$ children $x$ of $u$, with nonempty parts. Each $A_x$ is a cluster of $T'$, being a side of the edge $ux$, and a cluster of $T$ by hypothesis. Their union is not $[n]$, because $u\ne\rho$ and $\rho$ is internal: the components of $T'-\rho$ that do not contain $u$ have leaves outside $A_u$, and there are at least two such components, so $|A_u|\le n-2$. As $A_u\notin\clus(T)$ by assumption, $A_u\in\Nfam_{\deg(u)-1}(T)$. Finally $|A_u|\ge2$, so $A_u\in\cluso(T')$ by Lemma~\ref{lem:A-trees}(b).

For the existence, let $v$ be farthest from $\rho$ among the internal vertices $v\ne\rho$ with $A_v\notin\clus(T)$. A child $x$ of $v$ is either a leaf, and then $A_x$ is a singleton, a cluster of $T$, or an internal vertex farther from $\rho$, and then $A_x\in\clus(T)$ by the choice of $v$. So $u:=v$ has the required property.
\end{proof}

\begin{proof}[Proof of Lemma~\ref{lem:hubroot}]
Root $T'$ at its hub if it has one, and at any internal vertex otherwise. As $T'\in\Xn(s)$, every internal vertex other than the root has degree at most $s+1$. Every cluster of $T'$ is some $A_u$ or its complement, and the singletons are clusters of $T$. So if every $A_u$ with $u$ internal were a cluster of $T$, every cluster of $T'$ would be, contradicting $\clus(T')\not\subseteq\clus(T)$. Hence Lemma~\ref{lem:shortunion} gives an internal vertex $u\ne\rho$ with $A_u\in\Nfam_{\deg(u)-1}(T)\cap\cluso(T')$. As $u\ne\rho$, $\deg(u)\le s+1$, so $A_u\in\Nfam_s(T)$ by Lemma~\ref{lem:A-trees}(f).
\end{proof}

\subsection{Ranks of disjoint unions and the range of the bond dimension}

\begin{lemma}[Ranks of disjoint unions]\label{lem:A-rank}
Let $\varphi\in\Hil$, and let $Y_1,\dots,Y_k\subseteq[n]$ be pairwise disjoint. Then $r_\varphi(Y_1\cup\dots\cup Y_k)\le\prod_{j=1}^kr_\varphi(Y_j)$.
\end{lemma}

\begin{proof}
For $A\subseteq[n]$ let $V_A\subseteq\Hil_A$ be the column space of $\varphi^{(A)}$ and $\mathrm{pr}_A$ the orthogonal projector onto it. Writing $\varphi=\sum_\xi\varphi_\xi\otimes\ket\xi$ over the product basis $\{\ket\xi\}$ of $\Hil_{A^c}$, the vectors $\varphi_\xi$ are the columns of $\varphi^{(A)}$, so $(\mathrm{pr}_A\otimes1)\varphi=\varphi$. Let $Y:=Y_1\cup\dots\cup Y_k$, and order the tensor factors as $\Hil=\Hil_{Y_1}\otimes\dots\otimes\Hil_{Y_k}\otimes\Hil_{Y^c}$. The projectors $\mathrm{pr}_{Y_j}\otimes1$, $j\le k$, act on disjoint tensor factors, so they commute, and their product $(\bigotimes_j\mathrm{pr}_{Y_j})\otimes1$ fixes $\varphi$. Applying $1\otimes\bra\xi$ for the basis vectors $\ket\xi$ of $\Hil_{Y^c}$ shows that every column of $\varphi^{(Y)}$ lies in $\bigotimes_jV_{Y_j}$, whose dimension is $\prod_jr_\varphi(Y_j)$. Hence $r_\varphi(Y)\le\prod_jr_\varphi(Y_j)$.
\end{proof}

\begin{remark}[The range of the bond dimension]\label{rem:range}
Let $T\in\Xn$, $n\ge4$, $s\ge2$, and let $\{i,j\}\in\Nfam_2(T)\subseteq\Nfam_s(T)$ be a pair of leaves (Lemma~\ref{lem:A-trees}(f)). If $\chi\ge d^2$, then $r(\{i,j\})\le d^2\le\chi$, so $\gaps(\psi,T)=0$ for every unit vector $\psi\in\Hil$. If $\chi=1$ and $\psi\in\TTN(T,1)$, then $r(\{i,j\})\le r(\{i\})\,r(\{j\})\le1$ by Lemma~\ref{lem:A-rank}, so again $\gaps(\psi,T)=0$. For $\TTNo$ the gap can be positive at $\chi=1$: for $n=4$, $\cluso(T)=\{\{1,2\},\{3,4\}\}$ and $\psi=\phi_{12}\otimes\phi_{34}$ with both factors of Schmidt rank at least $2$, $\psi\in\TTNo(T,1)$ and every $A\in\Nfam_2(T)$, a pair $\{k,l\}$ with $k\in\{1,2\}$ and $l\in\{3,4\}$, has $r(\{k,l\})=r(\{k\})\,r(\{l\})\ge4$.
\end{remark}

In the example at the end of Remark~\ref{rem:range}, $n=4$, so by Lemma~\ref{lem:A-trees}(f) every member of $\Nfam_s(T)$ is a pair of leaves, the union of two singleton clusters. Hence $\Nfam_s(T)=\Nfam_2(T)$ for every $s\ge2$, and the gap is positive, since $r(A)\ge4>1$ gives $\varepsilon_A(1)>0$ for every $A\in\Nfam_2(T)$ (Section~\ref{sec:s2-states}).
\section{The one-sided rank test}\label{app:ranktest}
We prove the items of Lemma~\ref{lem:ranktest}, in the notation of Section~\ref{sec:s2-ranktest}, or give their exact sources, in the order (a) to (d). Lemma~\ref{lem:B-law} expresses the acceptance probability through the length of a longest strictly decreasing subsequence of a random word, and items (a) and (b) follow from it. Item (c) also needs Lemma~\ref{lem:B-grouping}, which groups the letters into intervals. Item (d) rests on Harrow's circuit for weak Schur sampling and on Lemma~\ref{lem:B-uncompute}, which treats the uncomputation and which Section~\ref{sec:s4-onebatch} cites. The proof of Theorem~\ref{thm:onebatch} in Appendix~\ref{app:onebatch} applies items (a) to (c).

Throughout, $D$, $m$, $\chi$ and $\rho$ are as in Lemma~\ref{lem:ranktest}, $\rho$ has the eigenvalues $\alpha_1\ge\dots\ge\alpha_D$, and we put $r:=\rank\rho$ and $\varepsilon:=\varepsilon(\rho;\chi)$. The eigenvalues form a probability vector $\alpha$ on $[D]$. The random word $w=w_1\cdots w_m\in[D]^m$ has independent letters with law $\alpha$, and $\mathrm{LDS}(w)$ is the length of a longest strictly decreasing subsequence of $w$ in the natural order of $[D]$. A strictly decreasing subsequence uses pairwise distinct letters.

\begin{lemma}[Law]\label{lem:B-law}
The random word $w$ satisfies $a_m(\rho)=\Pr[\mathrm{LDS}(w)\le\chi]$.
\end{lemma}

\begin{proof}
The law is stated in the paragraph before \cite[Proposition~6.1]{OW21}. The \emph{RSK shape} of $w$ is the common shape of the two Young tableaux that the Robinson--Schensted--Knuth correspondence assigns to $w$. By \cite[Proposition~2.23 and Remark~2.24]{OW21}, $\mathrm{WSS}_m$ on $\rho^{\otimes m}$ returns $\lambda$ with probability $f^\lambda s_\lambda(\alpha)$, where $s_\lambda$ is the Schur polynomial in $D$ variables, and this is the probability that the RSK shape of $w$ is $\lambda$. As $s_\lambda$ is symmetric, this holds for any order of the alphabet, in particular for the RSK shape of $w$ in the natural order of $[D]$. By Schensted's theorem~\cite{Sch61}, which \cite[Theorem~2.14]{OW21} extends, the number of rows of this shape is $\mathrm{LDS}(w)$. Hence $a_m(\rho)=\sum_{\lambda\vdash m,\ \ell(\lambda)\le\chi}f^\lambda s_\lambda(\alpha)=\Pr[\mathrm{LDS}(w)\le\chi]$. Lovitz and Lowe state the law as \cite[Lemma~2.7]{LL26}.
\end{proof}

\paragraph{Proof of Lemma~\ref{lem:ranktest}(a).} Item (a), the identity $a_m(\rho)=1$ for $\rank\rho\le\chi$, is part~(ii) of the proof of \cite[Proposition~6.1]{OW21}. It also follows from Lemma~\ref{lem:B-law}, since for $r\le\chi$ the word $w$ has almost surely at most $r$ distinct letters, and then $\mathrm{LDS}(w)\le r\le\chi$.

\paragraph{Proof of Lemma~\ref{lem:ranktest}(b).} Letters $i>r$ have probability $\alpha_i=0$, so $w\in[r]^m$ almost surely, and by Lemma~\ref{lem:B-law}
\begin{equation*}
a_m(\rho)=\sum_{u\in[r]^m:\ \mathrm{LDS}(u)\le\chi}\ \prod_{i=1}^m\alpha_{u_i},
\end{equation*}
which depends only on the nonzero eigenvalues $\alpha_1,\dots,\alpha_r$.

\paragraph{Proof of Lemma~\ref{lem:ranktest}(c).} We follow the proof of \cite[Lemma~6.2(i)]{OW21}, with a larger group mass than in their proof and with an exact binomial tail in place of their bound for sums of geometric variables \cite[Proposition~6.3]{OW21}.

\emph{Idea.} The first step, Lemma~\ref{lem:B-grouping}, groups the letters into $\chi+1$ consecutive intervals of mass at least $\mu$, and a word that visits them from the last to the first within $m$ letters has a strictly decreasing subsequence of length $\chi+1$. The time this takes is a sum of independent geometric variables, which a coupling bounds by the waiting time for $\chi+1$ successes in Bernoulli$(\mu)$ trials, and this gives the first inequality of (c). A union bound over blocks of trials and $1-\mu\le e^{-\mu}$ give the other two, and two choices of $\mu$, the second with a Chernoff bound, give the thresholds.

\begin{lemma}[Grouping]\label{lem:B-grouping}
Let $\alpha_1\ge\dots\ge\alpha_D\ge0$ with $\sum_i\alpha_i=1$, let $\varepsilon:=\sum_{i>\chi}\alpha_i>0$, and let $0<\mu\le\varepsilon/(2\chi+1)$. Then there are integers $0=c_0<c_1<\dots<c_{\chi+1}=D$ such that every interval $I_j:=\{c_{j-1}+1,\dots,c_j\}$ has mass $\nu_j:=\sum_{i\in I_j}\alpha_i\ge\mu$.
\end{lemma}

\begin{proof}
As $\varepsilon>0$, $D\ge\chi+1$. Let $t:=|\{i:\ \alpha_i\ge\mu\}|$, the number of heavy letters, which are $1,\dots,t$. If $t\ge\chi+1$, take $I_j=\{j\}$ for $j\le\chi$ and $I_{\chi+1}=\{\chi+1,\dots,D\}$, whose mass is at least $\alpha_{\chi+1}\ge\mu$.

Let $t\le\chi$. The light letters $t+1,\dots,D$ have mass less than $\mu$ each and total mass at least $\sum_{i>\chi}\alpha_i=\varepsilon$. Scan them in increasing order and put them into consecutive buckets, closing the current bucket as soon as its mass is at least $\mu$. A closed bucket has mass in $[\mu,2\mu)$, because its mass was below $\mu$ before its last letter, whose own mass is below $\mu$. The final open bucket, possibly empty, has mass below $\mu$. If $k$ buckets are closed, then $\varepsilon<2k\mu+\mu$, so $2k+1>\varepsilon/\mu\ge2\chi+1$ and $k\ge\chi+1$. Take $I_j=\{j\}$ for $j\le t$, let $I_{t+1},\dots,I_\chi$ be the first $\chi-t$ closed buckets, and let $I_{\chi+1}$ consist of the remaining $k-(\chi-t)\ge1$ closed buckets and the open one.
\end{proof}

\emph{Waiting times.} Let $\mu\in(0,\varepsilon/(2\chi+1)]$ and take the intervals of Lemma~\ref{lem:B-grouping} for the eigenvalues of $\rho$, whose tail $\sum_{i>\chi}\alpha_i$ is $\varepsilon$, positive by the hypotheses of (c). Every letter of $I_a$ is smaller than every letter of $I_b$ when $a<b$. Extend $w$ to an infinite sequence $w_1,w_2,\dots$ of independent letters with law $\alpha$, put $\tau_0:=0$, and for $j=1,\dots,\chi+1$ let $\tau_j$ be the first $i>\tau_{j-1}$ with $w_i\in I_{\chi+2-j}$. If $\tau_{\chi+1}\le m$, the letters $w_{\tau_1},\dots,w_{\tau_{\chi+1}}$ lie in $I_{\chi+1},I_\chi,\dots,I_1$, respectively, so they strictly decrease and $\mathrm{LDS}(w_1\cdots w_m)\ge\chi+1$. By Lemma~\ref{lem:B-law}, $a_m(\rho)\le\Pr[\tau_{\chi+1}>m]$.

\emph{Geometric gaps.} The gaps $g_j:=\tau_j-\tau_{j-1}$ are independent, and $g_j$ is geometric with success probability $\nu_{\chi+2-j}$. Indeed, let $x_1,\dots,x_{\chi+1}$ be positive integers. The event that $g_j=x_j$ for all $j$ is the event that, for every $j$, the $x_j-1$ letters after position $x_1+\dots+x_{j-1}$ avoid $I_{\chi+2-j}$ and the next letter lies in it. These conditions concern pairwise disjoint sets of positions, so the event has probability
\begin{equation*}
\prod_j(1-\nu_{\chi+2-j})^{x_j-1}\nu_{\chi+2-j}.
\end{equation*}

\emph{Binomial domination.} As $\nu_{\chi+2-j}\ge\mu$, $\Pr[g_j\le a]\ge1-(1-\mu)^a$ for every integer $a\ge0$. So the quantile coupling, with independent uniform variables, realizes the $g_j$ together with independent geometric variables $h_j$ with success probability $\mu$ such that $g_j\le h_j$ for all $j$. This coupling takes $g_j$ and $h_j$ as the quantiles of their laws at the same uniform variable. As $1-(1-\mu)^a=\Pr[h_j\le a]$, the distribution function of $g_j$ is at least that of $h_j$, so its quantiles are at most those of $h_j$. The sum $h_1+\dots+h_{\chi+1}$ has the law of the time of the $(\chi+1)$-st success in independent Bernoulli$(\mu)$ trials. This time exceeds $m$ exactly when at most $\chi$ of the first $m$ trials succeed. Therefore
\begin{equation*}
a_m(\rho)\le\Pr\Big[\sum_jg_j>m\Big]\le\Pr\Big[\sum_jh_j>m\Big]=\Pr[\mathrm{Bin}(m,\mu)\le\chi].
\end{equation*}

\emph{Block bound.} Let $\bar m:=\lfloor m/(\chi+1)\rfloor$, and split the first $(\chi+1)\bar m\le m$ trials into $\chi+1$ blocks of $\bar m$ consecutive trials. If every block contains a success, at least $\chi+1$ trials succeed, so a union bound over the blocks gives
\begin{equation*}
\Pr[\mathrm{Bin}(m,\mu)\le\chi]\le(\chi+1)(1-\mu)^{\bar m}\le(\chi+1)e^{-\mu\bar m}.
\end{equation*}

\emph{The threshold $m_R$.} Let $\bar m_R:=\lceil2(\chi+1)\ln((\chi+1)/\delta')/\varepsilon_0\rceil$, so that $m_R(\chi,\varepsilon_0,\delta')=(\chi+1)\bar m_R$, and $m\ge m_R$ gives $\bar m\ge\bar m_R$. With $\mu:=\varepsilon/(2(\chi+1))\le\varepsilon/(2\chi+1)$ and $\varepsilon\ge\varepsilon_0$, the steps \emph{Binomial domination} and \emph{Block bound} give
\begin{equation*}
a_m(\rho)\le(\chi+1)e^{-\varepsilon\bar m_R/(2(\chi+1))}\le(\chi+1)e^{-\varepsilon_0\bar m_R/(2(\chi+1))}\le(\chi+1)\cdot\frac{\delta'}{\chi+1}=\delta'.
\end{equation*}

\emph{The threshold $m_R^+$.} The value $\mu_0:=\varepsilon_0/(2(\chi+1))\le\varepsilon/(2(\chi+1))<\varepsilon/(2\chi+1)$ is admissible as $\mu$ in the steps \emph{Waiting times} to \emph{Binomial domination}, which need only $0<\mu\le\varepsilon/(2\chi+1)$. For $m\ge m_R^+(\chi,\varepsilon_0,\delta')=\lceil\max\{2\chi,8\ln(1/\delta')\}/\mu_0\rceil$ we have $m\mu_0\ge\max\{2\chi,8\ln(1/\delta')\}$, so $\chi\le m\mu_0/2$. The Chernoff bound $\Pr[\mathrm{Bin}(m,\mu_0)\le(1-\eta)m\mu_0]\le e^{-\eta^2m\mu_0/2}$ for $0<\eta<1$ \cite[Theorem~4.5]{MU05}, with $\eta=\frac12$, gives
\begin{equation*}
a_m(\rho)\le\Pr[\mathrm{Bin}(m,\mu_0)\le\chi]\le\Pr[\mathrm{Bin}(m,\mu_0)\le m\mu_0/2]\le e^{-m\mu_0/8}\le\delta'.
\end{equation*}
Finally, $\max\{a,b\}\le a+b$ for $a,b\ge0$ gives the upper bound on $m_R^+$ displayed in Section~\ref{sec:s2-ranktest}.

\paragraph{Proof of Lemma~\ref{lem:ranktest}(d).} The claims on uncomputation rest on the following lemma. It is the abstract form of the instrument construction \cite[Eqs.~(8.8)--(8.9)]{Har05}, and we prove it for completeness.

\begin{lemma}[Uncomputation]\label{lem:B-uncompute}
Let $\mathcal A$ and $\mathcal V$ be finite-dimensional Hilbert spaces, $\ket0\in\mathcal A$ a unit vector, $G$ a unitary on $\mathcal A\otimes\mathcal V$, and $\{F_x\}_{x\in X}$ a projective measurement on $\mathcal A$ with $X$ finite. Put $E_x:=(\bra0\otimes1)G^\dagger(F_x\otimes1)G(\ket0\otimes1)$. Then $E_x\ge0$, $\sum_xE_x=1$, and measuring $\{F_x\}$ on $G(\ket0\otimes\phi)$ gives $x$ with probability $\bra\phi E_x\ket\phi$, for every unit vector $\phi\in\mathcal V$. If every $E_x$ is an orthogonal projector, then for every $\phi\in\mathcal V$ and $x\in X$
\begin{equation*}
(F_x\otimes1)G(\ket0\otimes\phi)=G(\ket0\otimes E_x\phi)\qquad\text{and}\qquad G^\dagger(F_x\otimes1)G(\ket0\otimes\phi)=\ket0\otimes E_x\phi .
\end{equation*}
The same holds with $\mathcal V\otimes\Hil_{\mathrm{ref}}$, $G\otimes1$ and $E_x\otimes1$ in place of $\mathcal V$, $G$ and $E_x$, for any finite-dimensional $\Hil_{\mathrm{ref}}$. Thus, if every $E_x$ is an orthogonal projector, ``apply $G$, measure $\{F_x\}$, apply $G^\dagger$'' implements the Lüders instrument $\{E_x\}$, whose map for the outcome $x$ is $\omega\mapsto E_x\omega E_x$, and returns the ancilla to $\ket0$.
\end{lemma}

\begin{proof}
We have $\sum_xE_x=(\bra0\otimes1)G^\dagger G(\ket0\otimes1)=1$, each $E_x$ is a compression of $F_x\otimes1\ge0$, and $\lVert(F_x\otimes1)G(\ket0\otimes\phi)\rVert^2=\bra\phi E_x\ket\phi$ is the Born probability. Let the $E_x$ be orthogonal projectors. From $E_x=E_x\bigl(\sum_zE_z\bigr)E_x=E_x+\sum_{z\ne x}(E_zE_x)^\dagger(E_zE_x)$ we get $E_zE_x=0$ for $z\ne x$, so $\{E_x\}$ is a projective measurement. As $\sum_yE_y=1$, $\phi=\sum_yE_y\phi$. If $\phi'\in\operatorname{range}E_y$, then $\lVert(F_y\otimes1)G(\ket0\otimes\phi')\rVert^2=\bra{\phi'}E_y\ket{\phi'}=\lVert\phi'\rVert^2=\lVert G(\ket0\otimes\phi')\rVert^2$, so $G(\ket0\otimes\phi')$ lies in the range of the projector $F_y\otimes1$. As $F_xF_y=\delta_{xy}F_x$,
\begin{equation*}
(F_x\otimes1)G(\ket0\otimes\phi)=\sum_y(F_x\otimes1)(F_y\otimes1)G(\ket0\otimes E_y\phi)=G(\ket0\otimes E_x\phi),
\end{equation*}
and applying $G^\dagger$ gives the second identity. With a reference system, the same argument applies to $G\otimes1$, which induces the orthogonal projectors $E_x\otimes1$.
\end{proof}

\emph{Weak Schur sampling.} The formula for $\Pi_\lambda$ in Section~\ref{sec:s2-ranktest} is the standard projection onto an isotypic component, as in the proof of \cite[Proposition~2.23]{OW21}. Generalized phase estimation \cite[Section~8.1.1]{Har05} applies a unitary circuit $\hat C$ to label registers, prepared in a fixed state $\ket0$ that belongs to the trivial representation, and to the $m$ registers, and then measures the irrep label. For every input $\phi$, it returns $\lambda$ with probability $\bra\phi\Pi_\lambda\ket\phi$. It measures $\{\Pi_\lambda\}_{\lambda\vdash m}$ \cite[Section~8.1.1]{Har05} with two quantum Fourier transforms over $\mathfrak S_m$, of $\mathrm{poly}(m)$ gates each \cite{Bea97}, and one permutation of the $m$ registers controlled by a register that holds an element of $\mathfrak S_m$. Harrow states a total of $\mathrm{poly}(m)+O(m\log D)$ gates for this measurement and for the instruments of the form \cite[Eq.~(8.8)]{Har05}, which include $\mathrm{RT}(\chi,m)$ \cite[p.~160]{Har05}. For $D\ge2$ we have $\log D\ge\log2>0$, so this count is of the form $\mathrm{poly}(m)\cdot O(\log D)$ stated in (d). For $D=1$, $\mathrm{WSS}_m$ and $\mathrm{RT}(\chi,m)$ are trivial and need no gates.

\emph{The rank test.} Apply Lemma~\ref{lem:B-uncompute} with $G=\hat C$ and $\mathcal A$ the label registers. For the projector $F_\lambda$ onto the label $\lambda$, the induced operators satisfy $\bra\phi E_\lambda\ket\phi=\bra\phi\Pi_\lambda\ket\phi$ for all $\phi$ by the step \emph{Weak Schur sampling}, so $E_\lambda=\Pi_\lambda$. Hence $F_x:=\sum_{\lambda:\,[\ell(\lambda)\le\chi]=x}F_\lambda$, for $x\in\{0,1\}$, induces the projectors $E_1=\Pi_{\mathrm{acc}}$ and $E_0=\Pi_{\mathrm{rej}}$. The measurement $\{F_x\}$ is implemented by computing the bit $[\ell(\lambda)\le\chi]$ reversibly into a fresh qubit, measuring this qubit and uncomputing the bit. Followed by $\hat C^\dagger$, this implements the Lüders instrument $\mathrm{RT}(\chi,m)$ by Lemma~\ref{lem:B-uncompute}. It reveals only the accept bit, returns all ancillas to their initial states, and uses twice the gates of $\hat C$, the circuit of $\mathrm{WSS}_m$, plus those that compute and uncompute the bit. It is the instrument of \cite[Eqs.~(8.8)--(8.9)]{Har05} with operation elements $A^{(x)}_\lambda=1$ if $[\ell(\lambda)\le\chi]=x$ and $A^{(x)}_\lambda=0$ otherwise.

Without the final $\hat C^\dagger$, the same POVM $\{\Pi_{\mathrm{acc}},\Pi_{\mathrm{rej}}\}$ is measured, but the system is left in the normalized reduced state of $(F_x\otimes1)\hat C(\ket0\otimes\phi)$.

\emph{Gates on a cut.} On a cut, (b) lets the test act on the smaller side, as in Section~\ref{sec:s2-ranktest}. Then $D\le d^{\lfloor n/2\rfloor}$, so $\log D\le\lfloor n/2\rfloor\log d$ and a test costs $\mathrm{poly}(m)\cdot O(n\log d)$ gates. The gate counts in the proof of (d) treat the rotations of the quantum Fourier transform as exact gates. With a finite gate set the rotations are approximated, the circuit realizes the instrument only up to an error, and the gate count then depends on the precision. Lemma~\ref{lem:D-inexact} bounds the effect of such an error, measured in diamond distance, on the learner.

\section{The learner: the deterministic core and the counting lemma}\label{app:learner}
\begin{figure}[!hb]
\centering
\begin{tikzpicture}[x=0.85cm,y=0.85cm,
  iv/.style={circle,fill=black,inner sep=1.5pt},
  lf/.style={circle,draw=black,fill=white,inner sep=1.3pt},
  every label/.style={font=\footnotesize,inner sep=2pt}]
\begin{scope}
\node[iv,label=above:$u$] (u) at (0,0) {};
\node[iv,label=above:$w$] (w) at (1.7,0) {};
\node[iv,label=above left:$x$] (y) at (3.4,0) {};
\node[iv] (a) at (-0.85,0.8) {};
\node[iv] (b) at (1.7,-0.95) {};
\node[iv] (c) at (3.95,-0.9) {};
\node[lf,label=above:$1$] (l1) at (-1.5,1.55) {};
\node[lf,label=above:$2$] (l2) at (-0.35,1.6) {};
\node[lf,label=below:$3$] (l3) at (-0.85,-0.9) {};
\node[lf,label=below:$4$] (l4) at (1.15,-1.8) {};
\node[lf,label=below:$5$] (l5) at (2.25,-1.8) {};
\node[lf,label=above:$6$] (l6) at (3.75,1.0) {};
\node[lf,label=right:$7$] (l7) at (4.45,0.2) {};
\node[lf,label=below:$8$] (l8) at (3.45,-1.8) {};
\node[lf,label=below:$9$] (l9) at (4.6,-1.7) {};
\draw (u)--(w)--(y) (u)--(a) (a)--(l1) (a)--(l2) (u)--(l3) (w)--(b) (b)--(l4) (b)--(l5) (y)--(l6) (y)--(l7) (y)--(c) (c)--(l8) (c)--(l9);
\draw[dashed,rounded corners=5pt] (-1.95,0.5) rectangle (0.1,2.25);
\draw[dashed,rounded corners=5pt] (-1.25,-1.6) rectangle (-0.45,-0.6);
\draw[dashed,rounded corners=5pt] (0.75,-2.45) rectangle (2.65,-0.65);
\draw[dashed,rounded corners=5pt] (3.35,0.7) rectangle (4.15,1.7);
\draw[dashed,rounded corners=5pt] (4.15,-0.15) rectangle (5.1,0.55);
\draw[dashed,rounded corners=5pt] (3.05,-2.45) rectangle (5.0,-0.6);
\node[font=\footnotesize] at (-2.35,1.4) {$P_1$};
\node[font=\footnotesize] at (-1.65,-1.1) {$P_2$};
\node[font=\footnotesize] at (0.35,-1.55) {$P_3$};
\node[font=\footnotesize] at (4.55,1.2) {$P_4$};
\node[font=\footnotesize] at (5.5,0.2) {$P_5$};
\node[font=\footnotesize] at (5.4,-1.5) {$P_6$};
\node[font=\small] at (1.6,-3.2) {$T$ with the partition $\mathcal S=\{P_1,\dots,P_6\}$};
\end{scope}
\begin{scope}[shift={(9.2,0)}]
\node[iv,label={[label distance=3pt]above:$u$}] (u) at (0,0) {};
\node[iv,label=above:$w$] (w) at (1.7,0) {};
\node[iv,label={[label distance=3pt]above left:$x$}] (y) at (3.4,0) {};
\node[lf,label=above:$P_1$] (p1) at (-0.85,0.8) {};
\node[lf,label=below:$P_2$] (p2) at (-0.85,-0.9) {};
\node[lf,label=below:$P_3$] (p3) at (1.7,-0.95) {};
\node[lf,label=above:$P_4$] (p4) at (3.75,1.0) {};
\node[lf,label=right:$P_5$] (p5) at (4.45,0.2) {};
\node[lf,label=below:$P_6$] (p6) at (3.95,-0.9) {};
\draw (u)--(w)--(y) (u)--(p1) (u)--(p2) (w)--(p3) (y)--(p4) (y)--(p5) (y)--(p6);
\draw (u) circle (0.22);
\draw (y) circle (0.22);
\node[font=\small] at (1.7,-3.2) {$C=T|\mathcal S$};
\end{scope}
\end{tikzpicture}
\caption{A tree $T$ with a partition $\mathcal S$ into six clusters, and the contracted tree $C=T|\mathcal S$, in which every part is a leaf. Each dashed box encloses the subtree $C_P$ of a part $P$. The circled vertices $u$ and $x$ are the extremal vertices of $C$, since in $C$ each has exactly one internal neighbor, $w$. The parts adjacent to $u$ in $C$ form the cluster $D(w\to u)=P_1\cup P_2$ of $T$, as in Lemma~\ref{lem:C-cherry}(1). For $s=2$ the vertex $x$, of degree~$4$, is a hub, while $u$ and $w$ are not.}
\label{fig:C-contracted}
\end{figure}
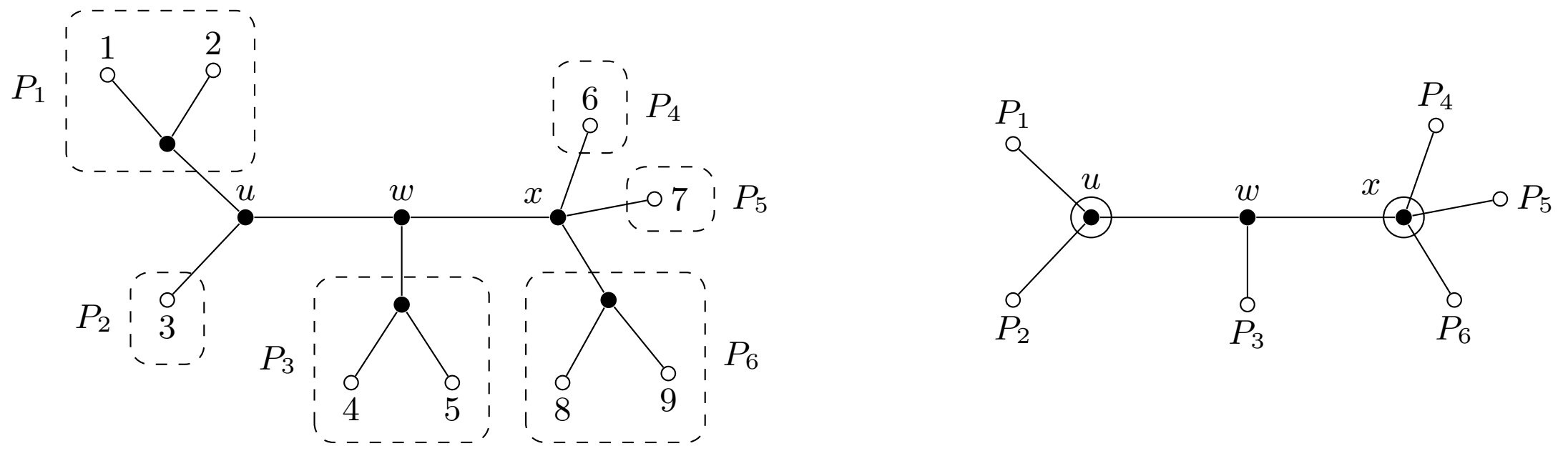

This appendix proves Theorem~\ref{thm:C-core}, the form of Theorem~\ref{thm:core} that covers every window, together with Lemma~\ref{lem:counting} and Corollary~\ref{cor:witness}. The proof of Theorem~\ref{thm:core} is outlined after its statement in Section~\ref{sec:learner}. Items (1) and (2) of Lemma~\ref{lem:C-cherry} are the facts on the contracted tree that this outline uses. Item (3) treats the star. Item (4) places at one vertex of $T$ a cluster that is a union of at least two parts and contains no smaller such cluster. Section~\ref{sec:s6-learner} and Appendix~\ref{app:onebatch} use Theorem~\ref{thm:C-core} for a general window. The proof of Lemma~\ref{lem:counting} is independent of Lemma~\ref{lem:C-cherry} and Theorem~\ref{thm:C-core} and charges each tested union to the part in it that was created last. Corollary~\ref{cor:witness} follows from Theorem~\ref{thm:core}. Table~\ref{tab:C-notation} collects the notation, and Figure~\ref{fig:C-contracted} shows a contracted tree.

For a partition $\mathcal S$ of $[n]$ into clusters of a tree $T\in\Xn$ with at least three parts, the edges $e_P=u_Pv_P$ and the subtrees $C_P$ are those defined before Lemma~\ref{lem:A-trees}, and $T|\mathcal S$ is the contracted tree of Lemma~\ref{lem:A-trees}(g). As in that item, a set of parts, such as a side or a branch in $T|\mathcal S$, is read as the union of its parts. By Lemma~\ref{lem:A-trees}(g), the internal vertices of $T|\mathcal S$ have the same degrees as in $T$, so they are hubs of $T|\mathcal S$ exactly when they are hubs of $T$. An \emph{extremal} vertex of an X-tree is an internal vertex adjacent to exactly one internal vertex.

\begin{table}[!hb]
\centering
\small
\begin{tabular}{@{}lp{8.4cm}l@{}}
\toprule
Symbol & Meaning & Defined in\\
\midrule
$T^*$, $T$ & in Theorem~\ref{thm:C-core}, a tree $T^*$ and its $k_0$-coarsening \mbox{$T=T^{*(k_0)}$}, which is $T^*$ itself for $k_0=1$. In Lemma~\ref{lem:C-cherry}, $T$ is any X-tree & Section~\ref{sec:s6-coarse}\\
$\mathcal S$, $P$ & a partition of $[n]$ into clusters of $T$ with at least three parts, and one of its parts. In the proof of Theorem~\ref{thm:C-core} it is the partition that the learner keeps & Section~\ref{sec:learner}\\
$C=T|\mathcal S$ & the contracted tree, in which each part is replaced by a leaf & Lemma~\ref{lem:A-trees}(g)\\
$e_P=u_Pv_P$, $C_P$ & the edge of $T$ that has $P$ as a side, with $v_P$ its end on the side of $P$, and the subtree of $T$ on that side & Appendix~\ref{app:trees}\\
$D(u\to v)$ & the set of leaves on the side of $v$ when the edge $uv$ is removed & Section~\ref{sec:s2-trees}\\
$\clus(T)$, $\cluso(T)$ & the clusters of $T$, which are the sides of its edges, and its internal clusters, the sides of its internal edges & Section~\ref{sec:s2-trees}\\
branch & a branch of an internal vertex $v$ is a set $D(v\to x)$ with $x$ a neighbor of $v$ & Section~\ref{sec:s2-trees}\\
$\Ufam_s(T)$, $\Nfam_s(T)$ & the sets other than $[n]$ that are unions of two to $s$ pairwise disjoint clusters of $T$, and those among them that are not clusters & Section~\ref{sec:s2-ttn}\\
$\Nfam^{(k_0)}_s(T)$ & the sets of $\Nfam_s(T)$ that have between $k_0+1$ and $n-k_0-1$ elements & Section~\ref{sec:s6-coarse}\\
window $k_0$ & \mbox{$k_0+1\le|W|\le n-k_0-1$} for every union $W$ that the learner tests or merges, and it skips every union outside this range & Section~\ref{sec:learner}\\
arity & the number of parts in a union & Section~\ref{sec:learner}\\
$\Gamma_P$ & the rooted tree with leaf set $P$ that records how the learner formed $P$ & Section~\ref{sec:learner}\\
hub & an internal vertex of degree at least $s+2$ & Section~\ref{sec:s2-trees}\\
$\hull{T}$ & the hull contraction, $T$ with all edges of its hub hull, the minimal subtree that contains its hubs, contracted & Section~\ref{sec:s2-trees}\\
$O$ & the output, in which a new vertex is joined to the roots of all trees $\Gamma_P$ & Section~\ref{sec:learner}\\
$N_T(s)$ & the bound of the counting lemma on the number of tests, which depends only on $n$ and $s$ & Lemma~\ref{lem:counting}\\
\bottomrule
\end{tabular}
\caption{Notation used in Appendix~\ref{app:learner}.}
\label{tab:C-notation}
\end{table}

\begin{lemma}[Extremal vertices of bounded degree]\label{lem:C-cherry}
Let $T\in\Xn$, let $\mathcal S$ be a partition of $[n]$ into clusters of $T$ with at least three parts, and let $C:=T|\mathcal S$.
\begin{enumerate}
\item[(1)] If $C$ is not a star, it has at least two extremal vertices. An extremal vertex $u$ with internal neighbor $w$ has the $\deg(u)-1$ parts adjacent to it as leaves, and their union $D(w\to u)$ is a cluster of $T$ whose complement contains at least two parts.
\item[(2)] If $C$ has at least two internal vertices and at most one hub, some union of $k$ parts with $2\le k\le\min\{s,|\mathcal S|-2\}$ is a cluster of $T$.
\item[(3)] If $C$ is a star, no union of between $2$ and $|\mathcal S|-2$ parts is a cluster of $T$.
\item[(4)] Let $J$ be a cluster of $T$ that is the union of $k\ge2$ parts $P_1,\dots,P_k$, with $k\le|\mathcal S|-2$, and such that no union of $j$ parts with $2\le j<k$ contained in $J$ is a cluster of $T$. Then $J=D(w\to u)$ for an edge $wu$ of $C$ with $u$ and $w$ internal, $P_1,\dots,P_k$ are exactly the branches of $u$ in $C$ other than the one containing $J^c$, and $\deg_C(u)=\deg_T(u)=k+1$. In $T$, the component of $T-wu$ containing $u$ consists of $u$ and the subtrees $C_{P_1},\dots,C_{P_k}$, attached to $u$ by the edges $e_{P_i}$ with $u_{P_i}=u$.
\end{enumerate}
\end{lemma}

\begin{proof}
(1) The leaves of $C$ are its parts, and its internal vertices induce a subtree, since a leaf of $C$ lies on a path only as an end. If $C$ is not a star, this subtree has at least two vertices, hence at least two vertices of degree one, which are the extremal vertices. All neighbors of an extremal vertex $u$ except $w$ are leaves of $C$, that is, parts, and $D(w\to u)$ is their union, a side of the edge $wu$ of $C$, which is a cluster of $T$ by Lemma~\ref{lem:A-trees}(g). The side $D(u\to w)$ contains the other branches of $w$, at least two since $\deg w\ge3$, each containing a part.

(2) Such a $C$ is not a star, so it has at least two extremal vertices by (1). As it has at most one hub, one of them, $u$, is not a hub. Then $3\le\deg(u)\le s+1$, so $k=\deg(u)-1$ satisfies $2\le k\le s$. By (1), the union of the $k$ parts adjacent to $u$ is a cluster of $T$ whose complement contains at least two parts, so $k\le|\mathcal S|-2$.

(3) The clusters of a star with leaf set $\mathcal S$ are the single parts and their complements. A union of between $2$ and $|\mathcal S|-2$ parts is neither, so by Lemma~\ref{lem:A-trees}(g) it is not a cluster of $T$.

(4) By Lemma~\ref{lem:A-trees}(g), $J$ is a cluster of $C$, so $J=D(w\to u)$ for an edge $wu$ of $C$, with $u$ internal since $J$ contains at least two parts. The subtree of $C$ on the side of $u$ has the $k$ parts as leaves. If it had an internal vertex $u'\ne u$, let $y$ be the neighbor of $u$ on the path to $u'$. The side $D(u\to y)$ is a cluster of $T$ by Lemma~\ref{lem:A-trees}(g). It is a union of at least two parts of $J$, as $u'$ has degree at least $3$, and of fewer than $k$ parts, as $u$ has a branch other than $D(u\to w)$ and $D(u\to y)$. This contradicts the minimality of $J$ in (4). So $u$ is the only internal vertex on that side, its branches other than $D(u\to w)=J^c$ are the $k$ parts, and $\deg_C(u)=k+1$, which is $\deg_T(u)$ by Lemma~\ref{lem:A-trees}(g). As $k\le|\mathcal S|-2$, the set $J^c$ contains at least two parts, so $w$ is internal as well, and $wu$ is an edge of $T$ by Lemma~\ref{lem:A-trees}(g). The last claim of (4) then restates how $T$ is recovered from $C$.
\end{proof}

For $k_0\ge1$ and $T^*\in\Xn$, the $k_0$-coarsening $T^{*(k_0)}$ is that of Section~\ref{sec:s6-coarse} with $k_0$ in place of $k$. It arises from $T^*$ by contracting every internal edge that has a side with at most $k_0$ leaves, so $T^{*(k_0)}\preceq T^*$ (Section~\ref{sec:s2-trees}). By Lemma~\ref{lem:A-trees}(e) and its proof, it is an X-tree whose clusters are those of $T^*$ other than the sides of the contracted edges. By Lemma~\ref{lem:A-trees}(b), its internal clusters are therefore the sets $A\in\cluso(T^*)$ with $k_0+1\le|A|\le n-k_0-1$, and $T^{*(1)}=T^*$, as both sides of an internal edge have at least two leaves. For $T\in\Xn$, $\Nfam^{(k_0)}_s(T)$ is the set of $A\in\Nfam_s(T)$ with $k_0+1\le|A|\le n-k_0-1$, as in Section~\ref{sec:s6-coarse}, and $\Nfam^{(1)}_s(T)=\Nfam_s(T)$ by Lemma~\ref{lem:A-trees}(f). Call a verdict on a set $W\in\cluso(T)\cup\Nfam^{(k_0)}_s(T)$ \emph{correct for $T$} if it is $\acc$ for $W\in\cluso(T)$ and $\rej$ for $W\in\Nfam^{(k_0)}_s(T)$. For $k_0=1$ and $T=T^*$, the verdicts correct for $T$ are the correct verdicts of Section~\ref{sec:learner}.

\begin{theorem}[Deterministic core, every window]\label{thm:C-core}
Let $T^*\in\Xn$, $s\ge2$ and $k_0\ge1$, put $T:=T^{*(k_0)}$, and run Algorithm~\ref{alg:gosgs} with window $k_0$, any rule and any source of verdicts such that every test of a set in $\cluso(T)\cup\Nfam^{(k_0)}_s(T)$ returns the verdict correct for $T$. Then every part of every round is a cluster of $T$, every tested set lies in $\cluso(T)\cup\Nfam^{(k_0)}_s(T)$, every merge of $k$ parts creates an internal vertex of $T$ of degree $k+1$, no set is tested twice, and the output is $O=\hull{T}$. For $k_0=1$ this is Theorem~\ref{thm:core}.
\end{theorem}

\begin{proof}
\emph{Idea.} The invariants (I1) and (I2) below say that the parts are clusters of $T$ and that the learner has built the subtree $C_P$ of $T$ for each part $P$. The steps \emph{Tested sets and verdicts} and \emph{A merge} show that a round keeps them, as Lemma~\ref{lem:C-cherry}(4) places the first accepted union at one vertex of $T$. The step \emph{The stop} identifies the output in two cases, according to whether the contracted tree is a star.

\emph{Invariants.} We show by induction over the rounds that
\begin{enumerate}
\item[(I1)] every part $P$ of the current partition $\mathcal S$ is a cluster of $T$, and
\item[(I2)] the rooted tree $\Gamma_P$ is the subtree $C_P$ of $T$ rooted at $v_P$, where each merge root, the root that a merge creates, is identified with an internal vertex of $T$ and the leaves keep their labels.
\end{enumerate}
Initially the parts are singletons, which are clusters by Lemma~\ref{lem:A-trees}(b), and $C_{\{i\}}=\{i\}$ is the one-vertex tree $\Gamma_{\{i\}}$. Let $\mathcal S$ be the current partition, with $|\mathcal S|\ge4$, assume (I1) and (I2) for $\mathcal S$, and let $C:=T|\mathcal S$.

\emph{Tested sets and verdicts.} By (I1), a union $W$ of $k$ parts with $2\le k\le|\mathcal S|-2$ is a set other than $[n]$ that is the union of $k$ pairwise disjoint clusters, so $W\in\Ufam_s(T)$ when $k\le s$. If $W\in\clus(T)$, then $2\le|W|\le n-2$, as $W$ and its complement each contain at least two parts, so $W\in\cluso(T)$ by Lemma~\ref{lem:A-trees}(b). Otherwise, for $k\le s$, $W\in\Nfam_s(T)$, and if $W$ is in the window, $W\in\Nfam^{(k_0)}_s(T)$. Hence every tested set lies in $\cluso(T)\cup\Nfam^{(k_0)}_s(T)$. Every examined union in the window has the verdict correct for $T$, also when this verdict was stored by an earlier test.

\emph{A merge.} We show that a merge keeps (I1) and (I2). Let the round merge $J=P_1\cup\dots\cup P_k$. The round examines unions by increasing arity, skips those outside the window, and merges at the first union with verdict $\acc$. By the previous step, this union is the first examined union in the window that is a cluster of $T$. So $J$ is a cluster of $T$, and no union of $j$ parts with $2\le j<k$ is one: those in the window were examined before $J$ and rejected. Such a union outside the window has at most $k_0$ or at least $n-k_0$ elements, while its complement contains at least two parts, as $j<k\le|\mathcal S|-2$. If it were a cluster of $T$, it would be an internal cluster with at most $k_0$ or at least $n-k_0$ elements. By the paragraph before Theorem~\ref{thm:C-core}, $T=T^{*(k_0)}$ has no such internal cluster.

Lemma~\ref{lem:C-cherry}(4) therefore applies to $J$, as the parts are clusters of $T$ by (I1) and $2\le k\le\min\{s,|\mathcal S|-2\}$. It gives an internal edge $wu$ of $C$ with $J=D(w\to u)$, such that $P_1,\dots,P_k$ are the branches of $u$ in $C$ other than the branch containing $J^c$, and $\deg_T(u)=k+1\le s+1$. By its last claim, the component of $T-wu$ containing $u$ consists of $u$ and the subtrees $C_{P_i}$, attached to $u$ by the edges $e_{P_i}$. The leaves of this component form $J$, so $e_J=wu$, with $u_J=w$ and $v_J=u$, and the component is $C_J$. So (I1) holds for $J$, and $\Gamma_J$, a new root with children the roots of $\Gamma_{P_1},\dots,\Gamma_{P_k}$, is $C_J$ rooted at $u$, by (I2) for the $P_i$. The other parts keep their trees, so (I1) and (I2) hold for the next partition, hence by induction for every partition of the run, the final one included.

\emph{The stop.} The run ends when a round stops, that is, merges nothing, or when three parts remain. Let $\mathcal S$ now be the final partition and $C:=T|\mathcal S$. By (I2), every internal vertex of $T$ inside some $C_P$ is a merge root, of degree at most $s+1$ in $T$ by the step \emph{A merge}. So no hub of $T$ lies inside a $C_P$, and all hubs of $T$ are internal vertices of $C$ (Lemma~\ref{lem:A-trees}(g)).

If three parts remain, $C$ has three leaves, so it is a star. If the round stops, no union of $k$ parts with $2\le k\le\min\{s,|\mathcal S|-2\}$ in the window is a cluster of $T$, as each was examined and rejected. Every internal cluster of $T$ is in the window, by the paragraph before Theorem~\ref{thm:C-core}, and such a union that is a cluster is internal. So no such union at all is a cluster, and by Lemma~\ref{lem:C-cherry}(2), $C$ is a star or has at least two hubs.

If $C$ is a star with center $w$, then $T$ consists of the subtrees $C_P$ attached to $w$ by the edges $e_P$ (Lemma~\ref{lem:A-trees}(g)). As the hubs of $T$ are internal vertices of $C$, $T$ has at most one hub. So $\hull{T}=T$, and the output $O$, which joins a new vertex to the roots $v_P$ (I2), is $T$ with the new vertex identified with $w$, that is, $O=\hull{T}$.

Otherwise the round stopped, since three parts give a star. Then $C$ has at least two internal vertices, and every extremal vertex $u$ of $C$ is a hub. If not, $\deg(u)\le s+1$, and by Lemma~\ref{lem:C-cherry}(1) the union of its $\deg(u)-1\le s$ adjacent parts is a cluster of $T$ whose complement contains at least two parts. So it is an internal cluster, in the window, of arity at most $\min\{s,|\mathcal S|-2\}$, which the round would have examined and accepted. Let $H$ be the subtree of $C$ induced by its internal vertices. Its edges are edges of $T$ by Lemma~\ref{lem:A-trees}(g). The leaves of $H$ are hubs, and $H$ contains all hubs of $T$. So $H$ contains the hub hull of $T$. Conversely, every vertex of $H$ lies on a path between two leaves of $H$, which are hubs, so it lies in the hub hull, and $H$ is the hub hull. Every $C_P$ hangs from the vertex $u_P\in H$ by the edge $e_P$. Contracting $H$ to a single vertex is therefore the hull contraction, and it gives the tree in which all $C_P$ hang from one vertex. By (I2), this tree is $O$, so $O=\hull{T}$.

Finally, a set is tested only when it has no stored verdict and receives one, so no set is tested twice.
\end{proof}

\begin{proof}[Proof of Lemma~\ref{lem:counting}]
A set is tested only if it has no verdict, and testing gives it one, so each set is tested at most once. Every tested set is the union of a $k$-subset of the parts of the partition current at the time of the test, $2\le k\le s$. This subset is unique, as subsets of one partition have distinct unions. Distinct tests test distinct sets, so they give distinct $k$-subsets. Charge each such $k$-subset to the most recently created part among its members. The initial partition contributes at most $\binom nk$ subsets. The $t$-th merge, $t\ge1$, creates one new part and leaves at most $n-t$ parts. At most $\binom{n-t-1}{k-1}$ subsets are charged to the new part. Indeed, the other members of such a subset were created before the new part and still exist when the union of the subset is tested, hence right after this merge. There are at most $n-3$ merges, as each reduces the number of parts by at least one and at least three parts remain. Summing, the number of $k$-subsets charged over the whole run is at most
\begin{equation*}
\binom nk+\sum_{t=1}^{n-3}\binom{n-t-1}{k-1}=\binom nk+\sum_{j=2}^{n-2}\binom j{k-1}\le\binom nk+\binom{n-1}k,
\end{equation*}
by the hockey-stick identity $\sum_{j=0}^{n-2}\binom j{k-1}=\binom{n-1}k$. Summing over $k$ gives $N_T(s)$. For $s=2$, $\binom n2+\binom{n-1}2=(n-1)^2$. The first inequality of the lemma holds as $\binom{n-1}k\le\binom nk$, and the last inequality of the lemma is the standard bound $\sum_{k\le s}\binom nk\le(en/s)^s$.
\end{proof}

Corollary~\ref{cor:witness} extends Theorem~\ref{thm:ident}(c) from $\Xn(s)$ to $\Xn$, with hull contractions in place of trees.

\begin{corollary}[Gapped witnesses]\label{cor:witness}
Let $s\ge2$, $\chi\ge1$, and let $\psi$ be a unit vector with $\psi\in\TTNo(T_i,\chi)$ and $\gaps(\psi,T_i)>0$ for two trees $T_1,T_2\in\Xn$. Then $\hull{T_1}=\hull{T_2}$.
\end{corollary}

\begin{proof}
Run Algorithm~\ref{alg:gosgs} with $k_0=1$, a fixed rule, and the verdict $\acc$ on a set $W$ exactly when $r_\psi(W)\le\chi$. This verdict is correct on $\cluso(T_i)\cup\Nfam_s(T_i)$ for $i=1,2$: internal clusters of $T_i$ have rank at most $\chi$, and the sets of $\Nfam_s(T_i)$ have a tail at least $\gaps(\psi,T_i)>0$ by Definition~\ref{def:gap}, hence a rank larger than $\chi$ (Section~\ref{sec:s2-states}). So the output of this one run is $\hull{T_1}$ and also $\hull{T_2}$, by Theorem~\ref{thm:core}.
\end{proof}
\section{\texorpdfstring{Proof of Theorem~\ref{thm:onebatch}}{Proof of the one-batch theorem}}\label{app:onebatch}
The proof of Theorem~\ref{thm:onebatch} is outlined in Section~\ref{sec:s4-onebatch}. Lemma~\ref{lem:D-history} describes the correct-verdict history, Lemma~\ref{lem:D-born} gives the probability that the run on one batch produces a given sequence of verdicts, and Lemma~\ref{lem:D-qub} is the quantum union bound. The proof of the theorem combines the three, and Lemma~\ref{lem:D-inexact} then treats tests that are implemented only approximately. Appendix~\ref{app:ranktest} refers to Lemma~\ref{lem:D-inexact} for finite gate sets, and the proof of Theorem~\ref{thm:largebond} applies this appendix to a window.

We use the notation of Section~\ref{sec:s4-onebatch}: $\Phi=\psi^{\otimes m}$, the projectors $\Pi^S_{\acc}$ and $\Pi^S_{\rej}$, the rank-correct verdicts $c(W)$, the correct-verdict history $\mathfrak h$, $\Nfam_s(\mathfrak h)$ and $L_{\mathfrak h}$. Table~\ref{tab:D-notation} collects the notation.

As in Appendix~\ref{app:learner}, this appendix applies to every window $k_0\ge1$. For a window $k_0$, put $T:=T^{*(k_0)}$, replace $T^*$ by $T$ in $\cluso(T^*)$, $\hull{T^*}$ and $\TTNo(T^*,\chi)$, and replace $\Nfam_s(T^*)$ by $\Nfam^{(k_0)}_s(T)$, also in $\Nfam_s(\mathfrak h)$ and $L_{\mathfrak h}$. Replace the promise on $\gaps(\psi,T^*)$ by the same promise on $\gapsk(\psi,T)$ with $k=k_0$, and the correct verdicts by the verdicts correct for $T$ of Appendix~\ref{app:learner}. Theorem~\ref{thm:C-core} replaces Theorem~\ref{thm:core}. Theorem~\ref{thm:onebatch} then means the theorem with window $k_0$ and these replacements. As $T$ arises from $T^*$ by contracting a set of internal edges, $T\preceq T^*$ (Section~\ref{sec:s2-trees}), so the hypothesis $\psi\in\TTNo(T^*,\chi)$ of Theorem~\ref{thm:largebond} implies $\psi\in\TTNo(T,\chi)$ (Section~\ref{sec:s2-ttn}). We write the case $k_0=1$.

\begin{table}[!htb]
\centering
\small
\begin{tabular}{@{}lp{8.6cm}l@{}}
\toprule
Symbol & Meaning & Defined in\\
\midrule
$\mathsf B$, $\Phi=\psi^{\otimes m}$ & the batch of $m$ copies of $\psi$ that the learner holds, and its initial state & Section~\ref{sec:s4-onebatch}\\
$\Pi^S_{\acc}$, $\Pi^S_{\rej}$ & the accept projector of $\RT(\chi,m)$ on the $S$-registers of the batch, and $1-\Pi^S_{\acc}$ & Section~\ref{sec:s4-onebatch}\\
(H$_{\mathrm{SB}}$) & every test is a bit-only Lüders instrument applied to the one batch & Definition~\ref{hyp:sb}\\
$c(W)$ & the rank-correct verdict, $\acc$ if $r(W)\le\chi$ and $\rej$ otherwise & Section~\ref{sec:s4-onebatch}\\
$\mathfrak h$ & the correct-verdict history, the run in which every test of a set $W$ returns $c(W)$ & Section~\ref{sec:s4-onebatch}\\
$\Nfam_s(\mathfrak h)$, $L_{\mathfrak h}$ & the sets of $\Nfam_s(T^*)$ that $\mathfrak h$ tests, and the sum of their acceptance probabilities & Section~\ref{sec:s4-onebatch}\\
$a_m(A)$ & the probability that $\RT(\chi,m)$ on $A$ accepts $m$ fresh copies of $\psi$ & Section~\ref{sec:s2-ranktest}\\
$N_T(s)$ & the bound of the counting lemma on the number of tests, which depends only on $n$ and $s$ & Lemma~\ref{lem:counting}\\
$\hull{T^*}$ & the hull contraction of $T^*$, which is $T^*$ itself if $T^*$ has at most one hub & Section~\ref{sec:s2-trees}\\
\bottomrule
\end{tabular}
\caption{Notation used in Appendix~\ref{app:onebatch}.}
\label{tab:D-notation}
\end{table}

\begin{lemma}[The correct-verdict history]\label{lem:D-history}
Let $\psi\in\TTNo(T^*,\chi)$ be a unit vector with $\gaps(\psi,T^*)>0$. Fix the rule. Then the tested sets of $\mathfrak h$ are pairwise distinct, there are at most $N_T(s)$ of them, every tested set $W$ lies in $\cluso(T^*)\cup\Nfam_s(T^*)$ with $c(W)=\acc$ if and only if $W\in\cluso(T^*)$, and the output of $\mathfrak h$ is $\hull{T^*}$. In particular $|\Nfam_s(\mathfrak h)|\le N_T(s)$.
\end{lemma}

\begin{proof}
If $A\in\cluso(T^*)$, then $r(A)\le\chi$, as $\psi\in\TTNo(T^*,\chi)$. If $A\in\Nfam_s(T^*)$, then $\varepsilon_A(\chi)\ge\gaps(\psi,T^*)>0$, so $r(A)>\chi$. Hence $c$ agrees on $\cluso(T^*)\cup\Nfam_s(T^*)$ with the correct verdicts of Section~\ref{sec:learner}. So the hypothesis of Theorem~\ref{thm:core} holds along $\mathfrak h$, and Theorem~\ref{thm:core} and Lemma~\ref{lem:counting} give all claims.
\end{proof}

The \emph{run fed the verdicts} $o_1,o_2,\dots$ is the run of Algorithm~\ref{alg:gosgs}, with the rule of the learner, whose $j$-th test returns $o_j$. The \emph{measured side} of a test of a set $W$ is its side $S\in\{W,W^c\}$ in (H$_{\mathrm{SB}}$)(ii).

\begin{lemma}[Verdict histories on one batch]\label{lem:D-born}
Assume (H$_{\mathrm{SB}}$). Let $o_1,\dots,o_K\in\{\acc,\rej\}$ be such that the run fed the verdicts $o_1,o_2,\dots$ performs at least $K$ tests, and let $A_k$ and $S_k$ be its $k$-th tested set and measured side, which are functions of $o_1,\dots,o_{k-1}$. Put $Q_k:=\Pi^{S_k}_{o_k}$. Then the first $K$ verdicts of the learner's run are $o_1,\dots,o_K$ with probability $\lVert Q_K\cdots Q_1\Phi\rVert^2$.
\end{lemma}

\begin{proof}
The instrument of test $k$ is fixed once $o_1,\dots,o_{k-1}$ are fixed. It acts on $\mathsf B$ alone, by (H$_{\mathrm{SB}}$)(ii), and every ancilla is returned to its initial state, by (H$_{\mathrm{SB}}$)(iii). So the law of each verdict and the state of $\mathsf B$ after it depend only on the instrument and the state of $\mathsf B$ before it. Put $v_0:=\Phi$ and $v_k:=Q_kv_{k-1}$. We argue by induction on $k$. For $k=0$ the empty sequence has probability $1=\lVert v_0\rVert^2$, and $\mathsf B$ is in the state $v_0=\Phi$ by (H$_{\mathrm{SB}}$)(i). Suppose that $o_1,\dots,o_{k-1}$ has probability $\lVert v_{k-1}\rVert^2>0$ and leaves $\mathsf B$ in the state $v_{k-1}/\lVert v_{k-1}\rVert$. By the Born rule and Lüders' rule, test $k$ then returns $o_k$ with conditional probability $\lVert Q_kv_{k-1}\rVert^2/\lVert v_{k-1}\rVert^2$ and leaves $v_k/\lVert v_k\rVert$. Multiplying gives the claim for $k$, which holds trivially if $v_{k-1}=0$.
\end{proof}

\begin{lemma}[Quantum union bound]\label{lem:D-qub}
Let $\Phi$ be a unit vector, $Q_1,\dots,Q_K$ orthogonal projectors, and $L:=\sum_{k=1}^K\bra\Phi(1-Q_k)\ket\Phi$. Then $\lVert Q_K\cdots Q_1\Phi\rVert^2\ge1-4L$.
\end{lemma}

\begin{proof}
This is Gao's quantum union bound~\cite[Theorem~1(1-b)]{Gao15} for the density matrix $\rho:=\ket\Phi\bra\Phi$, since $\Tr(Q_K\cdots Q_1\rho\,Q_1\cdots Q_K)=\lVert Q_K\cdots Q_1\Phi\rVert^2$. It also follows from~\cite[Theorem~1.3]{OV22}, and the constant $4$ cannot be lowered for general sequences of projectors~\cite[Section~3.6]{OV22}.
\end{proof}

\begin{proof}[Proof of Theorem~\ref{thm:onebatch}]
Part (a) is Lemma~\ref{lem:D-history}, which holds for every rule and does not involve $m$.

(b) Let $K$ be the number of tests of $\mathfrak h$, and let $A_k$ and $S_k$ be its $k$-th tested set and measured side, after the verdicts $c(A_1),\dots,c(A_{k-1})$. A run whose verdicts are all rank-correct follows $\mathfrak h$: by induction on $k$, its $k$-th test is that of $\mathfrak h$, since both are the same function of the same earlier verdicts. The same induction shows that a run whose first $K$ verdicts are $c(A_1),\dots,c(A_K)$ follows $\mathfrak h$, and so stops after them. So all verdicts are rank-correct exactly when the first $K$ verdicts are $c(A_1),\dots,c(A_K)$. The run fed these verdicts is $\mathfrak h$, so by Lemma~\ref{lem:D-born} this event has probability
\begin{equation*}
\lVert Q_K\cdots Q_1\Phi\rVert^2,
\end{equation*}
where $Q_k:=\Pi^{S_k}_{c(A_k)}$. By the quantum union bound (Lemma~\ref{lem:D-qub}), this probability is at least $1-4L$, where $L:=\sum_k\bra\Phi(1-Q_k)\ket\Phi$.

The reduced state of $\Phi$ on the $S_k$-registers is $\rho_{S_k}^{\otimes m}$, so $\bra\Phi\Pi^{S_k}_{\acc}\ket\Phi=a_m(\rho_{S_k})=a_m(A_k)$. For $S_k=A_k^c$ the second equality holds because $a_m$ depends only on the nonzero eigenvalues (Lemma~\ref{lem:ranktest}(b)), which $\rho_{A_k}$ and $\rho_{A_k^c}$ share with their multiplicities (Section~\ref{sec:s2-states}). If $A_k\in\cluso(T^*)$, then $r(A_k)\le\chi$, $c(A_k)=\acc$ and the $k$-th term of $L$ is $1-a_m(A_k)=0$ by Lemma~\ref{lem:ranktest}(a). If $A_k\in\Nfam_s(T^*)$, then $c(A_k)=\rej$ by Lemma~\ref{lem:D-history}, and the $k$-th term of $L$ is $a_m(A_k)$.

As the $A_k$ are distinct and lie in $\cluso(T^*)\cup\Nfam_s(T^*)$ by (a), $L=L_{\mathfrak h}$. A run that follows $\mathfrak h$ outputs $\hull{T^*}$ by (a), so $\Pr[O=\hull{T^*}]\ge1-4L_{\mathfrak h}$. Finally, $\Nfam_s(\mathfrak h)$ consists of at most $N_T(s)$ distinct sets of $\Nfam_s(T^*)$, so $L_{\mathfrak h}$ is at most each of the two terms of the minimum in (b).

(c) Put $\delta':=\delta/(4N_T(s))$. Then $\delta'\in(0,1)$ and $m^{(s)}_\star=m_R(\chi,\varepsilon_0,\delta')$. Every $A\in\Nfam_s(T^*)$ has $\varepsilon_A(\chi)\ge\gaps(\psi,T^*)\ge\varepsilon_0$, so $a_m(A)\le\delta'$ under either condition on $m$ by Lemma~\ref{lem:ranktest}(c). Hence the second term of the minimum in (b) is at most $N_T(s)\,\delta'=\delta/4$, which gives $\Pr[O=\hull{T^*}]\ge1-\delta$. By (H$_{\mathrm{SB}}$)(i), the learner uses only the $m$ copies of the batch. The bound on $m^{(s)}_\star$ uses $\ln(4(\chi+1)N_T(s)/\delta)\le\ln(8(\chi+1)/\delta)+s\ln(en/s)$, by Lemma~\ref{lem:counting} with $s\le n$, and $\lceil x\rceil\le x+1$. The bound on $m_R^+$ follows in the same way from the inequality for $m_R^+$ in Section~\ref{sec:s2-ranktest}, with $\ln(4N_T(s)/\delta)\le\ln(8/\delta)+s\ln(en/s)$. So both bounds are stated for $s\le n$, as explained after Lemma~\ref{lem:counting}.

(d) There are at most $N_T(s)$ tests by Lemma~\ref{lem:counting}, and each acts on all $m$ copies of $\mathsf B$ by (H$_{\mathrm{SB}}$)(ii). A tested set $W$ is the union of at least two nonempty parts, with at least two other parts outside $W$ (Algorithm~\ref{alg:gosgs}), so $2\le|W|\le n-2$ and $\min\{|W|,n-|W|\}\le\lfloor n/2\rfloor$.
\end{proof}

\paragraph{Inexact tests.}
Theorem~\ref{thm:onebatch} assumes exact tests. For two quantum instruments $\{\Phi_o\}_o$ and $\{\tilde\Phi_o\}_o$ from $\mathsf B$ to $\mathsf B$ with the same finite outcome set, as in Section~\ref{sec:s2-procedures}, let $\mathcal M(\omega):=\sum_o\ket o\bra o\otimes\Phi_o(\omega)$, and likewise $\tilde{\mathcal M}$. The \emph{diamond distance} of the two instruments is
\begin{equation*}
\frac12\sup_\omega\lVert((\tilde{\mathcal M}-\mathcal M)\otimes\mathrm{id})(\omega)\rVert_1,
\end{equation*}
with the trace norm $\lVert\cdot\rVert_1$, where $\omega$ runs over the states of $\mathsf B$ and a reference of any finite dimension~\cite[Section~3.3]{Wat18}. In Lemma~\ref{lem:D-inexact}, the \emph{exact learner} is the given learner with every test performed by the instrument of (H$_{\mathrm{SB}}$).

\begin{lemma}[Inexact tests]\label{lem:D-inexact}
Let a learner run Algorithm~\ref{alg:gosgs} on one batch as in Theorem~\ref{thm:onebatch}, with every test performed by an instrument on $\mathsf B$ at diamond distance at most $\eta$ from the instrument of (H$_{\mathrm{SB}}$). Then, for every input, the law of its output differs from that of the exact learner by at most $N_T(s)\,\eta$ in total variation. Under the conditions of Theorem~\ref{thm:onebatch}(c), it outputs $\hull{T^*}$ with probability at least $1-\delta-N_T(s)\,\eta$.
\end{lemma}

\begin{proof}
Put $\tau:=N_T(s)$. Every run makes at most $\tau$ tests (Lemma~\ref{lem:counting}), and we pad it with identity instruments to exactly $\tau$ steps. Step $j$ of the exact learner is a channel $\mathcal N_j$ on $\mathsf B$ and a classical register $\mathsf V$ of verdicts: it reads the earlier verdicts, applies to $\mathsf B$ the instrument of the test that they determine, and appends the new verdict. Let $\tilde{\mathcal N}_j$ be the same channel with the implemented instruments.

Let $\omega_0$ be the initial state and $\omega_j:=\tilde{\mathcal N}_j\cdots\tilde{\mathcal N}_1(\omega_0)$. Every step appends a classical verdict to $\mathsf V$, which starts empty, so each $\omega_{j-1}$ has the form $\sum_{\mathbf o}p_{\mathbf o}\ket{\mathbf o}\bra{\mathbf o}_{\mathsf V}\otimes\sigma_{\mathbf o}$, with $\mathbf o$ the earlier verdicts and $\sigma_{\mathbf o}$ a state of $\mathsf B$. On such a state the outputs of $\mathcal N_j$ and $\tilde{\mathcal N}_j$ differ in trace norm by $\sum_{\mathbf o}p_{\mathbf o}\lVert(\tilde{\mathcal M}^{\mathbf o}-\mathcal M^{\mathbf o})(\sigma_{\mathbf o})\rVert_1\le2\eta$, where $\mathcal M^{\mathbf o}$ and $\tilde{\mathcal M}^{\mathbf o}$ belong to the test determined by $\mathbf o$. The inequality follows from the bound $\eta$ on the diamond distance, with a reference of dimension one. Telescoping gives
\begin{equation*}
\tilde{\mathcal N}_\tau\cdots\tilde{\mathcal N}_1(\omega_0)-\mathcal N_\tau\cdots\mathcal N_1(\omega_0)=\sum_{j=1}^{\tau}\mathcal N_\tau\cdots\mathcal N_{j+1}\bigl((\tilde{\mathcal N}_j-\mathcal N_j)(\omega_{j-1})\bigr).
\end{equation*}
Channels do not increase the trace norm. So
\begin{equation*}
\bigl\lVert\tilde{\mathcal N}_\tau\cdots\tilde{\mathcal N}_1(\omega_0)-\mathcal N_\tau\cdots\mathcal N_1(\omega_0)\bigr\rVert_1\ \le\ \sum_{j=1}^{\tau}\bigl\lVert(\tilde{\mathcal N}_j-\mathcal N_j)(\omega_{j-1})\bigr\rVert_1\ \le\ 2\tau\eta .
\end{equation*}

The output is a function of the final verdicts, so its law, written as a diagonal state, is the image of the final state under a channel. The total variation distance of two laws is half the trace norm of the difference of the diagonal states. The two laws of the output therefore differ by at most $\tau\eta$ in total variation. The second claim follows from the first and Theorem~\ref{thm:onebatch}(c), since an event changes its probability by at most the total variation distance.
\end{proof}

Lemma~\ref{lem:D-inexact} models each implemented test as one instrument on $\mathsf B$, as (H$_{\mathrm{SB}}$) does. It does not cover an implementation whose ancilla keeps a state from one test to the next, for example an uncomputation that is only approximate and reuses its ancilla.
\section{\texorpdfstring{The hard family and the proof of Theorem~\ref{thm:lower}}{The hard family and the proof of the lower bound}}\label{app:lower}
We prove Theorem~\ref{thm:lower} with the hard family built over an arbitrary finite abelian group $G$ with a symmetric nondegenerate bicharacter. So the construction serves Section~\ref{sec:lower} with the group $\Z_q$ and Section~\ref{sec:largebond} with the group $\Z_d^2$. The argument is outlined after Theorem~\ref{thm:lower}.

After the group and the sites, the second part defines the hard family. Lemmas~\mbox{\ref{lem:E-block}--\ref{lem:E-tree}} describe the family: the block pair, the overlaps of the states $\Psi_x$ and the tree $T_x$. Lemmas~\ref{lem:E-isometry} and~\ref{lem:E-cut} are two tools for the spectrum across a cut. In the part on ranks and tails, Lemmas~\ref{lem:E-membership}--\ref{lem:E-gap} bound the ranks and the tails, which completes part~(a) of the theorem. Clusters are treated in Lemma~\ref{lem:E-membership}, the $\varepsilon$-type sets in Lemma~\ref{lem:E-epstype}, and all other sets in Lemma~\ref{lem:E-perturb}, which rests on Lemmas~\ref{lem:E-graphcut}, \ref{lem:E-startsets} and~\ref{lem:E-deletion}. In the part on discrimination, Lemmas~\ref{lem:E-povm}--\ref{lem:E-dual} lead to Proposition~\ref{prop:E-assouad}, which gives part~(b) from the overlaps and the distinct trees. In the last part, Lemma~\ref{lem:E-constants} gives the constant $c_0$ of part~(c), and Lemma~\ref{lem:E-corconst} gives the bound of Corollary~\ref{cor:rate} on $m^{(s)}_\star$.

\begin{figure}[!hb]
\centering
\begin{tikzpicture}[x=0.85cm,y=0.85cm,
  iv/.style={circle,fill=black,inner sep=1.5pt},
  lf/.style={circle,draw=black,fill=white,inner sep=1.3pt},
  every label/.style={font=\footnotesize,inner sep=2pt}]
\begin{scope}
\node[lf,label=below:$v_1$] (v1) at (0,0) {};
\node[lf,label=below:$v_2$] (v2) at (1.7,0) {};
\node[lf,label=below:$v_3$] (v3) at (3.4,0) {};
\node[lf,label=left:$\gamma_1$] (g1) at (-0.9,1.2) {};
\node[lf,label=right:$\gamma_2$] (g2) at (1.0,1.2) {};
\node[lf,label=above:$\alpha_1$] (a1) at (-1.45,2.3) {};
\node[lf,label=above:$\beta_1$] (b1) at (-0.4,2.3) {};
\node[lf,label=above:$\alpha_2$] (a2) at (0.5,2.3) {};
\node[lf,label=above:$\beta_2$] (b2) at (1.55,2.3) {};
\draw (v1)--(v2)--(v3) (g1)--(v1) (g2)--(v1) (g2)--(v2);
\draw[dashed] (g1)--(a1) (g1)--(b1) (g2)--(a2) (g2)--(b2);
\draw[densely dotted,rounded corners=4pt] (-1.95,0.8) rectangle (-0.05,2.95);
\draw[densely dotted,rounded corners=4pt] (0.15,0.8) rectangle (1.95,2.95);
\node[font=\footnotesize] at (-1.0,3.3) {$B_1$};
\node[font=\footnotesize] at (1.0,3.3) {$B_2$};
\node[font=\small] at (1.0,-1.5) {the graph $H^0_x$};
\end{scope}
\begin{scope}[shift={(7.8,0.5)}]
\node[iv,label=below right:$c_2$] (c2) at (0,0) {};
\node[iv,label=below:$c_3$] (c3) at (1.9,0) {};
\node[iv,label=above:$c_4$] (c4) at (3.8,0) {};
\node[iv,label=below left:$g'_1$] (h1) at (-0.9,1.0) {};
\node[iv,label=left:$g_1$] (k1) at (-0.9,2.0) {};
\node[lf,label=left:$\gamma_1$] (g1) at (-1.9,1.0) {};
\node[lf,label=above:$\alpha_1$] (a1) at (-1.4,2.8) {};
\node[lf,label=above:$\beta_1$] (b1) at (-0.4,2.8) {};
\node[lf,label=below:$v_1$] (v1) at (-0.7,-1.0) {};
\node[iv,label=left:$g'_2$] (h2) at (1.9,1.0) {};
\node[iv,label=left:$g_2$] (k2) at (1.9,2.0) {};
\node[lf,label=right:$\gamma_2$] (g2) at (2.9,1.0) {};
\node[lf,label=above:$\alpha_2$] (a2) at (1.4,2.8) {};
\node[lf,label=above:$\beta_2$] (b2) at (2.4,2.8) {};
\node[lf,label=below:$v_2$] (v2) at (3.8,-1.0) {};
\node[lf,label=right:$v_3$] (v3) at (4.7,0.4) {};
\draw (c2)--(c3)--(c4) (c2)--(h1) (h1)--(k1) (h1)--(g1) (k1)--(a1) (k1)--(b1) (c2)--(v1)
      (c3)--(h2) (h2)--(k2) (h2)--(g2) (k2)--(a2) (k2)--(b2) (c4)--(v2) (c4)--(v3);
\node[font=\small] at (1.9,-2.1) {the tree $T_x$};
\end{scope}
\end{tikzpicture}
\caption{The hard family for $K=2$ and $L=1$, so that $n=9$. Left: the graph $H^0_x$, with the blocks $B_1$ and $B_2$ in dotted boxes. Solid edges are the edges of $E_x$, which carry the gates $CZ_e$: the spine path $v_1v_2v_3$ and the edges from $\gamma_b$ to the attachment set $\mathcal A_b$, with $\mathcal A_1=\{v_1\}$ and $\mathcal A_2=\{v_1,v_2\}$. Dashed edges join $\gamma_b$ to $\alpha_b$ and $\beta_b$ inside the block $B_b$. Right: the tree $T_x$, which hangs the items $B_1$, $\{v_1\}$, $B_2$, $\{v_2\}$, $\{v_3\}$ from the path $c_2c_3c_4$. Open circles are sites, which are the leaves of $T_x$, and filled circles are the internal vertices $g_b$, $g'_b$ and $c_k$ of Lemma~\ref{lem:E-tree}. The value $x_b$ decides which of the sites $2_b$ and $3_b$ is $\beta_b$ and which is $\gamma_b$.}
\label{fig:E-family}
\end{figure}
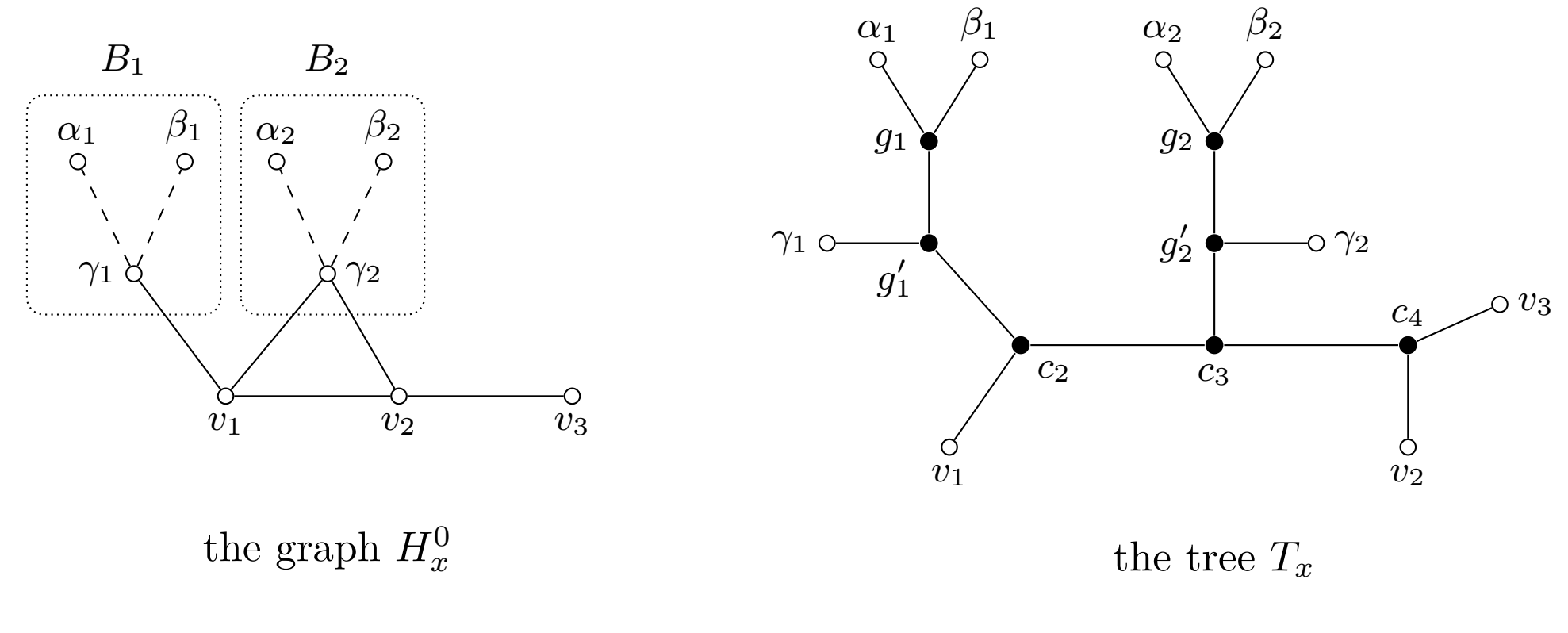

Appendix~\ref{app:largebond} applies Definition~\ref{def:E-hard} over $\Z_d^2$ and uses Lemmas~\ref{lem:E-isometry}, \ref{lem:E-connector}, \ref{lem:E-tree} and~\ref{lem:E-membership}--\ref{lem:E-perturb}, the equivalence displayed in the proof of Lemma~\ref{lem:E-gap}, Proposition~\ref{prop:E-assouad} and Lemma~\ref{lem:E-constants}. Table~\ref{tab:E-notation} collects the notation, and Figure~\ref{fig:E-family} shows the hard family for $K=2$ and $L=1$.

\subsection*{The group and the sites}
Throughout this appendix, $G$ is a finite abelian group of order $q\ge2$, written additively, with a \emph{bicharacter} $\langle\cdot,\cdot\rangle:G\times G\to\C^\times$ that is symmetric, $\langle g,h\rangle=\langle h,g\rangle$, multiplicative in each argument, $\langle g+g',h\rangle=\langle g,h\rangle\langle g',h\rangle$, and nondegenerate: $\langle g,h\rangle=1$ for all $h\in G$ only if $g=0$. Multiplicativity gives $\langle0,h\rangle=1$ and, as $qg=0$, $\langle g,h\rangle^q=\langle qg,h\rangle=1$, so the values are $q$-th roots of unity and $\overline{\langle g,h\rangle}=\langle-g,h\rangle$. Then $\sum_{t\in G}\langle c,t\rangle=q[c=0]$ for every $c\in G$, where a condition in brackets stands for $1$ if it holds and $0$ otherwise. Indeed, every term is $1$ if $c=0$. If $c\ne0$, nondegeneracy gives some $h$ with $\langle c,h\rangle\ne1$, and the shift $t\mapsto t+h$ multiplies the sum by $\langle c,h\rangle$ without changing it, so the sum is $0$. Schlingemann and Werner~\cite{SW01} build graph codes from a graph and such a group, with the phase $\langle g_u,g_v\rangle$ on each edge $uv$, as in the graph states over $G$ of Lemma~\ref{lem:E-graphcut}.

The two instances are $G=\Z_q$ with $\langle a,b\rangle:=\zeta_q^{ab}$, $\zeta_q:=e^{2\pi i/q}$, and $G=\Z_d^k$ with $\langle g,h\rangle:=\omega^{g\cdot h}$, $\omega:=e^{2\pi i/d}$, the latter with $k=2$ in Section~\ref{sec:largebond}.

The Hilbert space of a \emph{site} is $\C^G$, with the orthonormal computational basis $\ket g$, $g\in G$, and the operators
\begin{gather*}
X^g\ket h:=\ket{h+g},\qquad Z^g\ket h:=\langle g,h\rangle\ket h,\qquad\ket+:=q^{-1/2}\sum_{h\in G}\ket h,\\
CZ_{uv}\ket g_u\ket h_v:=\langle g,h\rangle\ket g_u\ket h_v,
\end{gather*}
and we put $Z_{\mathcal A}^g:=\prod_{v\in\mathcal A}Z^g_v$ for a set $\mathcal A$ of sites and $Z^c:=\prod_{v\in\mathcal A}Z^{c_v}_v$ for $c\in G^{\mathcal A}$. The vectors $Z^g\ket+$, $g\in G$, are orthonormal, since $\bra+Z^{g'-g}\ket+=q^{-1}\sum_h\langle g'-g,h\rangle=[g=g']$.

The gates $CZ_{uv}$ are diagonal in the computational basis, hence commute, and $CZ_{uv}=CZ_{vu}$ since the bicharacter is symmetric, so $CZ_e$ is defined for an edge $e=uv$. For a site $c$ and a set $N^\times\not\ni c$ of sites,
\begin{equation*}
\prod_{u\in N^\times}CZ_{cu}=\sum_{t\in G}\ket t\!\bra t_c\otimes Z_{N^\times}^t.
\end{equation*}

Throughout, $\chi$ is an integer with $q\le\chi\le q^2-1$. In Section~\ref{sec:lower}, $q=\min\{d,\chi\}$ with $2\le\chi\le d^2-1$, so that these inequalities hold, and $\C^G=\C^q$ is embedded in $\C^d$ as the span of $\ket0,\dots,\ket{q-1}$. In Section~\ref{sec:largebond}, $q=d^2\le\chi\le d^3-1$, and the code of Appendix~\ref{app:largebond} embeds each site into four physical qudits. For a unit vector $\phi$ and a set $A$ of its sites, $\varepsilon_A(\phi;\chi)$ is the $\chi$-tail of $\rho_A(\phi)$, which vanishes if $A$ is empty or contains all sites. The ranks $r_\phi(A)$, the sets $\TTN(T,\chi)$ and $\TTNo(T,\chi)$ and the gaps are defined as in Section~\ref{sec:setting}, with sites $\C^G$ in place of $\C^d$.

\subsection*{The hard family and its basic properties}
\begin{definition}[The hard family]\label{def:E-hard}
\begin{enumerate}
\item[(a)] The \emph{pair set} over $G$ is the set $P$ of all $(a,b)\in G^2$ with $a\ne b$, which has $M:=|P|=q(q-1)$ elements, and $M\ge\chi+1-q$ because $\chi\le q^2-1$. As in Definition~\ref{def:s5-hard}, with this $P$ and this $M$, the constants are $\theta:=(q+M-\chi)/M\in(0,1]$ and $\bar\varepsilon:=(1-\chi/q^2)/(\sqrt2+\sqrt\theta)^2$, and for $0<\varepsilon<1$ the block pair is
\begin{equation*}
\psi_1:=\sqrt{1-\varepsilon}\,\GHZ_q+\sqrt{\varepsilon/M}\sum_{(a,b)\in P}\ket{aabb}\qquad\text{and}\qquad\psi_2:=W_{(23)}\psi_1,
\end{equation*}
where $\GHZ_q:=q^{-1/2}\sum_{a\in G}\ket{aaaa}$ and $W_{(23)}$ exchanges the sites $2$ and $3$.
\item[(b)] Let $K\ge1$, $L\ge0$, $n:=4K+L$ and $m:=K+L$, and write $[n]$ as the disjoint union of the \emph{blocks} $B_b=\{1_b,2_b,3_b\}$, $b\in[K]$, and the \emph{spine} $S=\{v_1,\dots,v_m\}$. For $x\in\{1,2\}^K$ let $(\alpha_b,\beta_b,\gamma_b):=(1_b,2_b,3_b)$ if $x_b=1$ and $(1_b,3_b,2_b)$ if $x_b=2$, and
\begin{equation*}
\eta_a:=\sqrt{\tfrac{1-\varepsilon}{q}}\,\ket{aa}+\sqrt{\tfrac{\varepsilon}{M}}\sum_{c:\,(c,a)\in P}\ket{cc}\ \ \text{on }(\alpha_b,\beta_b),\qquad\tilde\psi_{x_b}:=\sum_{a\in G}\eta_a\otimes\ket a_{\gamma_b}.
\end{equation*}
With the attachment sets $\mathcal A_1:=\{v_1\}$ and $\mathcal A_b:=\{v_{b-1},v_b\}$ for $2\le b\le K$, and the edge set $E_x:=\{v_iv_{i+1}:i<m\}\cup\{\gamma_bv:b\in[K],\ v\in\mathcal A_b\}$, the \emph{hard states} are
\begin{equation*}
\Psi_x:=\prod_{e\in E_x}CZ_e\Bigl(\bigotimes_{b=1}^K\tilde\psi_{x_b}\otimes\ket+^{\otimes S}\Bigr)\in(\C^G)^{\otimes n}.
\end{equation*}
\item[(c)] Let $\ell:=2K+L$, and let the \emph{items} be $B_1$, $\{v_1\}$, $B_2$, $\{v_2\}$, \dots, $B_K$, $\{v_K\}$, $\{v_{K+1}\}$, \dots, $\{v_m\}$. The tree $T_x$ hangs the items, in this order, from a path $c_2\cdots c_{\ell-1}$, the first two at $c_2$ and the last two at $c_{\ell-1}$. For $\ell=2$, the two items are joined by an edge. An item $\{v\}$ is the leaf $v$, and an item $B_b$ is the rooted subtree $((\alpha_b,\beta_b),\gamma_b)$. The clusters of $T_x$ are the items, the unions of the first $k<\ell$ items, the sets $\{\alpha_b\}$, $\{\beta_b\}$, $\{\gamma_b\}$, $\{\alpha_b,\beta_b\}$, and their complements (Lemma~\ref{lem:E-tree}). The $4K$ \emph{$\varepsilon$-type sets} of $x$, forming the family $\mathcal E_x$, are $\{\alpha_b,\gamma_b\}$, $\{\beta_b,\gamma_b\}$ and their complements, $b\in[K]$.
\end{enumerate}
\end{definition}

Throughout, the symbols introduced in Definition~\ref{def:E-hard} keep their meaning, and $\varepsilon\in(0,1)$ unless stated otherwise.

For a set $D$ of items of Definition~\ref{def:E-hard}(c) we put $D':=\bigcup D$. Put $U_x:=\prod_{e\in E_x}CZ_e$ and $\Phi^0_x:=\bigotimes_b\tilde\psi_{x_b}\otimes\ket+^{\otimes S}$, so that $\Psi_x=U_x\Phi^0_x$, and let $H^0_x$ be the graph on $[n]$ with edge set $E_x\cup\{\gamma_b\alpha_b,\gamma_b\beta_b:b\in[K]\}$. At $\varepsilon=0$, with the formulas of Definition~\ref{def:E-hard}(b), $\Psi_x$ is the graph state $\ket{H^0_x}$ of Lemma~\ref{lem:E-graphcut} up to an inverse Fourier transform on each site $\alpha_b$ and $\beta_b$. The step \emph{Graph form at $\varepsilon=0$} of the proof of Lemma~\ref{lem:E-perturb} defines the Fourier transform and proves this claim. The superscript $0$ of $\Phi^0_x$ marks the vector before the gates, and $H^0_x$ does not depend on $\varepsilon$. A superscript $(0)$, as in the proof of Lemma~\ref{lem:E-perturb}, marks the value at $\varepsilon=0$.

By Lemma~\ref{lem:E-block}(vi), $\psi_1=\sum_{a\in G}\eta_a\otimes\ket a_3\otimes\ket a_4$ records the value $a$ on its sites $3$ and $4$, while the block vector $\tilde\psi_{x_b}=\sum_{a\in G}\eta_a\otimes\ket a_{\gamma_b}$ records it on $\gamma_b$ alone. By the identity for $\prod_{u\in N^\times}CZ_{cu}$, the gates $CZ_{\gamma_bv}$ with $v\in\mathcal A_b$ record it a second time, as $Z^a_{\mathcal A_b}$ on the spine. So the spine takes the role of the fourth sites of the $K$ block pairs (Lemma~\ref{lem:E-connector}(ii)). The spine site $v_j$ lies in $\mathcal A_i$ exactly for $i\in\{j,j+1\}\cap[K]$, which the proofs of Lemmas~\ref{lem:E-connector}(i) and~\ref{lem:E-startsets}(i) use.

In Definition~\ref{def:E-hard}(c), the Newick notation $((\alpha_b,\beta_b),\gamma_b)$ of phylogenetics denotes the rooted tree whose root has the children $\gamma_b$ and a vertex with the children $\alpha_b$ and $\beta_b$. For $\ell\ge3$, $T_x$ hangs the first item at $c_2$, the $k$-th item at $c_k$ for $2\le k\le\ell-1$ and the last item at $c_{\ell-1}$ (Lemma~\ref{lem:E-tree}).

\begin{table}[p]
\centering
\small
\begin{tabular}{@{}lp{7.9cm}l@{}}
\toprule
Symbol & Meaning & Defined in\\
\midrule
$G$, $q$, $\langle\cdot,\cdot\rangle$ & a finite abelian group of order $q$ with a symmetric nondegenerate bicharacter & start of Appendix~\ref{app:lower}\\
$X^g$, $Z^g$, $CZ_{uv}$, $\ket+$ & the shift, the phase and the controlled phase on sites $\C^G$, and the uniform superposition & start of Appendix~\ref{app:lower}\\
$P$, $M$, $\theta$, $\bar\varepsilon$ & the pair set, its size $q(q-1)$ and the two constants of the family & Definition~\ref{def:E-hard}(a)\\
$\psi_1$, $\psi_2$ & the block pair on four sites & Definition~\ref{def:E-hard}(a)\\
$B_b$, $S$ & the blocks $\{1_b,2_b,3_b\}$, $b\in[K]$, and the spine $\{v_1,\dots,v_m\}$ & Definition~\ref{def:E-hard}(b)\\
$(\alpha_b,\beta_b,\gamma_b)$ & the sites of $B_b$ in the order fixed by $x_b$ & Definition~\ref{def:E-hard}(b)\\
$\eta_a$, $\tilde\psi_{x_b}$ & the vectors on $(\alpha_b,\beta_b)$, and the block vector of $B_b$ before the gates & Definition~\ref{def:E-hard}(b)\\
$\mathcal A_b$, $E_x$ & the attachment set of the block $B_b$ in the spine, and the set of edges that carry gates & Definition~\ref{def:E-hard}(b)\\
$\Psi_x$, $T_x$ & the hard states and their trees, $x\in\{1,2\}^K$ & Definition~\ref{def:E-hard}\\
$\mathcal E_x$ & the family of the $4K$ $\varepsilon$-type sets & Definition~\ref{def:E-hard}(c)\\
items & the blocks and the singletons $\{v_j\}$ & Definition~\ref{def:E-hard}(c)\\
$D'$ & the union of a set $D$ of items & after Definition~\ref{def:E-hard}\\
$U_x$, $\Phi^0_x$ & the product of the gates $CZ_e$, $e\in E_x$, and the vector before the gates, so that $\Psi_x=U_x\Phi^0_x$ & after Definition~\ref{def:E-hard}\\
$H^0_x$ & the graph with the edges of $E_x$ and the edges $\gamma_b\alpha_b$, $\gamma_b\beta_b$ & after Definition~\ref{def:E-hard}\\
$I_k$, $\mathrm{Pre}_k$ & the items of $T_x$ in their order, and the union of the first $k$ items & Lemma~\ref{lem:E-tree}\\
$E^\times(A)$, $U^\times$, $U^A$ & the edges of $E_x$ with exactly one end in $A$, the product of their gates, and the product of the gates inside $A$ & Lemma~\ref{lem:E-cut}\\
$\partial A$, star set & the edges of $H^0_x$ with exactly one end in $A$. The set $A$ is a star set if they have a common vertex & Lemma~\ref{lem:E-startsets}\\
$\Psi_x[D]$ & the vector built like $\Psi_x$ from the items of $D$ alone & Lemma~\ref{lem:E-deletion}\\
$\PH(\sigma)$, $N_K(\varepsilon,\delta)$ & the best probability of telling apart two equally likely pure states with overlap $\sigma$, and the lower bound on the number of copies & Section~\ref{sec:lower}\\
$a_K$, $G_K(\delta)$, $c_0$ & the quantities that define $N_K(\varepsilon,\delta)$, and the numerical constant of Theorem~\ref{thm:lower}(c) & Section~\ref{sec:lower}\\
\bottomrule
\end{tabular}
\caption{Notation used in Appendix~\ref{app:lower}.}
\label{tab:E-notation}
\end{table}

\begin{lemma}[Isometries on one side of a cut]\label{lem:E-isometry}
Let $\phi\in\mathcal K_1\otimes\mathcal K_2$ be a unit vector, with finite-dimensional factors, and let $J_i:\mathcal K_i\to\mathcal K'_i$ be isometries. The reduced operators of $(J_1\otimes J_2)\phi$ on $\mathcal K'_1$ and of $\phi$ on $\mathcal K_1$ have the same nonzero eigenvalues, with the same multiplicities, and so do the two reduced operators of $\phi$. In particular ranks and tails of a pure state across a cut are the same for both sides. They do not change under isometries that each act within one side of the cut, such as the inclusion $\C^q\subseteq\C^d$ or the code of Appendix~\ref{app:largebond} applied to each site.
\end{lemma}

\begin{proof}
Take a Schmidt decomposition $\phi=\sum_k\sqrt{\mu_k}\,u_k\otimes w_k$ with orthonormal families $(u_k)$, $(w_k)$ and $\mu_k>0$. The reduced operators are $\sum_k\mu_k\ket{u_k}\!\bra{u_k}$ and $\sum_k\mu_k\ket{w_k}\!\bra{w_k}$, and $(J_1\otimes J_2)\phi=\sum_k\sqrt{\mu_k}\,J_1u_k\otimes J_2w_k$ is again a Schmidt decomposition, since isometries preserve orthonormality.
\end{proof}

A \emph{quartet tree} is a binary X-tree with four leaves, and the one with cherries $\{i,j\}$ and $\{k,l\}$ is written $ij|kl$~\cite[p.~116]{SS03}. As in Appendix~\ref{app:trees}, a cherry is a cluster with two elements, and $ij|kl$ also denotes the cut $\{i,j\}|\{k,l\}$ across its internal edge.

\begin{lemma}[The block pair]\label{lem:E-block}
Let $0<\varepsilon<1$.
\begin{enumerate}
\item[(i)] $\psi_1$ and $\psi_2$ are unit vectors, and $\braket{\psi_1}{\psi_2}=1-\varepsilon$.
\item[(ii)] $r_{\psi_1}(\{1,2\})\le q$ and $r_{\psi_1}(\{i\})\le q$ for every site $i$, hence $\psi_1\in\TTN(12|34,\chi)$.
\item[(iii)] Across $13|24$ and across $14|23$ the nonzero spectrum of $\psi_1$ consists of $q$ eigenvalues $(1-\varepsilon)/q$ and $M$ eigenvalues $\varepsilon/M$, so its rank $q+M$ exceeds $\chi$. Hence $12|34$ is the only quartet tree $T$ with $\psi_1\in\TTNo(T,\chi)$.
\item[(iv)] $\Nfam_2(12|34)$ consists of the pairs $\{1,3\}$, $\{1,4\}$, $\{2,3\}$, $\{2,4\}$, and if $\varepsilon\le M/(M+q)$, every member $A$ has $\varepsilon_A(\psi_1;\chi)=\theta\varepsilon$.
\item[(v)] (ii)--(iv) hold for $\psi_2$ with the sites $2$ and $3$ exchanged in every set and tree, which exchanges $12|34$ and $13|24$ and fixes $14|23$.
\item[(vi)] $\psi_1=\sum_{a\in G}\eta_a\otimes\ket a_3\otimes\ket a_4$, with $\eta_a$ of Definition~\ref{def:E-hard} on the sites $1,2$, and $\sum_a\lVert\eta_a\rVert^2=1$.
\end{enumerate}
\end{lemma}

\begin{proof}
(i) The basis vectors $\ket{aaaa}$ with $a\in G$ and $\ket{aabb}$, $\ket{abab}$ with $(a,b)\in P$ are pairwise distinct: the last two families have $a\ne b$, and $\ket{aabb}=\ket{a'b'a'b'}$ would force $a=a'=b'=b$. Hence $\lVert\psi_j\rVert^2=(1-\varepsilon)+M\varepsilon/M=1$ and $\braket{\psi_1}{\psi_2}=(1-\varepsilon)\braket{\GHZ_q}{\GHZ_q}=1-\varepsilon$.

(ii) Across $12|34$, with rows $x_1x_2$ and columns $x_3x_4$, all nonzero entries lie in the $q$ rows $aa$, and every one-site reduced state is supported on the $q$-dimensional space $\C^G$. The clusters of $12|34$ are the singletons, $\{1,2\}$, $\{3,4\}$ and their complements, which have the same ranks (Lemma~\ref{lem:E-isometry}). As $q\le\chi$, it follows that $\psi_1\in\TTN(12|34,\chi)$.

(iii) Across $13|24$, with rows $x_1x_3$ and columns $x_2x_4$, the term $\ket{aaaa}$ sits at $(aa,aa)$ and $\ket{aabb}$ at $(ab,ab)$. These $q+M$ entries occupy distinct rows and distinct columns, so up to permutations the matrix is diagonal with $q$ entries $\sqrt{(1-\varepsilon)/q}$ and $M$ entries $\sqrt{\varepsilon/M}$. Across $14|23$, with rows $x_1x_4$ and columns $x_2x_3$, the same holds. The rank $q+M\ge\chi+1$ violates the constraint on the internal edge of $13|24$ and of $14|23$, and with (ii) the claim follows.

(iv) The unions of two disjoint clusters of $12|34$ other than $[4]$ are the six pairs and the four triples, and the triples, $\{1,2\}$ and $\{3,4\}$ are clusters. This gives $\Nfam_2(12|34)$, and across each of its members the state has the spectrum of (iii) (Lemma~\ref{lem:E-isometry}). If $\varepsilon/M\le(1-\varepsilon)/q$, that is $\varepsilon\le M/(M+q)$, the $\chi$ largest eigenvalues are the $q$ values $(1-\varepsilon)/q$ and $\chi-q$ values $\varepsilon/M$, where $0\le\chi-q\le M-1$ because $q\le\chi$ and $M\ge\chi+1-q$. The tail is then $(M-\chi+q)\varepsilon/M=\theta\varepsilon$.

(v) $\psi_2=W_{(23)}\psi_1$, and the exchange of the sites $2,3$ exchanges $12|34$ and $13|24$, fixes $14|23$, and maps $\Nfam_2(12|34)$ to $\Nfam_2(13|24)$.

(vi) Collect the terms of $\psi_1$ by the value $a$ of site $3$: $\ket{aaaa}$ gives $\sqrt{(1-\varepsilon)/q}\,\ket{aa}_{12}$, and $\ket{ccaa}$ with $(c,a)\in P$ gives $\sqrt{\varepsilon/M}\,\ket{cc}_{12}$. In both, site $4$ equals site $3$, and $(a,a)\notin P$ excludes double counting. As the vectors $\ket a_3\ket a_4$ are orthonormal, it follows that $\sum_a\lVert\eta_a\rVert^2=\lVert\psi_1\rVert^2=1$.
\end{proof}

\begin{lemma}[Connector and overlaps]\label{lem:E-connector}
Let $\ket{H_S}:=\prod_{i<m}CZ_{v_iv_{i+1}}\ket+^{\otimes S}$ and $\mathcal W\ket{\mathbf a}:=\prod_bZ_{\mathcal A_b}^{a_b}\ket{H_S}$ for $\mathbf a\in G^K$. For $a\in G$ let $\phi^{x_b}_a:=\eta_a\otimes\ket a_{\gamma_b}\in(\C^G)^{\otimes B_b}$, with $\eta_a$ on $(\alpha_b,\beta_b)$.
\begin{enumerate}
\item[(i)] $\mathcal W:(\C^G)^{\otimes K}\to(\C^G)^{\otimes S}$ is an isometry.
\item[(ii)] $\Psi_x=\sum_{\mathbf a\in G^K}\bigl(\bigotimes_b\phi^{x_b}_{a_b}\bigr)\otimes\mathcal W\ket{\mathbf a}$ and $\lVert\Psi_x\rVert=1$. Equivalently, $\Psi_x=(1\otimes\mathcal W)\bigotimes_b\psi^{(b)}_{x_b}$, where $\psi^{(b)}_{x_b}:=\sum_a\phi^{x_b}_a\otimes\ket a_{\xi_b}$ is the state $\psi_{x_b}$ of Definition~\ref{def:E-hard}(a) on the sites $(1_b,2_b,3_b,\xi_b)$, and $\mathcal W$ acts on auxiliary sites $\xi_1,\dots,\xi_K$.
\item[(iii)] $\braket{\Psi_x}{\Psi_y}=\prod_b\braket{\psi^{(b)}_{x_b}}{\psi^{(b)}_{y_b}}=(1-\varepsilon)^{\dH(x,y)}$.
\end{enumerate}
\end{lemma}

\begin{proof}
(i) Put $z(\mathbf a):=\sum_ba_b\mathbf 1_{\mathcal A_b}\in G^S$, so $\mathcal W\ket{\mathbf a}=Z^{z(\mathbf a)}\ket{H_S}$. All amplitudes of $\ket{H_S}$ have modulus $q^{-m/2}$, as the gates multiply those of $\ket+^{\otimes S}$ by phases, and $Z^c$ multiplies the amplitude of $\ket y$ by $\prod_v\langle c_v,y_v\rangle$. So
\begin{equation*}
\braket{\mathcal W\mathbf a}{\mathcal W\mathbf a'}=q^{-m}\sum_{y\in G^S}\prod_v\langle z(\mathbf a')_v-z(\mathbf a)_v,y_v\rangle=[z(\mathbf a)=z(\mathbf a')].
\end{equation*}
As $v_j\in\mathcal A_i$ iff $i\in\{j,j+1\}\cap[K]$, $z(\mathbf a)_{v_K}=a_K$ and $z(\mathbf a)_{v_j}=a_j+a_{j+1}$ for $j<K$, which recovers $a_K,a_{K-1},\dots,a_1$. So $z$ is injective.

(ii) For a basis value $t$ of $\gamma_b$ and $\nu\in(\C^G)^{\otimes S}$, $\prod_{v\in\mathcal A_b}CZ_{\gamma_bv}(\ket t_{\gamma_b}\otimes\nu)=\ket t_{\gamma_b}\otimes Z^t_{\mathcal A_b}\nu$. Expanding $\Phi^0_x$ in the computational basis of the sites $\gamma_b$, and using that the spine gates commute with the $Z$'s, gives the first formula, and $\lVert\Psi_x\rVert=\lVert\Phi^0_x\rVert=1$ by Lemma~\ref{lem:E-block}(vi). By Lemma~\ref{lem:E-block}(vi) again, $\psi^{(b)}_{x_b}$ is $\psi_1$ on $(\alpha_b,\beta_b,\gamma_b,\xi_b)$, which is $\psi_1$ on $(1_b,2_b,3_b,\xi_b)$ if $x_b=1$ and $W_{(23)}\psi_1=\psi_2$ there if $x_b=2$.

(iii) By the first formula of (ii) and by (i),
\begin{equation*}
\braket{\Psi_x}{\Psi_y}=\sum_{\mathbf a}\prod_b\braket{\phi^{x_b}_{a_b}}{\phi^{y_b}_{a_b}}=\prod_b\sum_{a\in G}\braket{\phi^{x_b}_a}{\phi^{y_b}_a}=\prod_b\braket{\psi^{(b)}_{x_b}}{\psi^{(b)}_{y_b}},
\end{equation*}
and each factor is $1$ or $1-\varepsilon$ by Lemma~\ref{lem:E-block}(i).
\end{proof}

Lemma~\ref{lem:E-tree} lists the vertices and edges of the tree $T_x$ of Definition~\ref{def:E-hard}(c) and proves the list of its clusters.

\begin{lemma}[The tree]\label{lem:E-tree}
Let the items be $I_{2b-1}:=B_b$ and $I_{2b}:=\{v_b\}$ for $b\le K$ and $I_{2K+j}:=\{v_{K+j}\}$ for $j\le L$, and let $\mathrm{Pre}_k:=I_1\cup\dots\cup I_k$. Put $\mathrm{rt}(B_b):=g'_b$ and $\mathrm{rt}(\{v\}):=v$. The tree $T_x$ has the leaves $[n]$ and the internal vertices $g_b,g'_b$ for $b\in[K]$ and $c_2,\dots,c_{\ell-1}$. Its edges are $\alpha_bg_b$, $\beta_bg_b$, $g_bg'_b$ and $\gamma_bg'_b$ for $b\in[K]$, together with the edge $g'_1v_1$ if $\ell=2$, and, if $\ell\ge3$, with the edges $c_kc_{k+1}$ for $2\le k\le\ell-2$, $\mathrm{rt}(I_1)c_2$, $\mathrm{rt}(I_k)c_k$ for $2\le k\le\ell-1$ and $\mathrm{rt}(I_\ell)c_{\ell-1}$. It is a binary tree with leaf set $[n]$, and
\begin{align*}
\clus(T_x)={}&\{I_k,[n]\setminus I_k\}_{k\le\ell}\cup\{\mathrm{Pre}_k,[n]\setminus\mathrm{Pre}_k\}_{k<\ell}\\
&\cup\bigl\{Y,[n]\setminus Y:\ Y\in\{\{\alpha_b\},\{\beta_b\},\{\gamma_b\},\{\alpha_b,\beta_b\}\},\ b\in[K]\bigr\}.
\end{align*}
A cluster that meets some $B_b$ in a nonempty proper subset is one of $\{\alpha_b\}$, $\{\beta_b\}$, $\{\gamma_b\}$, $\{\alpha_b,\beta_b\}$ or a complement of one. The $2^K$ trees $T_x$ are pairwise distinct.
\end{lemma}

\begin{proof}
There are $n+2K+(\ell-2)=2n-2$ vertices and $2n-3$ edges, namely $4K+1$ for $\ell=2$ and $4K+(\ell-3)+\ell$ for $\ell\ge3$. Leaves have degree $1$. Each $g_b$ is adjacent to $\alpha_b$, $\beta_b$ and $g'_b$, and each $g'_b$ to $g_b$, $\gamma_b$ and the other end of its item edge. For $\ell\ge4$, $c_2$ is adjacent to $\mathrm{rt}(I_1)$, $\mathrm{rt}(I_2)$ and $c_3$, $c_{\ell-1}$ to $c_{\ell-2}$, $\mathrm{rt}(I_{\ell-1})$ and $\mathrm{rt}(I_\ell)$, and every other $c_k$ to $c_{k-1}$, $c_{k+1}$ and $\mathrm{rt}(I_k)$. For $\ell=3$ the single vertex $c_2$ is adjacent to $\mathrm{rt}(I_1)$, $\mathrm{rt}(I_2)$ and $\mathrm{rt}(I_3)$. The graph is connected, since every item hangs from the path $c_2\cdots c_{\ell-1}$, or from the edge $g'_1v_1$ if $\ell=2$, so with $|V|-1$ edges it is a tree, binary with leaf set $[n]$.

The edges $\alpha_bg_b$, $\beta_bg_b$, $g_bg'_b$, $\gamma_bg'_b$ have the sides $\{\alpha_b\}$, $\{\beta_b\}$, $\{\alpha_b,\beta_b\}$, $\{\gamma_b\}$, respectively. The item edge of $I_k$ has the side $I_k$, the edge $c_kc_{k+1}$ has the side $\mathrm{Pre}_k$, and for $\ell=2$ the edge $g'_1v_1$ has the sides $B_1=\mathrm{Pre}_1$ and $\{v_1\}$. Moreover, $\mathrm{Pre}_1=I_1$ and $\mathrm{Pre}_{\ell-1}=[n]\setminus I_\ell$. This gives $\clus(T_x)$. Items, prefixes and their complements meet each $B_b$ in $\emptyset$ or $B_b$, and so do the sets $Y\subseteq B_{b'}$ of the display with $b'\ne b$ and their complements. This gives the next claim.

If $x_b\ne y_b$, say $x_b=1$, then $\{1_b,2_b\}\in\clus(T_x)$. The clusters of $T_y$ meeting $B_b$ properly are $\{1_b\},\{3_b\},\{2_b\},\{1_b,3_b\}$, none of which is $\{1_b,2_b\}$, and their complements, which contain $v_1$. So $\{1_b,2_b\}\notin\clus(T_y)$, and $T_x\ne T_y$, as leaf-label-preserving isomorphisms preserve clusters.
\end{proof}

\begin{lemma}[CZ-cut]\label{lem:E-cut}
For $A\subseteq[n]$ let $E^\times(A)$ be the set of edges of $E_x$ with exactly one end in $A$, $U^\times:=\prod_{e\in E^\times(A)}CZ_e$, and $U^A$ the product of the $CZ_e$ over $e\in E_x$ with $e\subseteq A$. Then $\rho_A(\Psi_x)=U^A\rho_A(U^\times\Phi^0_x)(U^A)^\dagger$.
\end{lemma}

\begin{proof}
The gates commute, so $U_x=U^AU^{A^c}U^\times$, where $U^{A^c}$ is the product over the edges inside $A^c$. The partial trace over $A^c$ removes $U^{A^c}$ and conjugates by $U^A$.
\end{proof}
\subsection*{Ranks and tails}
\begin{lemma}[Membership]\label{lem:E-membership}
$r_{\Psi_x}(A)\le q\le\chi$ for every $A\in\clus(T_x)$. Hence $\Psi_x\in\TTN(T_x,\chi)\subseteq\TTNo(T_x,\chi)$ for every $\varepsilon\in(0,1)$, as a vector of $(\C^G)^{\otimes n}$ and, by Lemma~\ref{lem:E-isometry}, after any isometric embedding of the sites.
\end{lemma}

\begin{proof}
By Lemma~\ref{lem:E-isometry} it suffices to treat the clusters $Y$ listed in Lemma~\ref{lem:E-tree}, not their complements. For a singleton the reduced state is supported on the $q$-dimensional space $\C^G$. For $Y=\{\alpha_b,\beta_b\}$, no edge of $E_x$ touches $\alpha_b$ or $\beta_b$, so $E^\times(Y)=\emptyset$, and by Lemma~\ref{lem:E-cut} the spectrum is that of $\tilde\psi_{x_b}=\sum_{a\in G}\eta_a\otimes\ket a_{\gamma_b}$ across $\{\alpha_b,\beta_b\}|\{\gamma_b\}$, of rank at most $q$.

The other listed clusters are the items and the prefixes $\mathrm{Pre}_k$ with $k<\ell$. Each is a union of items, so $\Phi^0_x=\Xi_Y\otimes\Xi_{Y^c}$ with unit vectors $\Xi_Y$ and $\Xi_{Y^c}$ on $Y$ and $Y^c$. The edges of $E^\times(Y)$ share a vertex $c$, as the following list shows.
\begin{itemize}
\item For $Y=B_b$, $E^\times(Y)=\{\gamma_bv:v\in\mathcal A_b\}$ and $c=\gamma_b$.
\item For $Y=\{v_j\}$, $c=v_j$.
\item For $Y=\mathrm{Pre}_{2j}$ with $j\le K$, $E^\times(Y)\subseteq\{v_jv_{j+1},\gamma_{j+1}v_j\}$ and $c=v_j$, because $\gamma_i\in Y$ iff $i\le j$, in which case $\mathcal A_i\subseteq Y$, while for $i>j$ the set $\mathcal A_i$ meets $Y$ only if $i=j+1$.
\item For $Y=\mathrm{Pre}_{2j-1}$ with $2\le j\le K$, $E^\times(Y)=\{v_{j-1}v_j,\gamma_jv_j\}$ and $c=v_j$. The case $\mathrm{Pre}_1=B_1$ is that of a block.
\item For $Y=\mathrm{Pre}_k$ with $k>2K$, $E^\times(Y)$ is the single edge $v_{k-K}v_{k-K+1}$, and $c=v_{k-K}$.
\end{itemize}
If $c\in Y$, let $N^\times\subseteq Y^c$ be the set of other ends. Then $U^\times=\prod_{u\in N^\times}CZ_{cu}$. By the identity for this product, $U^\times=\sum_{t\in G}\ket t\!\bra t_c\otimes Z^t_{N^\times}$, so
\begin{equation*}
U^\times\Phi^0_x=\sum_t(\ket t\!\bra t_c\Xi_Y)\otimes(Z^t_{N^\times}\Xi_{Y^c})
\end{equation*}
has Schmidt rank at most $q$ across $Y|Y^c$. If $c\notin Y$, the same argument with $Y$ and $Y^c$ exchanged, where now $N^\times\subseteq Y$, gives the same bound. By Lemma~\ref{lem:E-cut}, $\rho_Y(\Psi_x)$ is a unitary conjugate of $\rho_Y(U^\times\Phi^0_x)$, so $r_{\Psi_x}(Y)\le q$.
\end{proof}

\begin{lemma}[$\varepsilon$-type sets]\label{lem:E-epstype}
Let $A\in\{\{\alpha_b,\gamma_b\},\{\beta_b,\gamma_b\}\}$. Then $A$ and $[n]\setminus A$ lie in $\Nfam_2(T_x)$, and the nonzero spectrum of $\rho_A(\Psi_x)$ consists of $q$ eigenvalues $(1-\varepsilon)/q$ and $M$ eigenvalues $\varepsilon/M$. Hence $\mathcal E_x\subseteq\Nfam_2(T_x)$, and if $\varepsilon\le M/(M+q)$, every $A\in\mathcal E_x$ has $\varepsilon_A(\Psi_x;\chi)=\theta\varepsilon$.
\end{lemma}

\begin{proof}
By the symmetry between $\alpha_b$ and $\beta_b$, let $A=\{\alpha_b,\gamma_b\}$.

\emph{Unions.} $A=\{\alpha_b\}\sqcup\{\gamma_b\}$ and $[n]\setminus A=([n]\setminus B_b)\sqcup\{\beta_b\}$ are unions of two disjoint clusters of $T_x$, different from $[n]$ (Lemma~\ref{lem:E-tree}).

\emph{Not clusters.} Both sets meet $B_b$ in a nonempty proper subset, so by Lemma~\ref{lem:E-tree} a cluster equal to one of them is $\{\alpha_b\}$, $\{\beta_b\}$, $\{\gamma_b\}$, $\{\alpha_b,\beta_b\}$ or a complement of one. Among these, only $[n]\setminus\{\beta_b\}$ meets $B_b$ in $A\cap B_b=\{\alpha_b,\gamma_b\}$, and it has $n-1\ne2=|A|$ elements, so $A$ is not a cluster. Likewise, only $\{\beta_b\}$ meets $B_b$ in $([n]\setminus A)\cap B_b=\{\beta_b\}$, and it has $1\ne n-2$ elements, so $[n]\setminus A$ is not a cluster.

\emph{Spectrum.} The edges of $E_x$ touching $A$ are the $\gamma_bv$ with $v\in\mathcal A_b\subseteq A^c$, so $U^A$ is the identity, and by Lemma~\ref{lem:E-cut} the spectrum is that of $\rho_A$ of
\begin{equation*}
U^\times\Phi^0_x=\Bigl[\sum_{a\in G}\eta_a\otimes\ket a_{\gamma_b}\otimes\tau_a\Bigr]\otimes\Xi,\qquad\tau_a:=Z^a_{\mathcal A_b}\ket+^{\otimes\mathcal A_b},
\end{equation*}
with $\Xi$ a unit vector on the remaining sites, all in $A^c$. As the vectors $Z^a\ket+$ are orthonormal and $\mathcal A_b\ne\emptyset$, the $\tau_a$ are orthonormal. So, for an auxiliary site $\xi$, $J:\ket a_\xi\mapsto\tau_a\otimes\Xi$ is an isometry into the space of the sites of $A^c\setminus\{\beta_b\}$, and $U^\times\Phi^0_x=(1\otimes J)\psi_1$ with $(\alpha_b,\beta_b,\gamma_b,\xi)$ in the roles of $(1,2,3,4)$ (Lemma~\ref{lem:E-block}(vi)). By Lemma~\ref{lem:E-isometry} the spectrum is that of $\psi_1$ across $13|24$, and for $\{\beta_b,\gamma_b\}$ across $14|23$. Lemma~\ref{lem:E-block}(iii),(iv) gives the spectrum and the tail, and complements have the same tails.
\end{proof}

The remaining nonempty proper sets, outside $\clus(T_x)\cup\mathcal E_x$, are treated in Lemma~\ref{lem:E-perturb} by a comparison with the graph states over $G$ of the next lemma. In Lemma~\ref{lem:E-graphcut}, $\Gamma[B,A]$ is the submatrix of $\Gamma$ with rows in $B$ and columns in $A$. It acts on $G^A$, a $\Z$-module as every abelian group is, by $(\Gamma[B,A]c)_w=\sum_{u\in A}\Gamma_{wu}c_u$ for $c\in G^A$ and $w\in B$.

\begin{lemma}[Cuts of graph states over $G$]\label{lem:E-graphcut}
Let $H$ be a graph on a finite set $V$ with adjacency matrix $\Gamma$, and let $\ket H:=\prod_{uv\in E(H)}CZ_{uv}\ket+^{\otimes V}\in(\C^G)^{\otimes V}$. For $\emptyset\ne A\subsetneq V$ and $B:=V\setminus A$ let $\Gamma_{BA}:G^A\to G^B$ be the homomorphism $c\mapsto\Gamma[B,A]c$, where the integer matrix acts on the $\Z$-module $G^A$. Then the Schmidt rank of $\ket H$ across $A|B$ is $r_{\ket H}(A)=|\Img\Gamma_{BA}|$, and $\rho_A(\ket H)=\Pi_A/r_{\ket H}(A)$ for an orthogonal projector $\Pi_A$. If there are distinct $u_1,u_2\in A$ and $w_1,w_2\in B$ such that $u_1w_1,u_2w_2\in E(H)$ and not both $u_1w_2,u_2w_1\in E(H)$, then $r_{\ket H}(A)\ge q^2$, and $\varepsilon_A(\ket H;\chi)=1-\chi/r_{\ket H}(A)\ge1-\chi/q^2$ for every $\chi<q^2$. For $G=\Z_d^k$ with $d$ prime, $|\Img\Gamma_{BA}|=d^{k\,\operatorname{rank}_{\F_d}\Gamma[B,A]}$.
\end{lemma}

\begin{proof}
\emph{Idea.} By the character sum, the nonzero entries of $\rho_A$ are exactly those between basis states of $A$ whose difference lies in $\ker\Gamma_{BA}$, each $q^{-|A|}$ times a phase from the gates inside $A$. Up to conjugation by these gates, $\rho_A$ is therefore a multiple of the projector onto the functions constant on the cosets of this kernel, whose rank is the number of cosets, $|\Img\Gamma_{BA}|$. Two edges as in the second claim give $|\Img\Gamma_{BA}|\ge q^2$, and for $G=\Z_d^k$ with $d$ prime the map $\Gamma_{BA}$ is linear over $\F_d$.

Write $y=(y_A,y_B)\in G^V$ and, by multiplicativity of the bicharacter,
\begin{equation*}
Q(y):=\prod_{uv\in E(H)}\langle y_u,y_v\rangle=Q_A(y_A)\,Q_B(y_B)\prod_{w\in B}\langle(\Gamma_{BA}y_A)_w,y_w\rangle,
\end{equation*}
where $Q_A$ and $Q_B$ are the products of $\langle y_u,y_v\rangle$ over the edges inside $A$ and inside $B$. With this $Q$, $\ket H=q^{-|V|/2}\sum_yQ(y)\ket y$. All values of $Q$ have modulus one, and the character sum $\sum_{y_w\in G}\langle g,y_w\rangle=q[g=0]$ gives
\begin{equation*}
\bra{y_A}\rho_A\ket{y'_A}=q^{-|A|}\,Q_A(y_A)\overline{Q_A(y'_A)}\,[\Gamma_{BA}(y_A-y'_A)=0].
\end{equation*}
So $\rho_A=q^{-|A|}CZ_A\bigl(\sum_{k\in\ker\Gamma_{BA}}S_k\bigr)CZ_A^\dagger$ with $CZ_A:=\sum_{y_A}Q_A(y_A)\ket{y_A}\!\bra{y_A}$, the product of the gates inside $A$, and $S_k\ket{y_A}:=\ket{y_A+k}$. The shifts $S_k$, $k\in\ker\Gamma_{BA}$, form a unitary representation of this finite abelian group, so $\Pi:=|\ker\Gamma_{BA}|^{-1}\sum_kS_k$ is the orthogonal projector onto the invariant vectors, the functions constant on the cosets of $\ker\Gamma_{BA}$. Its rank is the number of cosets, $q^{|A|}/|\ker\Gamma_{BA}|=|\Img\Gamma_{BA}|$. Hence $\rho_A=\Pi_A/|\Img\Gamma_{BA}|$ for the orthogonal projector $\Pi_A:=CZ_A\,\Pi\,CZ_A^\dagger$ of rank $|\Img\Gamma_{BA}|$, and $r_{\ket H}(A)=\rank\rho_A=|\Img\Gamma_{BA}|$.

For the second claim, let $u_1w_2\notin E(H)$, say. A vector $c\in G^A$ supported on $\{u_1,u_2\}$ is mapped to a vector whose coordinates at $w_1,w_2$ are $(c_{u_1}+\Gamma_{w_1u_2}c_{u_2},\ c_{u_2})$, and $(c_{u_1},c_{u_2})\mapsto(c_{u_1}+\Gamma_{w_1u_2}c_{u_2},c_{u_2})$ is a bijection of $G^2$. This map is the action on $G^2$ of the $2\times2$ block of $\Gamma[B,A]$ with rows $w_1,w_2$ and columns $u_1,u_2$, which is unitriangular and so unimodular. Hence $|\Img\Gamma_{BA}|\ge q^2$. By the first claim, the spectrum of $\rho_A$ is flat, that is, its nonzero eigenvalues are equal, with $r_{\ket H}(A)>\chi$ eigenvalues $1/r_{\ket H}(A)$, so the $\chi$-tail is $1-\chi/r_{\ket H}(A)\ge1-\chi/q^2$.

For $G=\Z_d^k$, the module $G^A$ is $\F_d^A\otimes\F_d^k$, on which $\Gamma_{BA}$ acts as $\Gamma[B,A]\otimes1$, whose image has dimension $k\operatorname{rank}_{\F_d}\Gamma[B,A]$ over $\F_d$.
\end{proof}

For $q=2$ the first claim of Lemma~\ref{lem:E-graphcut} is the Schmidt-rank formula of Hein, Eisert and Briegel~\cite[Proposition~3 and Eqs.~(41)--(43)]{HEB04}, and for cyclic $G$ it follows from the description of reduced states of stabilizer states by Looi and Griffiths~\cite[Sec.~II~C, Eqs.~(14)--(16)]{LG11}. For $\Z_q$ graph states, see also~\cite{LYGG08,HDD05}. For $G=\Z_q$ with composite $q$, the image $\Img\Gamma_{BA}$ is a subgroup rather than a subspace, and its order need not be a power of $q$, which is why the lemma is stated with $|\Img\Gamma_{BA}|$.

In the proof of Lemma~\ref{lem:E-perturb}, part~(iii) of the next lemma supplies the two edges that the second claim of Lemma~\ref{lem:E-graphcut} asks for. Lemma~\ref{lem:E-startsets} takes boundaries in $H^0_x$, the graph with the edges of $E_x$ and the edges $\gamma_b\alpha_b$ and $\gamma_b\beta_b$ inside the blocks. The set $E^\times(A)$ of Lemma~\ref{lem:E-cut} contains edges of $E_x$ only.

\begin{lemma}[Star sets]\label{lem:E-startsets}
For $\emptyset\ne A\subsetneq[n]$ let $\partial A$ be the set of edges of $H^0_x$ with exactly one end in $A$, and call $A$ a \emph{star set} if the edges in $\partial A$ have a common vertex.
\begin{enumerate}
\item[(i)] $H^0_x$ is connected and contains no cycle of length $4$.
\item[(ii)] The star sets are exactly the members of $\clus(T_x)\cup\mathcal E_x$.
\item[(iii)] If $\emptyset\ne A\subsetneq[n]$ and $A\notin\clus(T_x)\cup\mathcal E_x$, then $\partial A$ contains two disjoint edges $u_1w_1$, $u_2w_2$ with $u_1,u_2\in A$ and $w_1,w_2\notin A$ such that $u_1w_2$ and $u_2w_1$ are not both edges of $H^0_x$.
\end{enumerate}
\end{lemma}

\begin{proof}
\emph{Idea.} Part~(i) is read off from the neighborhoods in $H^0_x$, and (ii) first exhibits a common vertex of the boundary of each member. Conversely, a star set that contains a common vertex $c$ consists of $c$ and some, but not all, components of $H^0_x-c$, which are listed for each kind of vertex $c$. Part~(iii) finds two disjoint edges $u_1w_1$, $u_2w_2$ among three edges of $\partial A$ without a common vertex, each with exactly one end in $A$, and (i) excludes the $4$-cycle $u_1w_1u_2w_2$.

(i) The spine is a path, $\gamma_b$ is adjacent to $\mathcal A_b\ne\emptyset$, and $\alpha_b$, $\beta_b$ are adjacent to $\gamma_b$ only, so $H^0_x$ is connected. A $4$-cycle avoids the vertices $\alpha_b,\beta_b$ of degree $1$. If it passed through $\gamma_b$, both cycle neighbors of $\gamma_b$ would lie in $\mathcal A_b$, so $b\ge2$ and they would be $v_{b-1},v_b$, whose neighborhoods, $\{v_{b-2},v_b,\gamma_{b-1},\gamma_b\}$ and $\{v_{b-1},v_{b+1},\gamma_b,\gamma_{b+1}\}$ without the vertices that do not exist, meet only in $\gamma_b$. So a $4$-cycle avoids all $\gamma_b$ and lies in the spine, which has no cycle.

(ii) \emph{Members are star sets.} For the clusters that are unions of items, $\partial A=E^\times(A)$, as the edges of $H^0_x$ outside $E_x$ lie inside blocks. For the items and the prefixes, the proof of Lemma~\ref{lem:E-membership} finds a common vertex of these edges. The sets $\{\alpha_b\}$, $\{\beta_b\}$, $\{\gamma_b\}$ have their own vertex as a common vertex. The boundaries $\partial\{\alpha_b,\beta_b\}=\{\gamma_b\alpha_b,\gamma_b\beta_b\}$ and $\partial\{\alpha_b,\gamma_b\}=\{\gamma_b\beta_b\}\cup\{\gamma_bv:v\in\mathcal A_b\}$ have the common vertex $\gamma_b$, and so does $\partial\{\beta_b,\gamma_b\}$. Complements have the same boundary.

\emph{Star sets are members.} Let $A$ be a star set with a common vertex $c$. Both families are closed under complements, so we may assume $c\in A$. An edge of $H^0_x-c$ between $A\setminus\{c\}$ and $[n]\setminus A$ would lie in $\partial A$ and avoid $c$. So $A$ is the union of $\{c\}$ with some, but not all, components of $H^0_x-c$. Every vertex of $H^0_x$ is $\alpha_b$, $\beta_b$ or $\gamma_b$ for some $b$, or $v_j$ for some $j$, so the following three cases are exhaustive.

If $c\in\{\alpha_b,\beta_b\}$, then $c$ has degree $1$, $H^0_x-c$ is connected, and $A=\{c\}$.

If $c=\gamma_b$, the components are $\{\alpha_b\}$, $\{\beta_b\}$ and $[n]\setminus B_b$, the spine with the other blocks attached, which is connected. So $A$ is one of $\{\gamma_b\}$, $\{\alpha_b,\gamma_b\}$, $\{\beta_b,\gamma_b\}$, $B_b$, $[n]\setminus\{\alpha_b,\beta_b\}$, $[n]\setminus\{\beta_b\}$, $[n]\setminus\{\alpha_b\}$, all in $\clus(T_x)\cup\mathcal E_x$.

If $c=v_j$, let $\mathrm{Left}_j:=B_1\cup\dots\cup B_{\min\{j,K\}}\cup\{v_1,\dots,v_{j-1}\}$. Let $\mathrm{Right}_j:=B_{j+1}\cup\dots\cup B_K\cup\{v_{j+1},\dots,v_m\}$, so that $\mathrm{Left}_j\cup\mathrm{Right}_j=[n]\setminus\{v_j\}$. Here $\mathrm{Left}_j$ is nonempty and connected: it is $B_1$ for $j=1$, and otherwise $\gamma_1\sim v_1$ and $\gamma_i\sim v_{i-1}$ with $i-1\le j-1$. The set $\mathrm{Right}_j$ is connected, as $\gamma_i\sim v_i$ for $i>j$, and nonempty iff $j<m$. No edge joins the two sets, as $\mathcal A_i\subseteq\{v_1,\dots,v_j\}$ for $i\le j$ and $\mathcal A_i\subseteq\{v_j,\dots,v_m\}$ for $i>j$. Hence the components of $H^0_x-v_j$ are $\mathrm{Left}_j$ and, if $j<m$, $\mathrm{Right}_j$. So $A$ is $\{v_j\}$, $\{v_j\}\cup\mathrm{Left}_j$ or $\{v_j\}\cup\mathrm{Right}_j$, the last two only if they differ from $[n]$. Now, for $j\le K$, $\{v_j\}\cup\mathrm{Left}_j$ is $\mathrm{Pre}_{2j}$ and $\{v_j\}\cup\mathrm{Right}_j$ is the complement of $\mathrm{Pre}_{2j-1}$. For $j>K$, these sets are $\mathrm{Pre}_{K+j}$ and the complement of $\mathrm{Pre}_{K+j-1}$. When these sets differ from $[n]$, the index is at most $\ell-1$, and they are clusters by Lemma~\ref{lem:E-tree}, as is the item $\{v_j\}$.

(iii) $A$ is nonempty, proper, and by (ii) not a star set, so $\partial A\ne\emptyset$ by connectedness, and no vertex lies on all its edges. Take $e=uw\in\partial A$ with $u\in A$, an edge $f\in\partial A$ avoiding $u$ and an edge $g\in\partial A$ avoiding $w$. If $f$ avoids $w$, then $e$ and $f$ are disjoint. If $g$ avoids $u$, then $e$ and $g$ are disjoint. Otherwise, $f=u'w$ with $u'\in A\setminus\{u\}$ and $g=uw'$ with $w'\notin A\cup\{w\}$, and $f$ and $g$ are disjoint. In each case, write the two disjoint edges as $u_1w_1$ and $u_2w_2$ with $u_1,u_2\in A$ and $w_1,w_2\notin A$, which is possible as each edge of $\partial A$ has exactly one end in $A$. If both $u_1w_2$ and $u_2w_1$ were edges, $u_1w_1u_2w_2$ would be a $4$-cycle, contrary to (i).
\end{proof}
Recall that $D'$ is the union of a set $D$ of items.

\begin{lemma}[Deletion]\label{lem:E-deletion}
For a set $D$ of items let
\begin{equation*}
\Psi_x[D]:=\prod_{e\in E_x,\ e\subseteq D'}CZ_e\Bigl(\bigotimes_{B_b\in D}\tilde\psi_{x_b}\otimes\ket+^{\otimes(S\cap D')}\Bigr),
\end{equation*}
a unit vector on the sites of $D'$. Then $\varepsilon_A(\Psi_x;\chi)\ge\varepsilon_{A\cap D'}(\Psi_x[D];\chi)$ for every $A\subseteq[n]$.
\end{lemma}

\begin{proof}
\emph{Idea.} Measuring a site outside $A$ in its computational basis writes $\rho_A$ as a mixture over the outcomes, and a site inside $A$ is treated through the complement of $A$. The $\chi$-tail is concave by step (1), so by step (2) some outcome has a tail no larger than that of the mixture. Steps (3) and (4) measure all sites outside $D'$ and show that every outcome leaves $\Psi_x[D]$ on $D'$, up to a scalar and one-site unitaries.

(1) For a density operator $\rho$ with eigenvalues $\lambda_1\ge\lambda_2\ge\dots$ and an orthogonal projector $\Pi$ of rank at most $\chi$, $\Tr(\Pi\rho)=\sum_i\lambda_i\bra{e_i}\Pi\ket{e_i}$ for an eigenbasis $(e_i)$, with weights in $[0,1]$ summing to at most $\chi$. So $\Tr(\Pi\rho)\le\lambda_1+\dots+\lambda_\chi$, with equality for the projector onto $e_1,\dots,e_\chi$. This is Ky Fan's maximum principle~\cite{Fan49}. As a maximum of linear functionals, $\lambda_1+\dots+\lambda_\chi$ is convex in $\rho$, and the $\chi$-tail is concave.

(2) Let $\Phi$ be a unit vector on a finite set $V$ of sites with finite-dimensional spaces and computational bases, $A\subseteq V$, $i\in V\setminus A$ a site, $\Phi_t:=(\bra t_i\otimes1)\Phi$ and $p_t:=\lVert\Phi_t\rVert^2$. Tracing out site $i$ in its computational basis first gives
\begin{equation*}
\rho_A(\Phi)=\sum_{t:p_t>0}p_t\,\rho_A(\ket t_i\otimes\Phi_t/\sqrt{p_t}),
\end{equation*}
so by (1), $\varepsilon_A(\Phi;\chi)\ge\min_{p_t>0}\varepsilon_A(\ket t_i\otimes\Phi_t/\sqrt{p_t};\chi)$. For $i\in A$ apply this to $V\setminus A$ and use Lemma~\ref{lem:E-isometry}.

(3) Iterating (2) over the sites of $[n]\setminus D'$ gives $\varepsilon_A(\Psi_x;\chi)\ge\min_z\varepsilon_A(\ket z\otimes\Psi_z/\lVert\Psi_z\rVert;\chi)$, the minimum over $z\in G^{[n]\setminus D'}$ with $\Psi_z:=(\bra z\otimes1)\Psi_x\ne0$.

(4) Every item lies inside $D'$ or is disjoint from it. Under $\bra z$, the gates of the edges with both ends outside $D'$ become a phase. For an edge $uw$ with $u\notin D'$ and $w\in D'$, $(\bra t_u\otimes1)CZ_{uw}=Z^t_w(\bra t_u\otimes1)$, and the gates inside $D'$ commute with these $Z$'s. Finally, $\bra z\Phi^0_x$ is a scalar multiple of $\bigotimes_{B_b\in D}\tilde\psi_{x_b}\otimes\ket+^{\otimes(S\cap D')}$. Hence $\Psi_z=\kappa_zZ^y\Psi_x[D]$ for a scalar $\kappa_z$ and $y_w:=\sum_{u\notin D',\,uw\in E_x}z_u$. As $\ket z$ is a product state and $Z^y$ a product of one-site unitaries, the reduced state of $\ket z\otimes\Psi_z/\lVert\Psi_z\rVert$ on $A$ is a pure state on $A\setminus D'$ tensored with a unitary conjugate of $\rho_{A\cap D'}(\Psi_x[D])$. So its $\chi$-tail is $\varepsilon_{A\cap D'}(\Psi_x[D];\chi)$.
\end{proof}

\begin{lemma}[Perturbation]\label{lem:E-perturb}
Let $\emptyset\ne A\subsetneq[n]$ with $A\notin\clus(T_x)\cup\mathcal E_x$, and $0<\varepsilon\le\frac12(1-\chi/q^2)$. Then $\varepsilon_A(\Psi_x;\chi)\ge t_0(\varepsilon):=\bigl(\sqrt{1-\chi/q^2}-\sqrt{2\varepsilon}\bigr)^2$.
\end{lemma}

\begin{proof}
\emph{Idea.} The steps \emph{Choice of $D$} and \emph{Deletion} reduce the cut to the at most four items that meet the two disjoint edges of Lemma~\ref{lem:E-startsets}(iii). The steps \emph{Graph form at $\varepsilon=0$} and \emph{The tail at $\varepsilon=0$} use Lemma~\ref{lem:E-graphcut} to bound the tail across the reduced cut at $\varepsilon=0$ from below by $1-\chi/q^2$. The steps \emph{Distance} and \emph{Lipschitz step} carry the bound to $\varepsilon>0$.

\emph{Choice of $D$.} Take $u_1w_1$ and $u_2w_2$ from Lemma~\ref{lem:E-startsets}(iii), and let $D$ be the set of the at most four items that contain one of $u_1,w_1,u_2,w_2$. Every edge of $H^0_x$ meets at most one block: the edges $\gamma_b\alpha_b$, $\gamma_b\beta_b$ lie in a block, the edges $\gamma_bv$ join a block to the spine, and spine edges join spine sites. So $D$ contains $k_D\le2$ blocks.

\emph{Deletion.} Let $\Psi^{(\varepsilon)}_x[D]$ be the vector of Lemma~\ref{lem:E-deletion} at parameter $\varepsilon$, defined by the same formulas also for $\varepsilon=0$. Only the block vectors $\tilde\psi_{x_b}$ depend on $\varepsilon$. By Lemma~\ref{lem:E-deletion}, $\varepsilon_A(\Psi_x;\chi)\ge t_D(\varepsilon):=\varepsilon_{A\cap D'}(\Psi^{(\varepsilon)}_x[D];\chi)$.

\emph{Graph form at $\varepsilon=0$.} Here $\tilde\psi^{(0)}_{x_b}=q^{-1/2}\sum_{a\in G}\ket{aaa}$ on $(\alpha_b,\beta_b,\gamma_b)$. With the Fourier transform $\mathsf F\ket a:=q^{-1/2}\sum_{c\in G}\langle a,c\rangle\ket c$, which is unitary by nondegeneracy, expansion gives
\begin{equation*}
CZ_{\gamma\alpha}CZ_{\gamma\beta}\ket+_\alpha\ket+_\beta\ket+_\gamma=q^{-3/2}\sum_{a,c,c'}\langle a,c\rangle\langle a,c'\rangle\ket c_\alpha\ket{c'}_\beta\ket a_\gamma=q^{-1/2}\sum_a\mathsf F\ket a_\alpha\,\mathsf F\ket a_\beta\ket a_\gamma,
\end{equation*}
so $\tilde\psi^{(0)}_{x_b}=(\mathsf F^\dagger_{\alpha_b}\otimes\mathsf F^\dagger_{\beta_b})CZ_{\gamma_b\alpha_b}CZ_{\gamma_b\beta_b}\ket+^{\otimes B_b}$. No edge of $E_x$ touches $\alpha_b$ or $\beta_b$. So, for every set $D$ of items, $\Psi^{(0)}_x[D]=\bigotimes_{B_b\in D}(\mathsf F^\dagger_{\alpha_b}\otimes\mathsf F^\dagger_{\beta_b})\ket{H^0_x[D']}$, where $\ket{H^0_x[D']}$ is the graph state of the subgraph of $H^0_x$ induced on $D'$. The union of all items is $[n]$, so $\Psi_x$ at $\varepsilon=0$ is the graph state of $H^0_x$ up to these local unitaries.

\emph{The tail at $\varepsilon=0$.} Local unitaries do not change tails (Lemma~\ref{lem:E-isometry}). The cut $(A\cap D',D'\setminus A)$ of $H^0_x[D']$ is nonempty and proper, contains $u_1w_1$ and $u_2w_2$, and the adjacencies among these four vertices are those of $H^0_x$. So Lemma~\ref{lem:E-graphcut} gives $t_D(0):=\varepsilon_{A\cap D'}(\Psi^{(0)}_x[D];\chi)\ge1-\chi/q^2$.

\emph{Distance.} $\braket{\tilde\psi^{(0)}_{x_b}}{\tilde\psi^{(\varepsilon)}_{x_b}}=q^{-1/2}\sum_a\braket{aa}{\eta_a}=\sqrt{1-\varepsilon}$, because the terms $\ket{cc}$ of $\eta_a$ have $c\ne a$. The gates and the factors $\ket+$ are common to $\Psi^{(0)}_x[D]$ and $\Psi^{(\varepsilon)}_x[D]$, so their inner product is $(1-\varepsilon)^{k_D/2}\ge1-\varepsilon$, and $\lVert\Psi^{(\varepsilon)}_x[D]-\Psi^{(0)}_x[D]\rVert^2\le2\varepsilon$.

\emph{Lipschitz step.} By Lemma~\ref{lem:kyfan}, applied with $q$ in place of $d$ to the $|D'|\ge4$ sites of $D'$, the square root of a $\chi$-tail is $1$-Lipschitz in the unit vector. Hence $\sqrt{t_D(\varepsilon)}\ge\sqrt{t_D(0)}-\sqrt{2\varepsilon}\ge\sqrt{1-\chi/q^2}-\sqrt{2\varepsilon}\ge0$, the last inequality by the hypothesis $2\varepsilon\le1-\chi/q^2$. Squaring and the step \emph{Deletion} give $\varepsilon_A(\Psi_x;\chi)\ge t_D(\varepsilon)\ge t_0(\varepsilon)$.
\end{proof}

Lemma~\ref{lem:E-gap} assembles part~(a) of Theorem~\ref{thm:lower}, which for a general group $G$, with the sites $\C^G$, is read with $q$ in place of $d$. The hypotheses of Theorem~\ref{thm:lower} on $d$, $\chi$ and $q$ then become the standing ones, $q\le\chi\le q^2-1$, and the class of part~(a) is $\mathcal S^{q,\chi}_{n,s}(\theta\varepsilon)$.

\begin{lemma}[The total gap]\label{lem:E-gap}
Let $0<\varepsilon\le\bar\varepsilon$. For all $x,y\in\{1,2\}^K$, $\Psi_x$ is a unit vector in $\TTN(T_x,\chi)$ over the sites $\C^G$, $\braket{\Psi_x}{\Psi_y}=(1-\varepsilon)^{\dH(x,y)}$, the trees $T_x$ are pairwise distinct binary trees, and $\varepsilon_A(\Psi_x;\chi)\ge\theta\varepsilon$ for every nonempty proper $A\subseteq[n]$ that is not a cluster of $T_x$, with equality on $\mathcal E_x$ and, if $\varepsilon<\bar\varepsilon$, only there. Hence $\gaptot(\Psi_x,T_x)=\gaps(\Psi_x,T_x)=\theta\varepsilon$, $\Psi_x\in\mathcal S^{q,\chi}_{n,s}(\theta\varepsilon)$ and $T(\Psi_x)=T_x$ for every $s\ge2$. For $G=\Z_q$, with every site $\C^q$ embedded in $\C^d$ as in Section~\ref{sec:lower}, this is Theorem~\ref{thm:lower}(a).
\end{lemma}

\begin{proof}
The states $\Psi_x$ are unit vectors with the overlaps $(1-\varepsilon)^{\dH(x,y)}$ by Lemma~\ref{lem:E-connector}, they lie in $\TTN(T_x,\chi)$ by Lemma~\ref{lem:E-membership}, and the trees $T_x$ are pairwise distinct binary trees by Lemma~\ref{lem:E-tree}.

Since $(\sqrt2+\sqrt\theta)^2\ge2$ and $q\ge2$, we have
\begin{equation*}
\varepsilon\le\bar\varepsilon\le\frac12(1-\chi/q^2)\le\frac12\le1-1/q=M/(M+q).
\end{equation*}
So Lemma~\ref{lem:E-epstype} gives $\mathcal E_x\subseteq\Nfam_2(T_x)$ with tails $\theta\varepsilon$. Lemma~\ref{lem:E-perturb} applies to every nonempty proper $A\notin\clus(T_x)\cup\mathcal E_x$, and
\begin{equation*}
t_0(\varepsilon)\ge\theta\varepsilon\iff\sqrt{1-\chi/q^2}-\sqrt{2\varepsilon}\ge\sqrt{\theta\varepsilon}\iff\varepsilon\le\frac{1-\chi/q^2}{(\sqrt2+\sqrt\theta)^2},
\end{equation*}
which holds, and holds strictly if $\varepsilon<\bar\varepsilon$.

Hence every nonempty proper non-cluster $A$ has $\varepsilon_A(\Psi_x;\chi)\ge\theta\varepsilon$, with equality on $\mathcal E_x$ and, if $\varepsilon<\bar\varepsilon$, only there. So $\gaptot(\Psi_x,T_x)=\theta\varepsilon$, and $\gaps(\Psi_x,T_x)=\theta\varepsilon$ for every $s\ge2$, since $\mathcal E_x\subseteq\Nfam_2(T_x)\subseteq\Nfam_s(T_x)\subseteq\Nfam_\infty(T_x)$. As $T_x$ is binary, $T_x\in\Xn(s)$, and $\Psi_x\in\TTN(T_x,\chi)$, so $\Psi_x\in\mathcal S^{q,\chi}_{n,s}(\theta\varepsilon)$ and $T(\Psi_x)=T_x$ by Theorem~\ref{thm:ident}(c) with $q$ in place of $d$. In Section~\ref{sec:lower}, the embedding of the sites into $\C^d$ keeps inner products and does not change ranks or tails (Lemma~\ref{lem:E-isometry}). So there part~(a) holds as stated, with $\Psi_x\in\mathcal S^{d,\chi}_{n,s}(\theta\varepsilon)$.
\end{proof}

\subsection*{Discrimination}
\begin{lemma}[POVM reduction]\label{lem:E-povm}
For every $N$-copy procedure with output set $\mathcal O$ (Definition~\ref{def:procedures}) there are operators $Q_o\ge0$ on $\Hil^{\otimes N}$, $o\in\mathcal O$, with $\sum_oQ_o=1$ and $\Pr[\text{output }o]=\Tr(Q_o\varrho)$ whenever the $N$ copies are in the state $\varrho$.
\end{lemma}

\begin{proof}
For a record $\mathbf o=(o_1,\dots,o_R)$ let $\Phi_{\mathbf o}$ be the composition of the instrument maps along $\mathbf o$. It is linear and positive, and induction over the rounds, each instrument summing to a trace-preserving map, gives $\sum_{\mathbf o}\Tr\Phi_{\mathbf o}(X)=\Tr X$. For an orthonormal basis $(\ket\mu)$ of $\Hil^{\otimes N}$ put $Q_{\mathbf o}:=\sum_{\mu,\nu}\Tr\Phi_{\mathbf o}(\ket\mu\!\bra\nu)\,\ket\nu\!\bra\mu$. By linearity, $\Tr(Q_{\mathbf o}X)=\Tr\Phi_{\mathbf o}(X)$. Then $\bra\phi Q_{\mathbf o}\ket\phi=\Tr\Phi_{\mathbf o}(\ket\phi\!\bra\phi)\ge0$ for every $\phi$, so $Q_{\mathbf o}$ is Hermitian, by polarization, and positive semidefinite, and $\sum_{\mathbf o}Q_{\mathbf o}=1$ by the trace identity. Put $Q_o:=\sum_{\mathbf o:\,\mathrm{out}(\mathbf o)=o}Q_{\mathbf o}$.
\end{proof}

\begin{lemma}[Trace norm of two pure states]\label{lem:E-helstrom}
For unit vectors $\Phi_1,\Phi_2$,
\begin{equation*}
\bigl\lVert\ket{\Phi_1}\!\bra{\Phi_1}-\ket{\Phi_2}\!\bra{\Phi_2}\bigr\rVert_1=2\sqrt{1-|\braket{\Phi_1}{\Phi_2}|^2}.
\end{equation*}
\end{lemma}

\begin{proof}
This is twice the trace distance of two pure states~\cite[Eq.~(9.99)]{NC10}.
\end{proof}

\begin{lemma}[Product dual certificate]\label{lem:E-dual}
Let $\sigma\in[0,1)$, let $\Phi_x$, $x\in\{1,2\}^K$, be unit vectors with $\braket{\Phi_x}{\Phi_y}=\sigma^{\dH(x,y)}$, and let $F_x\ge0$ with $\sum_xF_x\le1$. Then $2^{-K}\sum_x\bra{\Phi_x}F_x\ket{\Phi_x}\le\PH(\sigma)^K$, and there is a projective measurement $\{F^\star_x\}_x\cup\{F^\star_\perp\}$ with $\bra{\Phi_x}F^\star_x\ket{\Phi_x}=\PH(\sigma)^K$ for every $x$.
\end{lemma}

\begin{proof}
\emph{Product model.} Let $\varphi_{1,2}:=\bigl(\sqrt{(1+\sigma)/2},\pm\sqrt{(1-\sigma)/2}\bigr)\in\R^2$ and $\Theta_x:=\bigotimes_b\varphi_{x_b}$. Then $\braket{\varphi_1}{\varphi_2}=\sigma$ and $\braket{\Theta_x}{\Theta_y}=\sigma^{\dH(x,y)}$. As $\sigma<1$, the $\Theta_x$ form a basis of $(\C^2)^{\otimes K}$, and the linear map $J$ with $J\Theta_x=\Phi_x$ preserves inner products, the Gram matrices being equal. So $J$ is an isometry.

\emph{Dual.} Let $\Delta:=\frac12(\ket{\varphi_2}\!\bra{\varphi_2}-\ket{\varphi_1}\!\bra{\varphi_1})=\Delta_+-\Delta_-$ with $\Delta_\pm\ge0$ of orthogonal supports, and $Y_1:=\frac12\ket{\varphi_1}\!\bra{\varphi_1}+\Delta_+$. Then $Y_1-\frac12\ket{\varphi_1}\!\bra{\varphi_1}=\Delta_+\ge0$ and $Y_1-\frac12\ket{\varphi_2}\!\bra{\varphi_2}=\Delta_-\ge0$, and since $\Tr\Delta=0$, $\Tr\Delta_+=\frac12\lVert\Delta\rVert_1=\frac12\sqrt{1-\sigma^2}$ by Lemma~\ref{lem:E-helstrom}, so $\Tr Y_1=\PH(\sigma)$. If $A_i\ge B_i\ge0$, then $A_1\otimes A_2-B_1\otimes B_2=(A_1-B_1)\otimes A_2+B_1\otimes(A_2-B_2)\ge0$. By induction, $Y:=Y_1^{\otimes K}\ge2^{-K}\ket{\Theta_x}\!\bra{\Theta_x}$ for every $x$. Hence $\tilde Y:=JYJ^\dagger\ge2^{-K}\ket{\Phi_x}\!\bra{\Phi_x}$, $\Tr\tilde Y=\PH(\sigma)^K$, and
\begin{equation*}
2^{-K}\sum_x\bra{\Phi_x}F_x\ket{\Phi_x}\le\sum_x\Tr(F_x\tilde Y)=\Tr\Bigl(\sum_xF_x\,\tilde Y\Bigr)\le\Tr\tilde Y=\PH(\sigma)^K.
\end{equation*}

\emph{Attainment.} With $\omega_{1,2}:=(1,\pm1)/\sqrt2$, $|\braket{\omega_j}{\varphi_j}|^2=\frac12\bigl(\sqrt{(1+\sigma)/2}+\sqrt{(1-\sigma)/2}\bigr)^2=\PH(\sigma)$ for $j=1,2$. The projectors $F^\star_y:=J\bigl(\bigotimes_b\ket{\omega_{y_b}}\!\bra{\omega_{y_b}}\bigr)J^\dagger$ and $F^\star_\perp:=1-JJ^\dagger$ sum to $1$, and $\bra{\Phi_x}F^\star_x\ket{\Phi_x}=\prod_b|\braket{\omega_{x_b}}{\varphi_{x_b}}|^2=\PH(\sigma)^K$.
\end{proof}

Lemma~\ref{lem:E-dual} is an explicit form of the multiplicativity of optimal guessing probabilities for product ensembles~\cite{KRS09}. The ensemble $\{\Psi_x^{\otimes N}\}$ is not itself a product ensemble, but $J$ maps the product ensemble $\{\Theta_x\}$ onto it.

\begin{proposition}[Assouad-type bound]\label{prop:E-assouad}
Let $K\ge1$, $\varepsilon,\delta\in(0,1)$, let $\Phi_x$, $x\in\{1,2\}^K$, be unit vectors in $\Hil$ with $\braket{\Phi_x}{\Phi_y}=(1-\varepsilon)^{\dH(x,y)}$, and let $o_x$, $x\in\{1,2\}^K$, be pairwise distinct outputs. If an $N$-copy procedure outputs $o_x$ with probability at least $1-\delta$ on $\Phi_x^{\otimes N}$ for every $x$, then $1-\delta\le\PH((1-\varepsilon)^N)^K$, hence $N\ge N_K(\varepsilon,\delta)$. Some $\lceil N_K(\varepsilon,\delta)\rceil$-copy procedure does this, so $\lceil N_K(\varepsilon,\delta)\rceil$ is the exact optimum.
\end{proposition}

\begin{proof}
By Lemma~\ref{lem:E-povm} the procedure has a POVM $\{Q_o\}$ on $\Hil^{\otimes N}$. Put $F_x:=Q_{o_x}$. The outputs $o_x$ are pairwise distinct, so $\sum_xF_x\le1$, and success on $x$ means $\bra{\Phi_x^{\otimes N}}F_x\ket{\Phi_x^{\otimes N}}\ge1-\delta$. Put $\sigma:=(1-\varepsilon)^N$. For $N\ge1$, $\braket{\Phi_x^{\otimes N}}{\Phi_y^{\otimes N}}=\sigma^{\dH(x,y)}$ with $\sigma<1$, and averaging over $x$, Lemma~\ref{lem:E-dual} gives $1-\delta\le\PH(\sigma)^K$. For $N=0$ the $F_x$ are numbers with $\sum_xF_x\le1$, so the average is at most $2^{-K}=\PH(1)^K$.

If $a_K\le0$, then $N\ge0=N_K(\varepsilon,\delta)$. If $a_K>0$, then
\begin{align*}
\PH(\sigma)^K\ge1-\delta&\iff\sqrt{1-\sigma^2}\ge a_K\iff\sigma^2\le1-a_K^2\\
&\iff2N\ln\tfrac1{1-\varepsilon}\ge\ln\tfrac1{1-a_K^2}\iff N\ge N_K(\varepsilon,\delta).
\end{align*}

Conversely, if $a_K>0$, the measurement of Lemma~\ref{lem:E-dual} on $N:=\lceil N_K(\varepsilon,\delta)\rceil\ge1$ copies, with output $o_y$ on $F^\star_y$ and any output on $F^\star_\perp$, outputs $o_x$ with probability $\PH((1-\varepsilon)^N)^K\ge1-\delta$ on every $\Phi_x^{\otimes N}$, by the same equivalences. If $a_K\le0$, that is $2^{-K}\ge1-\delta$, a uniformly random guess with no copies succeeds with probability $2^{-K}$. Both are one-round procedures of Definition~\ref{def:procedures}.
\end{proof}
\begin{proof}[Proof of Theorem~\ref{thm:lower}(b)]
The trees $T_x$ are pairwise distinct (Lemma~\ref{lem:E-tree}) and $\braket{\Psi_x}{\Psi_y}=(1-\varepsilon)^{\dH(x,y)}$ (Lemma~\ref{lem:E-connector}). The inclusion $\C^q\subseteq\C^d$ of each site embeds the $\Psi_x$ into $\Hil$ and keeps the overlaps. So Proposition~\ref{prop:E-assouad} applies with $\Phi_x=\Psi_x$ and $o_x=T_x$.
\end{proof}

\subsection*{Constants}
\begin{lemma}[Uniform constants]\label{lem:E-constants}
Let $K\ge1$ and $\delta\in(0,1)$.
\begin{enumerate}
\item[(i)] For $\delta<\frac12$, $G_1(\delta)=\ln\frac1{4\delta(1-\delta)}$. Moreover $G_K(\delta)$ is nondecreasing in $K$, and $G_K(\delta)\ge\ln\frac{K}{4\ln(1/(1-\delta))}$.
\item[(ii)] If $\delta\le\frac13$ and $n\in\{4K,\dots,4K+3\}$, then $G_K(\delta)\ge c_0\ln(n/\delta)$, with equality for $K=1$, $n=7$ and $\delta=\frac13$.
\end{enumerate}
\end{lemma}

\begin{proof}
(i) $a_1=1-2\delta$ and $1-a_1^2=4\delta(1-\delta)$. As $(1-\delta)^{1/K}$ increases with $K$, so do $a_K$ and $G_K(\delta)$. For the last bound let first $a_K>0$ and $\sigma:=(1-a_K^2)^{1/2}\in(0,1)$. Then $\PH(\sigma)=(1+a_K)/2=(1-\delta)^{1/K}$, so $K\ln\frac1{\PH(\sigma)}=\ln\frac1{1-\delta}$, and $\ln\frac1{\PH(\sigma)}\ge1-\PH(\sigma)=\frac12\bigl(1-\sqrt{1-\sigma^2}\bigr)\ge\frac{\sigma^2}4$, as $1-\sqrt{1-u}\ge u/2$ on $[0,1]$. Hence $\sigma^2\le4\ln\frac1{1-\delta}/K$, that is $G_K(\delta)=\ln\frac1{\sigma^2}\ge\ln\frac{K}{4\ln(1/(1-\delta))}$. If $a_K\le0$, then $(1-\delta)^{1/K}\le\frac12$, so $K/4\le K\ln2\le\ln\frac1{1-\delta}$, and the right side is at most $0=G_K(\delta)$.

(ii) \emph{The case $K=1$.} Here $n\le7$. Write $G_1(\delta)=R(\delta)\ln\frac7\delta$ with $R(\delta):=1-\ln(28(1-\delta))/\ln\frac7\delta$. The derivative of $\ln(28(1-\delta))/\ln\frac7\delta$ has the sign of $\Upsilon(\delta):=(1-\delta)\ln(28(1-\delta))-\delta\ln\frac7\delta$, and $\Upsilon(\frac12)=0$ and $\Upsilon'(\delta)=-\ln\frac{196(1-\delta)}{\delta}<0$ on $(0,\frac12)$. So $\Upsilon>0$ there, and $R$ decreases on $(0,\frac12)$. For $\delta\le\frac13$, $G_1(\delta)\ge R(\frac13)\ln\frac7\delta\ge R(\frac13)\ln\frac n\delta$, and $R(\frac13)=\ln\frac98/\ln21=c_0$. Equality holds at $n=7$, $\delta=\frac13$.

\emph{The case $K\ge2$.} $\ln\frac1{1-\delta}/\delta$ increases with $\delta$, so $\ln\frac1{1-\delta}\le3\ln\frac32\cdot\delta$ for $\delta\le\frac13$, and (i) gives $G_K(\delta)\ge\mathfrak s-\mathfrak a$ with $\mathfrak s:=\ln K+\ln\frac1\delta\ge\ln6$ and $\mathfrak a:=\ln(12\ln\frac32)$. Also $n\le4K+3\le\frac{11}2K$, so $\ln\frac n\delta\le\mathfrak s+\mathfrak b$ with $\mathfrak b:=\ln\frac{11}2$. We use $\frac u{1+u}\le\ln(1+u)\le u-\frac{u^2}2+\frac{u^3}3$ for $u\ge0$, which gives $\frac13\le\ln\frac32\le\frac5{12}$, $\ln\frac65\ge\frac16$ and $\ln\frac98<\frac18$, and $e^2>7$, $e^3<21$, which give $\ln33<4$ and $\ln21>3$. Thus $4\le12\ln\frac32\le5$, so $0<\mathfrak a<\ln6$, and as $(\mathfrak s-\mathfrak a)/(\mathfrak s+\mathfrak b)$ increases with $\mathfrak s$,
\begin{equation*}
\frac{G_K(\delta)}{\ln(n/\delta)}\ \ge\ \frac{\ln6-\mathfrak a}{\ln6+\mathfrak b}=\frac{\ln\bigl(1/(2\ln\frac32)\bigr)}{\ln33}\ \ge\ \frac{\ln\frac65}{\ln33}\ >\ \frac1{24}\ >\ \frac{\ln\frac98}{\ln21}=c_0.\qedhere
\end{equation*}
\end{proof}

\begin{proof}[Proof of Theorem~\ref{thm:lower}(c)]
Put $\varepsilon:=\varepsilon_0/\theta\in(0,\bar\varepsilon]$, $K:=\lfloor n/4\rfloor\ge1$ and $L:=n-4K\in\{0,1,2,3\}$. By (a), every $\Psi_x$ lies in $\mathcal S^{d,\chi}_{n,s}(\varepsilon_0)\subseteq\mathcal S^{d,\chi,\circ}_{n,s}(\varepsilon_0)$ with $T(\Psi_x)=T_x$, so a procedure that succeeds on either class outputs $T_x$ with probability at least $1-\delta$ on every $\Psi_x$, and (b) gives $N\ge N_K(\varepsilon_0/\theta,\delta)$. For $\delta\le\frac13$, Lemma~\ref{lem:E-constants}(ii) gives $N\ge G_K(\delta)/(2\ln\frac1{1-\varepsilon})\ge c_0\ln(n/\delta)/(2\ln\frac1{1-\varepsilon})$, and $\ln\frac1{1-\varepsilon}\le\frac{\varepsilon}{1-\varepsilon}\le\frac{\varepsilon_0}{\theta(1-\bar\varepsilon)}$ gives~\eqref{eq:s5-class}.
\end{proof}

The values of $\varepsilon_{d,\chi}$ quoted after Theorem~\ref{thm:lower} follow from the definitions. For $q=\chi$, that is $d\ge\chi$, we have $\theta=1$ and $\bar\varepsilon=(1-\chi/q^2)/(\sqrt2+1)^2=(1-1/\chi)(3-2\sqrt2)$. So $\varepsilon_{d,\chi}=\theta\bar\varepsilon=\bar\varepsilon$, which is $(3-2\sqrt2)/2$ for $\chi=2$. For $d=2$ and $\chi=3$: $q=2$, $M=2$, $\theta=\frac12$ and $\bar\varepsilon=\frac14\big/(\sqrt2+\sqrt{1/2})^2=\frac1{18}$, so $\varepsilon_{2,3}=\frac1{36}$.

Lemma~\ref{lem:E-corconst} bounds the number $m^{(s)}_\star$ of copies of Section~\ref{sec:s4-onebatch} for Corollary~\ref{cor:rate}, and $N_T(s)$ in its formula is the bound of Lemma~\ref{lem:counting} on the number of tests.

\begin{lemma}[Constants of Corollary~\ref{cor:rate}]\label{lem:E-corconst}
Let $n\ge4$, $\chi\ge1$, $s\ge2$, $0<\delta\le\frac13$, $0<\varepsilon_0\le1$ and $m^{(s)}_\star:=(\chi+1)\bigl\lceil2(\chi+1)\ln(4(\chi+1)N_T(s)/\delta)/\varepsilon_0\bigr\rceil$. Then $m^{(s)}_\star\le C_{\chi,s}\ln(n/\delta)/\varepsilon_0$, with $C_{\chi,s}$ as in Corollary~\ref{cor:rate}.
\end{lemma}

\begin{proof}
As $n\ge4$ and $\delta\le\frac13$, $\ln(n/\delta)\ge\ln12$. By Lemma~\ref{lem:counting}, $\ln N_T(s)\le\ln2+s\ln\frac{en}s$ for $s\le n$, where $s\ln\frac{en}s\le s\ln\frac{en}2\le s\ln\frac n\delta$ since $\frac e2\le3\le\frac1\delta$. For $s>n$, $\ln N_T(s)=\ln N_T(n)\le\ln2+n\le\ln2+s\ln\frac n\delta$. So $\ln\frac{4(\chi+1)N_T(s)}\delta\le\ln(8(\chi+1))+\ln\frac1\delta+s\ln\frac n\delta$ in both cases. Here $\ln\frac1\delta\le\ln\frac n\delta$ and $\ln(8(\chi+1))\le\frac{\ln(8(\chi+1))}{\ln12}\ln\frac n\delta$. So the logarithm is at most $\bigl(1+s+\frac{\ln(8(\chi+1))}{\ln12}\bigr)\ln\frac n\delta$.

Put $y:=2(\chi+1)\ln(4(\chi+1)N_T(s)/\delta)/\varepsilon_0$. The argument of the logarithm is at least $4\cdot2\cdot9\cdot3=216>12$, as $\chi+1\ge2$, $N_T(s)\ge N_T(2)=(n-1)^2\ge9$ and $1/\delta\ge3$, so $y\ge\ln12/\varepsilon_0$ and $\lceil y\rceil\le y+1\le y(1+\varepsilon_0/\ln12)\le y(1+1/\ln12)$. Multiplying out gives
\begin{align*}
m^{(s)}_\star=(\chi+1)\lceil y\rceil&\le2(\chi+1)^2\bigl(1+s+\ln(8(\chi+1))/\ln12\bigr)\bigl(1+1/\ln12\bigr)\frac{\ln(n/\delta)}{\varepsilon_0}\\
&=C_{\chi,s}\frac{\ln(n/\delta)}{\varepsilon_0}.\qedhere
\end{align*}
\end{proof}
\section{Large bond dimension: the encoded hard family}\label{app:largebond}
This appendix proves Theorem~\ref{thm:lower-k}. Throughout, $d$ is a prime, $d^2\le\chi\le d^3-1$, $G=\Z_d^2$ with $\langle g,h\rangle=\omega^{g\cdot h}$, where $\omega:=e^{2\pi i/d}$, and $q=d^2$. The coarse family $\Psi_x$, $T_x$, $\mathcal E_x$ and the code $E$ are those of Section~\ref{sec:largebond}, built on $n'=4K+L$ super-sites, with the graph $H^0_x$ of Appendix~\ref{app:lower}. Here $K\ge1$ and $L\ge0$ are integers, as in Definition~\ref{def:E-hard}, so that $n'\ge4$.

On the coarse level, a super-site is a site of the coarse family, and Appendix~\ref{app:lower} applies with $n'$ in place of $n$. On the physical level, a super-site $s$ is the set of the four physical qudits $(s,1),\dots,(s,4)$, so that $n=4n'$. For a set $A$ of physical qudits and a super-site $s$ we write $P_s:=A\cap s$, and $A$ is \emph{coarse} if every $P_s$ is empty or all of $s$, and \emph{fine} otherwise. We read a coarse set as the set of super-sites it contains, so that clusters of $T_x$ and members of $\mathcal E_x$ are coarse sets. In Lemma~\ref{lem:F-transfer}, its proof and the proof of Theorem~\ref{thm:lower-k}, $s$ is instead the hub parameter of Section~\ref{sec:s2-trees}.

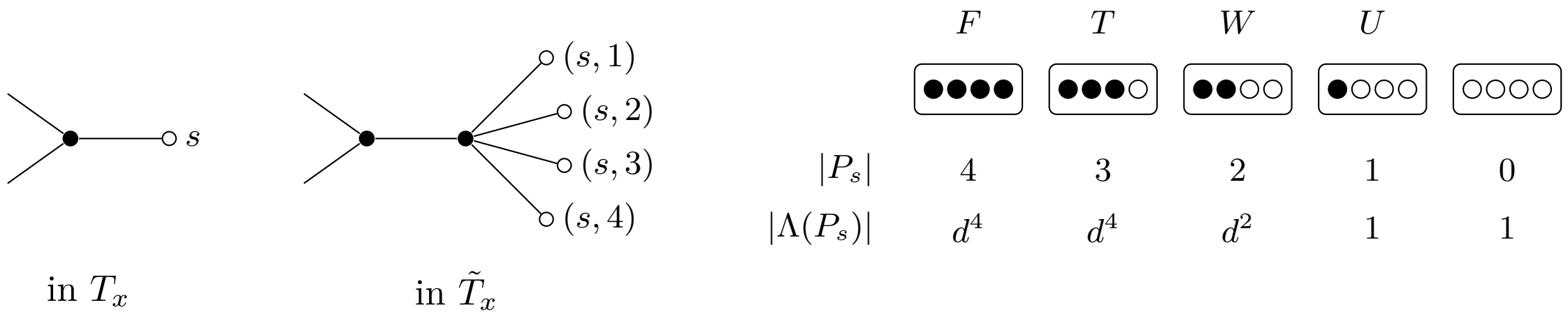
\begin{figure}[!hb]
\centering
\begin{tikzpicture}[x=0.85cm,y=0.85cm,
  iv/.style={circle,fill=black,inner sep=1.5pt},
  lf/.style={circle,draw=black,fill=white,inner sep=1.3pt},
  qin/.style={circle,draw=black,fill=black,inner sep=1.7pt},
  qout/.style={circle,draw=black,fill=white,inner sep=1.7pt},
  every label/.style={font=\footnotesize,inner sep=2pt}]
\begin{scope}
\node[iv] (p) at (0,0) {};
\draw (p)--(-0.7,0.5) (p)--(-0.7,-0.5);
\node[lf,label=right:$s$] (s) at (1.1,0) {};
\draw (p)--(s);
\node[font=\small] at (0.2,-1.7) {in $T_x$};
\end{scope}
\begin{scope}[shift={(3.3,0)}]
\node[iv] (p) at (0,0) {};
\draw (p)--(-0.7,0.5) (p)--(-0.7,-0.5);
\node[iv] (s) at (1.1,0) {};
\node[lf,label=right:{$(s,1)$}] (q1) at (2.0,0.9) {};
\node[lf,label=right:{$(s,2)$}] (q2) at (2.2,0.3) {};
\node[lf,label=right:{$(s,3)$}] (q3) at (2.2,-0.3) {};
\node[lf,label=right:{$(s,4)$}] (q4) at (2.0,-0.9) {};
\draw (p)--(s) (s)--(q1) (s)--(q2) (s)--(q3) (s)--(q4);
\node[font=\small] at (1.0,-1.7) {in $\tilde T_x$};
\end{scope}
\begin{scope}[shift={(10.0,0.55)}]
\foreach \px/\qa/\qb/\qc/\qd in {0/qin/qin/qin/qin, 1.5/qin/qin/qin/qout, 3.0/qin/qin/qout/qout, 4.5/qin/qout/qout/qout, 6.0/qout/qout/qout/qout} {
  \draw[rounded corners=2pt] (\px-0.6,-0.28) rectangle (\px+0.6,0.28);
  \node[\qa] at (\px-0.39,0) {};
  \node[\qb] at (\px-0.13,0) {};
  \node[\qc] at (\px+0.13,0) {};
  \node[\qd] at (\px+0.39,0) {};
}
\node[font=\footnotesize] at (0,0.75) {$F$};
\node[font=\footnotesize] at (1.5,0.75) {$T$};
\node[font=\footnotesize] at (3.0,0.75) {$W$};
\node[font=\footnotesize] at (4.5,0.75) {$U$};
\node[font=\footnotesize,anchor=east] at (-0.9,-0.9) {$|P_s|$};
\foreach \px/\val in {0/4, 1.5/3, 3.0/2, 4.5/1, 6.0/0} { \node[font=\footnotesize] at (\px,-0.9) {$\val$}; }
\node[font=\footnotesize,anchor=east] at (-0.9,-1.55) {$|\Lambda(P_s)|$};
\foreach \px/\val in {0/d^4, 1.5/d^4, 3.0/d^2, 4.5/1, 6.0/1} { \node[font=\footnotesize] at (\px,-1.55) {$\val$}; }
\end{scope}
\end{tikzpicture}
\caption{Left: a leaf $s$ of $T_x$, and the vertex of degree $5$ that replaces it in $\tilde T_x$, with the four physical qudits of the super-site $s$ as leaves. Right: the parts $P_s=A\cap s$ of a set $A$ of physical qudits, shown by filled dots, in five super-sites. The proof of Lemma~\ref{lem:F-fine0} writes $F$, $T$, $W$, $U$ for the sets of super-sites with $|P_s|=4,3,2,1$, and $A$ is fine if at least one of $T$, $W$, $U$ is nonempty. The last row gives the size of $\Lambda(P_s)$ by Lemma~\ref{lem:F-code}.}
\label{fig:F-supersite}
\end{figure}

The argument is outlined in the proof idea of Theorem~\ref{thm:lower-k} in Section~\ref{sec:largebond}. Lemmas~\ref{lem:F-code} and~\ref{lem:F-stabrank} are tools: the properties of the code, and the ranks of stabilizer states. Lemma~\ref{lem:F-transfer} collects what transfers from the coarse family: the unit vectors, the membership, the trees, the overlaps and the tails across coarse sets. Lemmas~\ref{lem:F-counting}--\ref{lem:F-fine0} concern the \emph{unperturbed} states, those at $\varepsilon=0$, which are stabilizer states. They end with the lower bound $d^3$ for the rank of the physical state across the fine sets $A$ with $3\le|A|\le n-3$. Lemma~\ref{lem:F-fine} treats these sets for the \emph{perturbed} states, those at $\varepsilon>0$, by comparison with $\varepsilon=0$, with Lemma~\ref{lem:F-ghz} as a tool. The theorem is proved at the end. As in Appendix~\ref{app:lower}, a superscript $(0)$ marks the value at $\varepsilon=0$, while $H^0_x$ does not depend on $\varepsilon$. Table~\ref{tab:F-notation} collects the notation, and Figure~\ref{fig:F-supersite} shows a super-site in the tree $\tilde T_x$ and the parts of a set in the super-sites.

\begin{table}[!hb]
\centering
\small
\begin{tabular}{@{}lp{7.9cm}l@{}}
\toprule
Symbol & Meaning & Defined in\\
\midrule
super-site, $n'$, $n$ & a site $s$ of the coarse family, encoded into the four physical qudits $(s,1),\dots,(s,4)$. There are $n'$ super-sites and $n=4n'$ qudits & Section~\ref{sec:largebond}\\
$E$, $\ket{\varphi_a}$ & the code $\C^G\to(\C^d)^{\otimes4}$ and its basis codewords & Section~\ref{sec:largebond}\\
$\Psi_x$, $T_x$, $\mathcal E_x$ & the coarse states, their trees and the family of the $\varepsilon$-type sets, over $G$ & Definition~\ref{def:E-hard}\\
$H^0_x$ & the graph on the $n'$ super-sites with the edges of $E_x$ and the edges $\gamma_b\alpha_b$, $\gamma_b\beta_b$ & after Definition~\ref{def:E-hard}\\
$\tilde\Psi_x$, $\tilde T_x$ & the physical states $E^{\otimes n'}\Psi_x$ and their trees & Section~\ref{sec:largebond}\\
$P_s$, coarse, fine & the part $A\cap s$ of a set $A$ of physical qudits in the super-site $s$. The set $A$ is coarse if every $P_s$ is empty or all of $s$, and fine otherwise & start of Appendix~\ref{app:largebond}\\
exponent, supported & the Pauli operator $\omega^kX^uZ^v$ has the exponent $(u,v)$, and it is supported in $A$ if $u$ and $v$ vanish outside $A$ & before Lemma~\ref{lem:F-code}\\
$\mathbf 1$, $\mathbf c$, $\bar X^g$, $\bar Z^g$ & the vectors $(1,1,1,1)$ and $(-1,-1,1,1)$ of the code stabilizers $X^{\mathbf 1}$ and $Z^{\mathbf c}$, and the operators $\bar X^g=X^{(g_1,g_2,g_1+g_2,0)}$ and $\bar Z^g=Z^{(g_1,g_2,0,-g_1-g_2)}$, which satisfy $\bar X^gE=EX^g$ and $\bar Z^gE=EZ^g$ & Lemma~\ref{lem:F-code}\\
$\Lambda(P)$ & the pairs $(x,z)\in G^2$ for which some element of the coset of $\bar X^x\bar Z^z$ modulo the code stabilizers is supported in $P$ & Lemma~\ref{lem:F-code}\\
$\mathcal S_A$, $\dim\mathcal S_A$ & the elements of a stabilizer group $\mathcal S$ that are supported in $A$, and the integer with $|\mathcal S_A|=d^{\dim\mathcal S_A}$ & Lemma~\ref{lem:F-stabrank}\\
$\Psi^{(0)}_x$, $\tilde\Psi^{(0)}_x$, $S$, $\tilde S$ & the coarse and the physical state at $\varepsilon=0$, and their stabilizer groups & Lemma~\ref{lem:F-counting}\\
$e=(e_s)$ & the coarse exponent of a Pauli operator on the $2n'$ qudits of the super-sites, with $e_s=(x_s,z_s)\in G^2$ & Lemma~\ref{lem:F-counting}\\
$f$, $\Lambda_A$, $S\cap\Lambda_A$ & the number of super-sites contained in $A$, the product of the sets $\Lambda(P_s)$, and the elements of $S$ with $e_s\in\Lambda(P_s)$ for all $s$ & Lemma~\ref{lem:F-counting}\\
$\tilde\ell(A)$ & the exponent of the rank $d^{\tilde\ell(A)}$ of $\tilde\Psi^{(0)}_x$ across $A$ & Lemma~\ref{lem:F-counting}\\
$\ell(C)$, $S(C)$ & for a coarse set $C$: the number $2\operatorname{rank}_{\F_d}\Gamma[C^c,C]$, with $\Gamma$ the adjacency matrix of $H^0_x$, and the elements of $S$ with $e_s=0$ for $s\notin C$ & Lemma~\ref{lem:F-counting}\\
cheap, center $\hat s$ & a coarse set $C$ is cheap if $\ell(C)\le2$. A center of a nonempty proper cheap set is a vertex shared by the edges of $H^0_x$ between the set and its complement & Lemma~\ref{lem:F-cheap}\\
$F$, $T$, $W$, $U$, $A_\bullet$, $A^*$ & the sets of super-sites $s$ with $|P_s|=4,3,2,1$, of sizes $f$, $t$, $w$, $u$, and the coarse sets $A_\bullet=F\cup T\cup W$ and $A^*=F\cup T$ & proof of Lemma~\ref{lem:F-fine0}\\
\bottomrule
\end{tabular}
\caption{Notation used in Appendix~\ref{app:largebond}. Here $S$ is a stabilizer group, while in Appendix~\ref{app:lower} it is the spine. Table~\ref{tab:E-notation} lists the notation of the coarse family.}
\label{tab:F-notation}
\end{table}

\subsection*{The code and stabilizer states}
The space $\C^G$ of a super-site is identified with $(\C^d)^{\otimes2}$, on which $X^g=X^{g_1}\otimes X^{g_2}$ and $Z^g=Z^{g_1}\otimes Z^{g_2}$ for $g=(g_1,g_2)\in G$, with the single-qudit operators $X\ket t=\ket{t+1}$ and $Z\ket t=\omega^t\ket t$. For $v\in\Z_d^r$ we write $X^v:=\bigotimes_iX^{v_i}$ and $Z^v:=\bigotimes_iZ^{v_i}$ on $r$ qudits. As $d$ is prime, $\Z_d$ is the field $\F_d$. A \emph{Pauli operator} on a set $V$ of qudits is an operator $\omega^kX^uZ^v$ with $k\in\Z_d$ and $u,v\in\F_d^V$, its \emph{exponent} is $(u,v)$, and it is \emph{supported} in $A\subseteq V$ if $u$ and $v$ vanish outside $A$. Its \emph{weight} is the number of qudits $i$ with $(u_i,v_i)\ne0$. Recall that $E\ket a=\ket{\varphi_a}=d^{-1/2}\sum_{t\in\Z_d}\ket{t+a_1,t+a_2,t+a_1+a_2,t}$ for $a=(a_1,a_2)\in G$ (Section~\ref{sec:largebond}). Lemma~\ref{lem:F-code} writes the stabilizers $X^{\otimes4}$ and $Z^{-1}\otimes Z^{-1}\otimes Z\otimes Z$ of the code as $X^{\mathbf 1}$ and $Z^{\mathbf c}$, the \emph{code stabilizers}. A spectrum is flat if its nonzero eigenvalues are equal, as in the proof of Lemma~\ref{lem:E-graphcut}.

\begin{lemma}[The code]\label{lem:F-code}
Put $\mathbf 1:=(1,1,1,1)$ and $\mathbf c:=(-1,-1,1,1)$ in $\Z_d^4$, and for $g\in G$ put $\bar X^g:=X^{(g_1,g_2,g_1+g_2,0)}$ and $\bar Z^g:=Z^{(g_1,g_2,0,-g_1-g_2)}$. For $\emptyset\ne P\subsetneq[4]$ let $\Lambda(P)\subseteq G^2$ be the set of pairs $(x,z)$ such that some element of the coset $\bar X^x\bar Z^z\{X^{c\mathbf 1}Z^{c'\mathbf c}:c,c'\in\Z_d\}$ is supported in $P$, let $\Lambda(\emptyset):=\{0\}$ and $\Lambda([4]):=G^2$.
\begin{enumerate}
\item[(a)] $\{\ket{\varphi_a}\}_{a\in G}$ is orthonormal, so $E$ is an isometry, and $X^{\mathbf 1}$ and $Z^{\mathbf c}$ fix every $\ket{\varphi_a}$.
\item[(b)] $\bar X^gE=EX^g$ and $\bar Z^gE=EZ^g$ as maps $\C^G\to(\C^d)^{\otimes4}$, that is, $\bar X^g\ket{\varphi_a}=\ket{\varphi_{a+g}}$ and $\bar Z^g\ket{\varphi_a}=\langle g,a\rangle\ket{\varphi_a}$.
\item[(c)] For every $\emptyset\ne P\subsetneq[4]$ and every $a\in G$, the Schmidt rank of $\ket{\varphi_a}$ across $P|[4]\setminus P$ is $d$, with a flat spectrum.
\item[(d)] $\Lambda(P)=\{0\}$ if $|P|\le1$, $\Lambda(P)=\F_dg_P\times\F_dh_P$ for some nonzero $g_P,h_P\in G$ if $|P|=2$, and $\Lambda(P)=G^2$ if $|P|=3$. For $P\subsetneq[4]$ and $(x,z)\in\Lambda(P)$, exactly one element of the coset of $(x,z)$ is supported in $P$, up to a phase. For $(x,z)=0$ it is the identity.
\end{enumerate}
\end{lemma}

\begin{proof}
\emph{Idea.} Parts (a)--(c) are read off from the $d$ basis vectors that make up a codeword. For (d), the Pauli operators that commute with the code stabilizers $X^{\mathbf 1}$ and $Z^{\mathbf c}$ are, up to a phase, exactly the elements of the cosets, and on one qudit only scalars do so. Two elements of one coset differ by a product of code stabilizers, which acts on all four qudits unless it is the identity, and this gives the uniqueness.

(a) The $d$ strings $(t+a_1,t+a_2,t+a_1+a_2,t)$, $t\in\Z_d$, of $\ket{\varphi_a}$ are distinct, and for $a\ne a'$ the strings of $\ket{\varphi_a}$ and $\ket{\varphi_{a'}}$ are disjoint, since two strings agree in the last coordinate only if they have the same $t$, and then they agree only if $a=a'$. $X^{\mathbf 1}$ permutes the strings of each codeword, and $\mathbf c\cdot(t\mathbf 1+(a_1,a_2,a_1+a_2,0))=0$, so $Z^{\mathbf c}$ acts trivially.

(b) $X^{(g_1,g_2,g_1+g_2,0)}$ maps the string with parameters $(t,a)$ to the string with parameters $(t,a+g)$, and $(g_1,g_2,0,-g_1-g_2)\cdot(t\mathbf 1+(a_1,a_2,a_1+a_2,0))=g\cdot a$.

(c) The $d$ strings of $\ket{\varphi_a}$ restricted to $P$ are distinct, as any single coordinate $t+a_1$, $t+a_2$, $t+a_1+a_2$ or $t$ determines $t$, and likewise on the complement, so the flattening is a partial permutation matrix with $d$ entries $d^{-1/2}$.

(d) \emph{Commuting operators.} For Pauli operators $X^uZ^v$ and $X^{u'}Z^{v'}$ on $r$ qudits,
\begin{equation*}
X^uZ^vX^{u'}Z^{v'}=\omega^{v\cdot u'-v'\cdot u}X^{u'}Z^{v'}X^uZ^v.
\end{equation*}
So a Pauli operator with exponent $(w,y)$ on the four qudits commutes with $X^{\mathbf 1}$ iff $\mathbf 1\cdot y=0$ and with $Z^{\mathbf c}$ iff $\mathbf c\cdot w=0$. An $X$-exponent $w\in\F_d^4$ with $\mathbf c\cdot w=0$ is $(x_1,x_2,x_1+x_2,0)+c\mathbf 1$ exactly for $c=w_4$ and $x_i=w_i-w_4$, $i=1,2$. A $Z$-exponent $y\in\F_d^4$ with $\mathbf 1\cdot y=0$ is $(z_1,z_2,0,-z_1-z_2)+c'\mathbf c$ exactly for $c'=y_3$ and $z_i=y_i+y_3$, $i=1,2$. Conversely, $\mathbf c\cdot(x_1,x_2,x_1+x_2,0)=\mathbf 1\cdot(z_1,z_2,0,-z_1-z_2)=\mathbf c\cdot\mathbf 1=0$. So the Pauli operators that commute with $X^{\mathbf 1}$ and $Z^{\mathbf c}$ are, up to a phase, exactly the elements of the cosets, each in exactly one coset.

\emph{At most one qudit.} By the first step, a single-qudit Pauli $X^uZ^v$ on qudit $i$ commutes with $X^{\mathbf 1}$ only if $v=0$, and with $Z^{\mathbf c}$ only if $\mathbf c_iu=0$, that is $u=0$, as $\mathbf c_i=\pm1$ is a unit of $\F_d$. So no element of weight one lies in a coset, and a scalar lies only in the coset of $0$, which gives $\Lambda(P)=\{0\}$ for $|P|\le1$.

\emph{Uniqueness.} Two elements of one coset supported in $P\subsetneq[4]$ differ by an element $X^{c\mathbf 1}Z^{c'\mathbf c}$ supported in $P$, which has weight $4$ unless $c=c'=0$. This gives the uniqueness claim, and the coset of $0$ is the group $\{X^{c\mathbf 1}Z^{c'\mathbf c}\}$, whose only element supported in a proper subset is the identity.

\emph{Two qudits.} For $P=\{i,j\}$, a Pauli $X_i^{u}Z_i^{v}X_j^{u'}Z_j^{v'}$ supported in $P$ commutes with $X^{\mathbf 1}$ iff $v+v'=0$ and with $Z^{\mathbf c}$ iff $\mathbf c_iu+\mathbf c_ju'=0$, so the commuting ones form a two-dimensional space of exponents, which meets the exponents of the stabilizers $\{X^{c\mathbf 1}Z^{c'\mathbf c}\}$ only in $0$. The map that sends such a Pauli to the pair $(x,z)$ of its coset is linear on exponents and, by the first step, vanishes only on the exponents of the stabilizers. So it is injective, and its image $\Lambda(P)$ is a two-dimensional subspace of $G^2$. The $X$-type elements $X_i^uX_j^{u'}$ with $\mathbf c_iu+\mathbf c_ju'=0$ form one line of the space of commuting exponents, mapped to pairs $(x,0)$. The $Z$-type elements $Z_i^vZ_j^{-v}$ form the other, mapped to pairs $(0,z)$. By the injectivity, $x\ne0$ for $u\ne0$ and $z\ne0$ for $v\ne0$. So $\Lambda(P)=\F_dg_P\times\F_dh_P$ with $g_P,h_P\ne0$.

\emph{Three qudits.} For $|P|=3$, $P=[4]\setminus\{p\}$, choose $c$ and $c'$ to cancel the $p$-th coordinates of the element $X^{(x_1,x_2,x_1+x_2,0)+c\mathbf 1}Z^{(z_1,z_2,0,-z_1-z_2)+c'\mathbf c}$ of the coset of $(x,z)$, which is possible since $\mathbf 1_p$ and $\mathbf c_p$ are units of $\F_d$. This element then acts trivially on $p$.
\end{proof}

A unit vector that spans the common fixed space of a group $\mathcal S$ as in Lemma~\ref{lem:F-stabrank} is called a \emph{stabilizer state} with \emph{stabilizer group} $\mathcal S$.

\begin{lemma}[Schmidt ranks of stabilizer states]\label{lem:F-stabrank}
Let $V$ be a finite set of qudits of prime dimension $d$, and let $\mathcal S$ be an abelian group of Pauli operators on $V$ of order $d^{|V|}$ that contains no scalar $\omega^k1$ with $k\ne0$. Then the common fixed space of $\mathcal S$ is spanned by a unit vector $\phi$, and $\ket\phi\bra\phi=d^{-|V|}\sum_{P\in\mathcal S}P$. For $A\subseteq V$ let $\mathcal S_A$ be the subgroup of the elements of $\mathcal S$ supported in $A$. Then $|\mathcal S_A|=d^{\dim\mathcal S_A}$ for an integer $\dim\mathcal S_A$, the spectrum of $\rho_A(\phi)$ is flat, and $r_\phi(A)=d^{|A|-\dim\mathcal S_A}$.
\end{lemma}

\begin{proof}
\emph{Idea.} The average of the elements of $\mathcal S$ is the projector onto their common fixed space, and its trace is $1$. The partial trace over $A^c$ annihilates every element that is not supported in $A$. So $\rho_A(\phi)$ is a multiple of the average of the elements of $\mathcal S_A$, as operators on $A$, which is again a projector.

The exponent map $\omega^kX^uZ^v\mapsto(u,v)$ is a homomorphism from $\mathcal S$ to $\F_d^{2|V|}$, since the product of two Pauli operators has the sum of their exponents up to a phase, and its kernel consists of the scalars in $\mathcal S$, which is trivial. So $\mathcal S$ is isomorphic to a subgroup, hence a subspace, of $\F_d^{2|V|}$, and so is every subgroup $\mathcal S_A$, which gives $|\mathcal S_A|=d^{\dim\mathcal S_A}$.

The elements of $\mathcal S$ form a finite group of unitaries, so $\Pi:=|\mathcal S|^{-1}\sum_{P\in\mathcal S}P$ is the orthogonal projector onto their common fixed space. As $\Tr(X^uZ^v)=\prod_{i\in V}\sum_t\omega^{v_it}[u_i=0]=d^{|V|}[u=v=0]$, and the identity is the only element of $\mathcal S$ with exponent $0$, $\Tr\Pi=|\mathcal S|^{-1}d^{|V|}=1$, so $\Pi=\ket\phi\bra\phi$ for a unit vector $\phi$.

For $A\subseteq V$, write $P=P|_A\otimes P|_{A^c}$ with Pauli operators on $A$ and $A^c$ up to a phase. Then $\Tr_{A^c}P=P|_A\Tr(P|_{A^c})$ vanishes unless $P$ is supported in $A$, in which case $P|_{A^c}$ is the identity and $\Tr_{A^c}P=d^{|A^c|}P|_A$. Hence
\begin{equation*}
\rho_A(\phi)=d^{-|A|}\sum_{P\in\mathcal S_A}P|_A.
\end{equation*}
The operators $P|_A$, $P\in\mathcal S_A$, form an abelian group of unitaries on $A$ without nontrivial scalars, isomorphic to $\mathcal S_A$, so $|\mathcal S_A|^{-1}\sum_{P\in\mathcal S_A}P|_A$ is an orthogonal projector, whose rank is $|\mathcal S_A|^{-1}d^{|A|}$ by the same trace computation. Therefore $\rho_A(\phi)$ is $|\mathcal S_A|d^{-|A|}$ times a projector of rank $d^{|A|}/|\mathcal S_A|=d^{|A|-\dim\mathcal S_A}$.
\end{proof}

For qubits this is the argument of Fattal et al.~\cite{FCYBC04}, and for qudits see~\cite{LG11,HDD05}.
\subsection*{Coarse cuts}
\begin{lemma}[What transfers from the coarse family]\label{lem:F-transfer}
Let $0<\varepsilon\le\bar\varepsilon$. Then $\tilde\Psi_x$ is a unit vector, $\tilde\Psi_x\in\TTN(\tilde T_x,\chi)$, $\tilde T_x^{(2)}=\tilde T_x$, the trees $\tilde T_x$ are pairwise distinct, $\tilde T_x\in\Xn(s)$ for every $s\ge4$, and $\braket{\tilde\Psi_x}{\tilde\Psi_y}=(1-\varepsilon)^{\dH(x,y)}$. For every coarse set $A$, $\varepsilon_A(\tilde\Psi_x;\chi)=\varepsilon_A(\Psi_x;\chi)$, which is $0$ if $A$ is a cluster of $T_x$, exactly $\theta\varepsilon$ if $A\in\mathcal E_x$, and at least $(\sqrt{1-\chi/q^2}-\sqrt{2\varepsilon})^2\ge\theta\varepsilon$ if $A$ is a nonempty proper non-cluster set outside $\mathcal E_x$.
\end{lemma}

\begin{proof}
$E^{\otimes n'}$ is an isometry acting on each super-site, so inner products and the Schmidt coefficients of every coarse cut are those of $\Psi_x$ (Lemma~\ref{lem:E-isometry}), and the claims about inner products and coarse sets are Lemmas~\ref{lem:E-connector}, \ref{lem:E-membership}, \ref{lem:E-epstype} and~\ref{lem:E-perturb} over $G=\Z_d^2$, whose hypotheses $q\le\chi\le q^2-1$ and $\varepsilon\le\bar\varepsilon\le\frac12(1-\chi/q^2)\le M/(M+q)$ hold. The bound $(\sqrt{1-\chi/q^2}-\sqrt{2\varepsilon})^2\ge\theta\varepsilon$ holds for $\varepsilon\le\bar\varepsilon$ by the equivalence displayed in the proof of Lemma~\ref{lem:E-gap}.

The clusters of $\tilde T_x$ are the physical singletons, of rank at most $d\le\chi$, the coarse clusters, of rank at most $q\le\chi$ by Lemma~\ref{lem:E-membership}, and complements. Every internal edge of $\tilde T_x$ has both sides of size at least $4>2$, so nothing is contracted in $\tilde T_x^{(2)}$. The internal vertices of $\tilde T_x$ have degree $3$ or $5$, so none is a hub when $s\ge4$. Distinctness is Lemma~\ref{lem:E-tree}, as the coarse clusters of $\tilde T_x$ determine $T_x$.
\end{proof}

\subsection*{Fine cuts of the unperturbed states}
At $\varepsilon=0$, the formulas of Definition~\ref{def:E-hard}(b) give $\eta_a=q^{-1/2}\ket{aa}$ and $\tilde\psi_{x_b}=\GHZ_G$ on $(\alpha_b,\beta_b,\gamma_b)$, where $\GHZ_G:=q^{-1/2}\sum_{a\in G}\ket{aaa}$, a three-site state, unlike the four-site $\GHZ_q$ of Definition~\ref{def:E-hard}(a). As in the proof of Lemma~\ref{lem:E-perturb}, $\Psi^{(0)}_x$ is the coarse state given by these formulas, and $\tilde\Psi^{(0)}_x:=E^{\otimes n'}\Psi^{(0)}_x$. A Pauli operator $\omega^k\prod_sX^{x_s}_sZ^{z_s}_s$ on the $2n'$ qudits of the super-sites, with $x_s,z_s\in G$, has the \emph{coarse exponent} $e=(e_s)_{s\in[n']}$, $e_s=(x_s,z_s)\in G^2$, and the \emph{encoded element} $\omega^k\prod_s\bar X^{x_s}_s\bar Z^{z_s}_s$, with $\bar X^g$ and $\bar Z^g$ of Lemma~\ref{lem:F-code}.

For a set $A$ of physical qudits let $f:=|\{s:P_s=s\}|$, the number of super-sites contained in $A$, and $\Lambda_A:=\prod_s\Lambda(P_s)$. Let $\Gamma$ be the adjacency matrix of $H^0_x$, and for a coarse set $C$ put $\ell(C):=2\operatorname{rank}_{\F_d}\Gamma[C^c,C]$. As in the proof of Lemma~\ref{lem:F-stabrank}, the exponents of a group $\mathcal S$ of Pauli operators without nontrivial scalars form a subspace, whose dimension $\dim\mathcal S$ satisfies $|\mathcal S|=d^{\dim\mathcal S}$.

\begin{lemma}[The stabilizer groups and the counting identity]\label{lem:F-counting}
Let $\varepsilon=0$. The coarse state $\Psi^{(0)}_x$ is a stabilizer state on the $2n'$ qudits of the super-sites: its stabilizer group $S$ is generated by the elements $Z^g_{\alpha_b}Z^{-g}_{\gamma_b}$, $Z^g_{\beta_b}Z^{-g}_{\gamma_b}$ and $X^g_{\alpha_b}X^g_{\beta_b}X^g_{\gamma_b}Z^g_{\mathcal A_b}$ for $b\in[K]$, and $X^g_v\prod_{u\sim v}Z^g_u$ for every spine site $v$, with $g\in G$, where $u\sim v$ runs over the neighbors of $v$ in $H^0_x$. It has order $d^{2n'}$, and an element of $S$ is determined by its coarse exponent. The physical state $\tilde\Psi^{(0)}_x$ is a stabilizer state on the $n$ physical qudits whose group $\tilde S$ is generated by the encoded elements of the generators of $S$, together with the code stabilizers $X^{\mathbf 1}_s$ and $Z^{\mathbf c}_s$, $s\in[n']$. For a set $A$ of physical qudits let $S\cap\Lambda_A$ denote the subgroup of the elements of $S$ whose coarse exponent satisfies $e_s\in\Lambda(P_s)$ for all $s$. Then
\begin{equation*}
\dim\tilde S_A=\dim(S\cap\Lambda_A)+2f,\qquad r_{\tilde\Psi^{(0)}_x}(A)=d^{\tilde\ell(A)},\qquad\tilde\ell(A):=\sum_s|P_s|-2f-\dim(S\cap\Lambda_A),
\end{equation*}
with a flat spectrum. For a coarse set $C$, read as a set of super-sites, $\tilde\ell(C)=\ell(C)$, and the subgroup $S(C)$ of the elements of $S$ with $e_s=0$ for $s\notin C$ has $\dim S(C)=2|C|-\ell(C)$.
\end{lemma}

\begin{proof}
\emph{Idea.} The steps \emph{The coarse group} and \emph{The physical group} check that the listed operators fix the two states, commute and are independent, so that Lemma~\ref{lem:F-stabrank} applies to both groups. The step \emph{Counting} counts the elements of $\tilde S$ supported in $A$ one super-site at a time, using Lemma~\ref{lem:F-code}(d) on the super-sites not inside $A$. In the step \emph{Coarse sets}, the rank across a coarse set is that of the graph state of $H^0_x$, which Lemma~\ref{lem:E-graphcut} gives.

\emph{The coarse group.} The listed operators fix $\Psi^{(0)}_x=\prod_{uv\in E_x}CZ_{uv}\bigl(\bigotimes_b\GHZ_G(\alpha_b\beta_b\gamma_b)\otimes\bigotimes_{j\le m}\ket+_{v_j}\bigr)$. The $Z$-type pairs fix each $\GHZ_G$ and commute with the gates. The product $X^g_{\alpha_b}X^g_{\beta_b}X^g_{\gamma_b}$ fixes the block vector, and conjugation by $CZ_{\gamma_bv}$ turns $X^g_{\gamma_b}$ into $X^g_{\gamma_b}Z^g_v$, since $CZ_{uv}X^g_uCZ_{uv}^\dagger=X^g_uZ^g_v$. Finally, $X^g_v$ fixes $\ket+$, and conjugation by the gates at $v$ attaches $Z^g$ on the neighbors. The listed operators commute pairwise, as two Pauli operators commute up to a power of $\omega$ (proof of Lemma~\ref{lem:F-code}(d)), and this power is $1$ when both fix $\Psi^{(0)}_x\ne0$. They are independent: a product of generators with coefficient vector $c$ has $X$-exponent $c_b$ on $\alpha_b$ for the block $X$-generator and $c_v$ on $v$ for the spine generators, so a trivial $X$-part forces all $X$-coefficients to vanish, and then the $Z$-exponents on $\alpha_b$ and $\beta_b$ force the $Z$-coefficients to vanish. Hence the group $S$ they generate has order $d^{2(3K+m)}=d^{2n'}$, it contains no nontrivial scalar, since every element fixes $\Psi^{(0)}_x$, and Lemma~\ref{lem:F-stabrank} on the $2n'$ qudits shows that $\Psi^{(0)}_x$ spans its fixed space. The coarse exponent determines the element, since two elements with the same exponent differ by a scalar in $S$. From now on we identify an element of $S$ with its coarse exponent. So $S$ is a subspace of $(G^2)^{n'}=\F_d^{4n'}$ of dimension $2n'$, and $S\cap\Lambda_A$ is its intersection with $\Lambda_A$, a subspace by Lemma~\ref{lem:F-code}(d).

\emph{The physical group.} By Lemma~\ref{lem:F-code}(b), $\bigl(\prod_s\bar X^{x_s}_s\bar Z^{z_s}_s\bigr)E^{\otimes n'}=E^{\otimes n'}\prod_sX^{x_s}_sZ^{z_s}_s$, so the encoded element of every element of $S$ fixes $\tilde\Psi^{(0)}_x=E^{\otimes n'}\Psi^{(0)}_x$, and so do the code stabilizers, which fix every codeword $\ket{\varphi_a}$ by Lemma~\ref{lem:F-code}(a). The physical exponents of $\bar X^g$ and $\bar Z^g$ on a super-site are $(g_1,g_2,g_1+g_2,0)$ and $(g_1,g_2,0,-g_1-g_2)$, and the symplectic product of $\bar X^g$ with $\bar Z^h$ is $g\cdot h$, the same as that of $X^g$ with $Z^h$ on $\C^G$, so encoded elements of commuting coarse elements commute, and the code stabilizers commute with all of them.

On each super-site the $X$-exponents $(1,0,1,0)$, $(0,1,1,0)$ and $\mathbf 1$ are linearly independent, and so are the $Z$-exponents $(1,0,0,-1)$, $(0,1,0,-1)$ and $\mathbf c$. Hence a physical exponent vector determines the coarse exponent $e$ and the coefficients $c_s,c'_s$ of the code stabilizers. Counted over $\F_d$, $S$ has $2n'$ independent generators, two in each of its $n'$ families indexed by $g\in G=\F_d^2$, and with the $2n'$ code stabilizers the $2n'+2n'=n$ generators of $\tilde S$ are independent. So $\tilde S$ has order $d^n$, it contains no nontrivial scalar, since every element fixes $\tilde\Psi^{(0)}_x$, and $\tilde\Psi^{(0)}_x$ spans its fixed space by Lemma~\ref{lem:F-stabrank}.

\emph{Counting.} An element of $\tilde S$ is, up to a phase, $\prod_sR_s$, where $R_s$ lies in the coset $\bar X^{x_s}\bar Z^{z_s}\{X^{c\mathbf 1}Z^{c'\mathbf c}\}$ of $e_s$ for a unique $e\in S$. It is supported in $A$ if and only if every $R_s$ is supported in $P_s$. For $P_s=s$ there is no condition, and all $d^2$ elements of the coset are allowed. For $P_s\ne s$, by Lemma~\ref{lem:F-code}(d), the coset of $e_s$ contains an element supported in $P_s$ if and only if $e_s\in\Lambda(P_s)$, and then exactly one. Hence $|\tilde S_A|=|S\cap\Lambda_A|\cdot d^{2f}$, which is the first identity, and Lemma~\ref{lem:F-stabrank} gives the rank and the flat spectrum.

\emph{Coarse sets.} For a coarse $C$, the physical rank equals the coarse rank by Lemma~\ref{lem:E-isometry}. Up to the local Fourier transforms on the super-sites $\alpha_b,\beta_b$, the coarse state is the graph state of $H^0_x$ over $G$ (step \emph{Graph form at $\varepsilon=0$} of the proof of Lemma~\ref{lem:E-perturb}), whose rank across a nonempty proper $C$ is $d^{2\operatorname{rank}_{\F_d}\Gamma[C^c,C]}$ by Lemma~\ref{lem:E-graphcut}. For $C\in\{\emptyset,[n']\}$ the rank is $1=d^{\ell(C)}$, as $\Gamma[C^c,C]$ has no entries. Finally, $S(C)$ is the subgroup of the elements of $S$ supported in the $2|C|$ qudits of $C$, so Lemma~\ref{lem:F-stabrank} on the coarse system gives $d^{\ell(C)}=d^{2|C|-\dim S(C)}$.
\end{proof}

\begin{lemma}[Cheap coarse sets are star sets]\label{lem:F-cheap}
Call a coarse set $C$ \emph{cheap} if $\ell(C)\le2$, that is, if $\Psi^{(0)}_x$ has rank at most $\chi$ across $C$. The cheap coarse sets are exactly $\emptyset$, $[n']$, the clusters of $T_x$ and the $\varepsilon$-type sets, and every nonempty proper cheap set is a star set: all edges of $H^0_x$ between $C$ and its complement share a vertex $\hat s$, a \emph{center} of $C$. Every other coarse set has $\ell(C)\ge4$, and $\ell(C)=0$ exactly for $C\in\{\emptyset,[n']\}$.
\end{lemma}

\begin{proof}
$\ell(C)\le2$ means $\operatorname{rank}_{\F_d}\Gamma[C^c,C]\le1$. As $\ell(C)$ is even and $d^2\le\chi<d^4$, it also means that the rank $d^{\ell(C)}$ of $\Psi^{(0)}_x$ across $C$ (Lemma~\ref{lem:F-counting}) is at most $\chi$. Call the edges of $H^0_x$ between $C$ and $C^c$ the \emph{crossing edges}. For a nonempty proper $C$ they form the set $\partial C$ of Lemma~\ref{lem:E-startsets}. A nonempty proper set that is not a star set has, by Lemma~\ref{lem:E-startsets}(ii) and~(iii), two disjoint crossing edges $u_1w_1,u_2w_2$ with $u_1,u_2\in C$ and $w_1,w_2\notin C$ such that $u_1w_2,u_2w_1$ are not both edges. So the $2\times2$ minor of $\Gamma[C^c,C]$ on these rows and columns has determinant $\pm1$, and the rank is at least $2$. A star set has all crossing edges at one vertex, so $\Gamma[C^c,C]$ has a single nonzero row or column and rank at most $1$. The identification of star sets with $\clus(T_x)\cup\mathcal E_x$ is Lemma~\ref{lem:E-startsets}(ii). The rank is $0$ exactly when there are no crossing edges, which for the connected graph $H^0_x$ means $C\in\{\emptyset,[n']\}$.
\end{proof}
\begin{lemma}[Fine sets are far at $\varepsilon=0$]\label{lem:F-fine0}
Let $A$ be a fine set of physical qudits with $3\le|A|\le n-3$. Then $\tilde\ell(A)\ge3$: the Schmidt rank of $\tilde\Psi^{(0)}_x$ across $A$ is at least $d^3>\chi$, and $\varepsilon_A(\tilde\Psi^{(0)}_x;\chi)=1-\chi/d^{\tilde\ell(A)}\ge1-\chi/d^3$.
\end{lemma}

\begin{proof}
\emph{Idea.} Two bounds compare $\tilde\ell(A)$ with $\ell$ of two coarse sets, the super-sites that $A$ meets in at least two qudits and those it meets in at least three. With Lemma~\ref{lem:F-cheap} and $3\le|A|\le n-3$, they settle every case except one: $A$ meets every super-site in $0$, $2$ or $4$ qudits, and both coarse sets are cheap. Then the conditions $e_s\in\Lambda(P_s)$ lower the dimension of the subgroup of $S$ supported in the first set by at least one if its $\ell$ is $2$, and by at least three if it is $0$.

As $\rho_A(\tilde\Psi^{(0)}_x)$ has a flat spectrum with $d^{\tilde\ell(A)}$ nonzero eigenvalues (Lemma~\ref{lem:F-counting}), it suffices to show $\tilde\ell(A)\ge3$: the rank is then at least $d^3>\chi$, and the $\chi$-tail is $1-\chi/d^{\tilde\ell(A)}\ge1-\chi/d^3$.

Write $F$, $T$, $W$, $U$ for the sets of super-sites $s$ with $|P_s|=4,3,2,1$ and $f,t,w,u$ for their sizes, and put $A_\bullet:=F\cup T\cup W$ and $A^*:=F\cup T$, two coarse sets. By Lemma~\ref{lem:F-counting} and $\sum_s|P_s|=4f+3t+2w+u$, $\tilde\ell(A)=2f+3t+2w+u-\dim(S\cap\Lambda_A)$, and $\dim S(C)=2|C|-\ell(C)$ for every coarse $C$.

\emph{Two bounds.} Both bounds use only Lemmas~\ref{lem:F-code}(d) and~\ref{lem:F-counting}, so they hold for every set $A$ of physical qudits. An element of $S\cap\Lambda_A$ has $e_s=0$ whenever $|P_s|\le1$, as $\Lambda(P_s)=\{0\}$ then, so $S\cap\Lambda_A\subseteq S(A_\bullet)$ and
\begin{equation}\label{eq:F-bound1}
\tilde\ell(A)\ \ge\ 2f+3t+2w+u-\bigl(2(f+t+w)-\ell(A_\bullet)\bigr)\ =\ \ell(A_\bullet)+t+u.
\end{equation}
The map $e\mapsto(e_s)_{s\in W}$ from $S\cap\Lambda_A$ to the space $\prod_{s\in W}\Lambda(P_s)$, of dimension $2w$ by Lemma~\ref{lem:F-code}(d), has kernel contained in $S(A^*)$, so
\begin{equation}\label{eq:F-bound2}
\tilde\ell(A)\ \ge\ 2f+3t+2w+u-\bigl(2(f+t)-\ell(A^*)+2w\bigr)\ =\ \ell(A^*)+t+u.
\end{equation}

\emph{Cases.} If $t+u\ge3$, then~\eqref{eq:F-bound1} gives the claim. If $A_\bullet$ or $A^*$ is not cheap, then~\eqref{eq:F-bound1} or~\eqref{eq:F-bound2} gives $\tilde\ell(A)\ge4$ by Lemma~\ref{lem:F-cheap}. So assume that $t+u\le2$ and that both $A_\bullet$ and $A^*$ are cheap.

Suppose first that $t+u\ge1$. If one of $\ell(A_\bullet)$, $\ell(A^*)$ equals $2$, the claim follows from~\eqref{eq:F-bound1} or~\eqref{eq:F-bound2}. If both are $0$, then $A^*\subseteq A_\bullet$ both lie in $\{\emptyset,[n']\}$ by Lemma~\ref{lem:F-cheap}: $A^*=A_\bullet=\emptyset$ forces $f=t=w=0$ and $|A|=u\le2$. The case $A^*=A_\bullet=[n']$ forces $w=u=0$ and $|A^c|=t\le2$. The case $A^*=\emptyset$, $A_\bullet=[n']$ forces $f=t=0$, $w=n'$ and $u=0$, so $t+u=0$. The first two contradict $3\le|A|\le n-3$, and the third contradicts $t+u\ge1$.

It remains to treat $t=u=0$. Then $A$ consists of the whole super-sites of $C:=A^*=F$ and two qudits from each super-site of $W$, with $w\ge1$ since $A$ is fine, and both $C$ and $C\cup W=A_\bullet$ are cheap. Here
\begin{equation*}
\tilde\ell(A)=2|C|+2|W|-\kappa,\qquad\kappa:=\dim\{e\in S(C\cup W):\ e_s\in\Lambda(P_s)\ \text{for all}\ s\in W\},
\end{equation*}
and we must show $\kappa\le2|C\cup W|-3$. As $C\cup W$ is cheap and $\ell$ takes even values, $\ell(C\cup W)$ is $2$ or $0$.

\emph{The graph picture.} Up to the local Fourier transforms on the super-sites $\alpha_b,\beta_b$, the coarse state is the graph state of $H^0_x$ over $G$ (step \emph{Graph form at $\varepsilon=0$} of the proof of Lemma~\ref{lem:E-perturb}), whose stabilizer elements are $\prod_s\bigl(X^{c_s}_s\prod_{s'\sim s}Z^{c_s}_{s'}\bigr)$, $c\in G^{n'}$, with exponents $e_s=(c_s,(\Gamma c)_s)$, $(\Gamma c)_s:=\sum_{s'\sim s}c_{s'}$. The Fourier transform $\mathsf F\ket a:=q^{-1/2}\sum_{y\in G}\langle a,y\rangle\ket y$ of that step satisfies $\mathsf FX^g\mathsf F^\dagger=Z^g$ and $\mathsf FZ^g\mathsf F^\dagger=X^{-g}$, so it exchanges the $X$- and $Z$-exponents up to sign, which maps a subgroup $\F_dg\times\F_dh$ of $G^2$ to one of the same form. So we may compute in the graph picture, where the constraint $e_s\in\Lambda(P_s)$ reads $c_s\in\F_dg_s$ and $(\Gamma c)_s\in\F_dh_s$ for nonzero $g_s,h_s\in G$ that depend on $s$ and $P_s$, and $S(C\cup W)$ consists of the $c$ supported in $C\cup W$ with $(\Gamma c)_s=0$ for all $s\notin C\cup W$.

\emph{Case $\ell(C\cup W)=2$.} By Lemma~\ref{lem:F-cheap}, $C\cup W$ is a nonempty proper star set with a center $\hat s$, and $\dim S(C\cup W)=2|C\cup W|-2$. It suffices to exhibit one element of $S(C\cup W)$ that violates a constraint at some $s\in W$, since then the constrained subspace is proper and $\kappa\le2|C\cup W|-3$.

(i) If some $s\in W$ has no neighbor outside $C\cup W$, take $c:=g\delta_s$ with $g\in G\setminus\F_dg_s$: it is supported in $C\cup W$, $\Gamma c$ is supported on the neighbors of $s$, all inside $C\cup W$, and $c_s\notin\F_dg_s$.

Otherwise every $s\in W$ lies on a crossing edge of $C\cup W$, hence equals $\hat s$ or is adjacent to $\hat s$ with $\hat s\notin C\cup W$.

(ii) Suppose $\hat s\in C\cup W$. Then $W=\{\hat s\}$, and $C\ne\emptyset$ because $|A|=4|C|+2\ge3$. As $C$ is cheap, nonempty and not $[n']$, $\ell(C)=2$. Let $\pi:S(C\cup\{\hat s\})\to G^2$, $e\mapsto e_{\hat s}$. Its kernel consists of the elements supported in $C$, which is $S(C)$, so
\begin{equation*}
\dim\Img\pi=\dim S(C\cup\{\hat s\})-\dim S(C)=(2|C|+2-2)-(2|C|-2)=2.
\end{equation*}
The vertex $\hat s$ has a neighbor $s'$ outside $C\cup W$, as the set is proper and $H^0_x$ is connected, and $s'$ has no other neighbor in $C\cup W$ by the star property, so $(\Gamma c)_{s'}=c_{\hat s}$, which membership in $S(C\cup W)$ forces to vanish. Hence $\Img\pi\subseteq\{0\}\times G$, and by dimension $\Img\pi=\{0\}\times G$. Thus
\begin{equation*}
\kappa=\dim S(C)+\dim\bigl((\{0\}\times G)\cap(\F_dg_{\hat s}\times\F_dh_{\hat s})\bigr)=2|C|-2+1=2|C\cup W|-3.
\end{equation*}

(iii) Suppose $\hat s\notin C\cup W$, so every $s\in W$ is adjacent to $\hat s$. Let $N':=(C\cup W)\cap N(\hat s)$, where $N(\hat s)$ is the neighborhood of $\hat s$. The crossing edges of $C\cup W$ are exactly $s\hat s$, $s\in N'$, and a vector $c$ supported in $C\cup W$ lies in $S(C\cup W)$ if and only if $\sum_{s\in N'}c_s=0$, since no other vertex outside $C\cup W$ has a neighbor inside. Fix $s\in W$. If $N'\ne\{s\}$, pick $s'\in N'\setminus\{s\}$ and $g\in G\setminus\F_dg_s$ and take $c:=g\delta_s-g\delta_{s'}$: it lies in $S(C\cup W)$ and violates the constraint at $s$. If $N'=\{s\}$, every vertex of $C\cup W$ other than $s$ has no neighbor outside $C\cup W$. If $s$ has a neighbor $s'\in C\cup W$, take $c:=h\delta_{s'}$ with $h\in G\setminus\F_dh_s$: it is supported in $C\cup W$, $\Gamma c$ is supported on $N(s')\subseteq C\cup W$, so $c\in S(C\cup W)$, and $(\Gamma c)_s=h\notin\F_dh_s$. If $s$ has no neighbor in $C\cup W$, then $(C\cup W)\setminus\{s\}$ has no edge to $s$ and none to the outside, so it is a union of connected components of the connected graph $H^0_x$ that misses $s$, hence empty. Then $C=\emptyset$ and $|A|=2$, which is excluded.

\emph{Case $\ell(C\cup W)=0$.} Then $C\cup W=[n']$, $|A^c|=2w\ge3$, so $w\ge2$, and $\kappa=2n'-r$, where $r$ is the rank of the system of constraints $c_s\in\F_dg_s$, $(\Gamma c)_s\in\F_dh_s$, $s\in W$, on $c\in G^{n'}\cong\F_d^{2n'}$, and we need $r\ge3$. The constraints $c_s\in\F_dg_s$, $s\in W$, are $w$ independent linear equations on disjoint coordinate blocks. If $w\ge3$ we are done.

If $W=\{s_1,s_2\}$, then $C=[n']\setminus W\ne\emptyset$ since $n'\ge4$, and some $s_i$ has a neighbor $s'\in C$, for otherwise $\{s_1,s_2\}$ would be a union of components of the connected graph $H^0_x$. The constraint $(\Gamma c)_{s_i}\in\F_dh_{s_i}$ is one linear equation $\lambda((\Gamma c)_{s_i})=0$ for a functional $\lambda$ on $G$ with kernel $\F_dh_{s_i}$, and it involves $c_{s'}$ through $\lambda(c_{s'})$, a nonzero functional of a coordinate block that the two equations on $c_{s_1},c_{s_2}$ do not involve. So $r\ge3$.
\end{proof}
\subsection*{Fine cuts of the perturbed states}
Lemma~\ref{lem:F-ghz}, a tool for Lemma~\ref{lem:F-fine}, treats the encoded block $E^{\otimes3}\GHZ_G$ at $\varepsilon=0$, where $\GHZ_G$ is the three-site state of the setup before Lemma~\ref{lem:F-counting}.

\begin{lemma}[The encoded block]\label{lem:F-ghz}
Let $\phi:=E^{\otimes3}\GHZ_G$ on the twelve qudits of three super-sites. For every nonempty proper subset $Q$ of these qudits, the spectrum of $\rho_Q(\phi)$ is flat with rank at least $d$.
\end{lemma}

\begin{proof}
The state $\GHZ_G=q^{-1/2}\sum_{a\in G}\ket{aaa}$ is a stabilizer state on six qudits with group $S_{\GHZ}=\{\prod_{s\le3}X^{x}_sZ^{z_s}_s:\ x,z_1,z_2,z_3\in G,\ z_1+z_2+z_3=0\}$, of order $d^6$, generated by $X^gX^gX^g$ and the pairs $Z^g_sZ^{-g}_{s'}$. Put $Q_s:=Q\cap s$, which plays the role of $P_s$. The arguments of Lemma~\ref{lem:F-counting} apply verbatim to $\phi$, with the counting identity $\tilde\ell(Q)=\sum_s|Q_s|-2f-\dim(S_{\GHZ}\cap\Lambda_Q)$ and the flat spectrum. The coarse rank of $\GHZ_G$ across a nonempty proper union $C$ of super-sites is $q=d^2$, so $\ell_{\GHZ}(C)=2$ for such $C$ and $0$ for $C\in\{\emptyset,[3]\}$, and $\dim S_{\GHZ}(C)=2|C|-\ell_{\GHZ}(C)$.

The two bounds~\eqref{eq:F-bound1} and~\eqref{eq:F-bound2} hold with $\ell_{\GHZ}$ in place of $\ell$, with $Q_\bullet$, $Q^*$ and $t,u,w,f$ defined from $Q$.

\emph{Case $t+u\ge1$.} When $Q_\bullet$ is nonempty and proper, \eqref{eq:F-bound1} gives $\tilde\ell(Q)\ge3$. When $Q_\bullet=\emptyset$, $f=t=w=0$ and $\tilde\ell(Q)\ge u\ge1$. When $Q_\bullet=[3]$, $u=0$ and $t\ge1$, and \eqref{eq:F-bound2} gives $\tilde\ell(Q)\ge\ell_{\GHZ}(Q^*)+t\ge1$.

\emph{Case $t=u=0$.} Then every super-site meets $Q$ in $0$, $2$ or $4$ qudits. If $w=0$, then $Q$ is a nonempty proper union of super-sites, and $\tilde\ell(Q)=\ell_{\GHZ}(F)=2$. Let $w\ge1$. If $f\ge1$, then $Q^*=F$ is nonempty and proper, as $w\ge1$, and \eqref{eq:F-bound2} gives $\tilde\ell(Q)\ge2$. If $f=0$ and $w\le2$, then $Q_\bullet=W$ is nonempty and proper and \eqref{eq:F-bound1} gives $\tilde\ell(Q)\ge2$. If $f=0$ and $w=3$, then $|Q_s|=2$ and $\Lambda(Q_s)=\F_dg_s\times\F_dh_s$ for each $s$ by Lemma~\ref{lem:F-code}(d). Hence $S_{\GHZ}\cap\Lambda_Q$ consists of the elements with $x\in\bigcap_s\F_dg_s$, a space of dimension at most $1$, and $z_s\in\F_dh_s$ with $z_1+z_2+z_3=0$, a space of dimension at most $2$, so $\tilde\ell(Q)\ge6-3=3$.

In all cases $\tilde\ell(Q)\ge1$.
\end{proof}

For a real number $\tau$ we write $\tau_+:=\max\{\tau,0\}$.

\begin{lemma}[Fine sets at $\varepsilon>0$]\label{lem:F-fine}
Let $0<\varepsilon<1$ and let $A$ be a fine set with $3\le|A|\le n-3$. Then
\begin{equation*}
\varepsilon_A(\tilde\Psi_x;\chi)\ \ge\ \bigl(\sqrt{1-\chi/d^3}-\sqrt{3\varepsilon}\bigr)_+^2.
\end{equation*}
\end{lemma}

\begin{proof}
Write $\tilde B_b$ for the twelve physical qudits of block $b$ and call $b$ \emph{split} by $A$ if $\emptyset\ne A\cap\tilde B_b\ne\tilde B_b$.

\emph{The block isometries.} The map $V_b:\ket a\mapsto\sqrt q\,\eta_a\otimes\ket a_{\gamma_b}$ is an isometry from $\C^G$ into the coarse space $(\C^G)^{\otimes B_b}$ of $B_b$. Indeed, the vectors $\ket a_{\gamma_b}$ are orthonormal, and $\lVert\eta_a\rVert^2=(1-\varepsilon)/q+(q-1)\varepsilon/M=1/q$, as each $a\in G$ lies in $q-1$ pairs $(c,a)$ of the pair set of Definition~\ref{def:E-hard}. The state $\psi^{(b)}_{x_b}$ of Lemma~\ref{lem:E-connector}(ii) is $(V_b\otimes1)\ket{\Phi_q}$, with $\ket{\Phi_q}:=q^{-1/2}\sum_a\ket a\ket a$ on an auxiliary register $B'_b$ and the site $\xi_b$. With the connector $\mathcal W$ of that lemma,
\begin{equation*}
\tilde\Psi_x=\Bigl(\bigotimes_bE^{\otimes3}V_b\ \otimes\ E^{\otimes m}\mathcal W\Bigr)\bigotimes_b\ket{\Phi_q}_{B'_b\xi_b}.
\end{equation*}
Only the $V_b$ depend on $\varepsilon$, and $\bra{\Phi_q}(V_b^{(0)\dagger}V_b^{(\varepsilon)}\otimes1)\ket{\Phi_q}=q^{-1}\sum_aq\braket{\eta^{(0)}_a}{\eta^{(\varepsilon)}_a}=\sqrt{1-\varepsilon}$, since the terms $\ket{cc}$ of $\eta_a$ with $c\ne a$ are orthogonal to $\ket{aa}$.

\emph{At most two split blocks.} For a block $b$ that is not split, the isometry $E^{\otimes3}V_b$ acts on one side of the cut $A|A^c$, and removing it, that is, replacing $\tilde B_b$ by the register $B'_b$ on the same side, changes no Schmidt coefficient (Lemma~\ref{lem:E-isometry}). Let $\Theta^{(\varepsilon)}$ be the vector obtained by removing the isometries of all non-split blocks and $A'$ the corresponding set. Then $\varepsilon_A(\tilde\Psi_x;\chi)=\varepsilon_{A'}(\Theta^{(\varepsilon)};\chi)$ for every $\varepsilon$, including $\varepsilon=0$, where the same removal applies. The super-sites of the spine need no removal, even if $A$ splits them, since their factor $E^{\otimes m}\mathcal W$ does not depend on $\varepsilon$ and Lemma~\ref{lem:F-fine0} holds for every fine $A$ with $3\le|A|\le n-3$.

The vectors $\Theta^{(\varepsilon)}$ and $\Theta^{(0)}$ differ only in the isometries of the $j\le2$ split blocks, so $\braket{\Theta^{(0)}}{\Theta^{(\varepsilon)}}=(1-\varepsilon)^{j/2}$ and $\lVert\Theta^{(\varepsilon)}-\Theta^{(0)}\rVert^2=2(1-(1-\varepsilon)^{j/2})\le2\varepsilon$. As each register $B'_b$ carries $\C^G=(\C^d)^{\otimes2}$, two qudits, Lemma~\ref{lem:kyfan} applies to $\Theta^{(\varepsilon)}$ and $\Theta^{(0)}$: the square root of a $\chi$-tail is $1$-Lipschitz, so
\begin{equation*}
\sqrt{\varepsilon_A(\tilde\Psi_x;\chi)}\ge\sqrt{\varepsilon_{A'}(\Theta^{(0)};\chi)}-\sqrt{2\varepsilon}=\sqrt{\varepsilon_A(\tilde\Psi^{(0)}_x;\chi)}-\sqrt{2\varepsilon}\ge\sqrt{1-\chi/d^3}-\sqrt{2\varepsilon}
\end{equation*}
by Lemma~\ref{lem:F-fine0}.

\emph{At least three split blocks.} Choose three split blocks $b_1,b_2,b_3$ and let $D$ consist of these three items, so that $D'$ consists of their nine super-sites. Measure every physical qudit outside $D'$ in the computational basis. Iterating step~(2) of the proof of Lemma~\ref{lem:E-deletion} over these qudits, as its step~(3) does, gives $\varepsilon_A(\tilde\Psi_x;\chi)\ge\min_w\varepsilon_A(\ket w\otimes\tilde\Psi_w/\lVert\tilde\Psi_w\rVert;\chi)$, the minimum over the strings $w$ on these qudits with $\tilde\Psi_w:=(\bra w\otimes1)\tilde\Psi_x\ne0$. On a super-site $s\notin D'$, a string $(t+a_1,t+a_2,t+a_1+a_2,t)$ occurs in exactly one codeword, $\ket{\varphi_a}$, and every other string in none (proof of Lemma~\ref{lem:F-code}(a)). So the part $w_s$ of $w$ on $s$ has $\bra{w_s}E=d^{-1/2}\bra a$ or $\bra{w_s}E=0$. Hence, as $\tilde\Psi_w\ne0$, $\tilde\Psi_w=d^{-(n'-9)/2}E^{\otimes D'}\Psi_z$, where $z\in G^{[n']\setminus D'}$ collects these values $a$ and $\Psi_z$ is the vector of step~(3) of the proof of Lemma~\ref{lem:E-deletion}. By its step~(4), $\Psi_z$ is a multiple of $Z^y\Psi_x[D]$ for a vector $y\in G^{D'}$. Here $\Psi_x[D]=\bigotimes_{i\le3}\tilde\psi_{x_{b_i}}$, as $D'$ holds no spine site and every edge of $E_x$ has an end in the spine. By Lemma~\ref{lem:F-code}(b), $E^{\otimes D'}Z^y=\bar Z^yE^{\otimes D'}$, where $\bar Z^y$ acts as $\bar Z^{y_s}$ on each super-site $s$ of $D'$, a product of single-qudit unitaries. So $\tilde\Psi_w/\lVert\tilde\Psi_w\rVert$ is, up to a phase, $\bar Z^y\tilde\Psi_x[D]$ with the unit vector $\tilde\Psi_x[D]:=\bigotimes_{i\le3}E^{\otimes3}\tilde\psi_{x_{b_i}}$ on the $36$ qudits of $D'$. As $\ket w$ is a product vector, the reduced state of $\ket w\otimes\bar Z^y\tilde\Psi_x[D]$ on $A$ is a pure state on $A\setminus D'$ tensored with a unitary conjugate of $\rho_{A\cap D'}(\tilde\Psi_x[D])$, so $\varepsilon_A(\tilde\Psi_x;\chi)\ge\varepsilon_{A\cap D'}(\tilde\Psi_x[D];\chi)$.

Across $A\cap D'$ the Schmidt coefficients of a product are the products of the Schmidt coefficients of the factors across $A\cap\tilde B_{b_i}$. At $\varepsilon=0$ each factor is $E^{\otimes3}\GHZ_G$, whose spectrum across the nonempty proper subset $A\cap\tilde B_{b_i}$ is flat of rank at least $d$ by Lemma~\ref{lem:F-ghz}. So the product $\tilde\Psi_x[D]$ at $\varepsilon=0$ has a flat spectrum across $A\cap D'$ of rank at least $d^3$ and $\chi$-tail at least $1-\chi/d^3$.

Finally $\lVert\bigotimes_iE^{\otimes3}\tilde\psi^{(\varepsilon)}_{x_{b_i}}-\bigotimes_iE^{\otimes3}\tilde\psi^{(0)}_{x_{b_i}}\rVert^2=2(1-(1-\varepsilon)^{3/2})\le3\varepsilon$, since $\braket{\tilde\psi^{(0)}_{x_b}}{\tilde\psi^{(\varepsilon)}_{x_b}}=\sqrt{1-\varepsilon}$ by step \emph{Distance} of the proof of Lemma~\ref{lem:E-perturb}. So the reduction to $\tilde\Psi_x[D]$ and Lemma~\ref{lem:kyfan}, on the $36$ qudits of $D'$, give $\sqrt{\varepsilon_A(\tilde\Psi_x;\chi)}\ge\sqrt{\varepsilon_{A\cap D'}(\tilde\Psi_x[D];\chi)}\ge\sqrt{1-\chi/d^3}-\sqrt{3\varepsilon}$.

In both cases $\sqrt{\varepsilon_A(\tilde\Psi_x;\chi)}\ge\bigl(\sqrt{1-\chi/d^3}-\sqrt{3\varepsilon}\bigr)_+$, as $\sqrt{2\varepsilon}\le\sqrt{3\varepsilon}$ and square roots are nonnegative. Squaring gives the claim.
\end{proof}

\begin{proof}[Proof of Theorem~\ref{thm:lower-k}]
(a) Lemma~\ref{lem:F-transfer} gives the unit vectors, the membership, $\tilde T_x^{(2)}=\tilde T_x\in\Xn(s)$ for $s\ge4$, the distinct trees, the overlaps, and everything about coarse sets: clusters have tail $0$, the $\varepsilon$-type sets have tail exactly $\theta\varepsilon$, and the other nonempty proper non-cluster coarse sets have tail at least $\theta\varepsilon$, because $\varepsilon\le\bar\varepsilon'\le\bar\varepsilon$.

A fine set $A$ with $3\le|A|\le n-3$ has tail at least $\bigl(\sqrt{1-\chi/d^3}-\sqrt{3\varepsilon}\bigr)_+^2$ by Lemma~\ref{lem:F-fine}, as $\varepsilon\le\bar\varepsilon'<1$. This bound is at least $\theta\varepsilon$ if and only if $\sqrt{1-\chi/d^3}\ge\sqrt{3\varepsilon}+\sqrt{\theta\varepsilon}$, that is, if and only if $\varepsilon\le(1-\chi/d^3)/(\sqrt3+\sqrt\theta)^2$, which holds.

The family $\Nfam^{(2)}_s(\tilde T_x)$ consists of non-cluster sets with $3\le|A|\le n-3$, and it contains $\mathcal E_x$. Indeed, the members of $\mathcal E_x$ are unions of two super-sites or complements thereof, with $8$ or $n-8$ elements. They lie in $\Nfam_2(T_x)$ by Lemma~\ref{lem:E-epstype}, hence in $\Nfam_2(\tilde T_x)$, as the coarse clusters of $\tilde T_x$ are the clusters of $T_x$ (construction of $\tilde T_x$ in Section~\ref{sec:largebond}). Lemma~\ref{lem:A-trees}(f) gives $\Nfam_2(\tilde T_x)\subseteq\Nfam_s(\tilde T_x)$. So $\gapsk(\tilde\Psi_x,\tilde T_x)=\theta\varepsilon$ for every $s\ge2$. Hence $\tilde\Psi_x$ is in the class $\mathcal S^{d,\chi}_{n,s,2}(\theta\varepsilon)$ with target $\tilde T_x$ for $s\ge4$.

(b) Put $n':=n/4\ge4$ and $\varepsilon:=\varepsilon_0/\theta\le\bar\varepsilon'$. Then part (a) puts every $\tilde\Psi_x$ in the class with target $\tilde T_x$, which the procedure must output with probability at least $1-\delta$, and the trees $\tilde T_x$ are pairwise distinct. The vectors $\tilde\Psi_x$ have $\braket{\tilde\Psi_x}{\tilde\Psi_y}=(1-\varepsilon)^{\dH(x,y)}$, so Proposition~\ref{prop:E-assouad} gives $1-\delta\le\PH((1-\varepsilon)^N)^K$ and $N\ge N_K(\varepsilon,\delta)$, with $K=\lfloor n'/4\rfloor=\lfloor n/16\rfloor\ge1$. For $\delta\le\frac13$, Lemma~\ref{lem:E-constants}(ii) with $n'\in\{4K,\dots,4K+3\}$ gives $N\ge c_0\ln(n'/\delta)/(2\ln\frac1{1-\varepsilon})$, and $\ln(n'/\delta)=\ln(n/\delta)-\ln4\ge\frac12\ln(n/\delta)$ because $n/\delta\ge16$. Finally $\ln\frac1{1-\varepsilon}\le\varepsilon/(1-\varepsilon)\le\varepsilon_0/(\theta(1-\bar\varepsilon'))$. So
\begin{equation*}
N\ \ge\ \frac{c_0\ln(n'/\delta)}{2\ln\frac1{1-\varepsilon}}\ \ge\ \frac{c_0\cdot\frac12\ln(n/\delta)}{2\varepsilon_0/(\theta(1-\bar\varepsilon'))}\ =\ \frac{c_0\,\theta\,(1-\bar\varepsilon')}{4}\cdot\frac{\ln(n/\delta)}{\varepsilon_0}.
\end{equation*}
\end{proof}

\section*{Acknowledgments}
This research received no external funding.

\section*{Author contributions}
The author conceived the research, developed the theoretical framework, and derived the results. Large language models were used for literature research, text generation, and editorial refinement. Remote viewing techniques were used as a personal exploratory practice to inform problem selection, research direction, and decision-making. All mathematical results were established through rigorous mathematical reasoning and are independently verifiable.

\section*{Data availability}
This work is theoretical and uses no data.
 
\bibliographystyle{quantum}

\end{document}